\documentclass[10pt,
nofootinbib,           
amsmath,amssymb,
aps,
prb,
twocolumn,
floatfix, 
superscriptaddress]{revtex4-2}
\usepackage{color}
\usepackage{mathtools}
\usepackage[table]{xcolor}
\usepackage{minitoc}
\usepackage{times}
\usepackage{cases}
\usepackage{gensymb}
\usepackage{booktabs}
\usepackage{graphicx}
\usepackage{placeins}
\usepackage{dcolumn}
\usepackage{bm}
\usepackage[T1]{fontenc}
\usepackage{multirow}
\usepackage{hyperref}
\usepackage{natbib}

\hypersetup{
    colorlinks=true,
    linkcolor=red,      
    citecolor=blue,      
    urlcolor=blue       
}

\hypersetup{linktocpage,colorlinks=true,citecolor=blue}

\makeatletter
\AtBeginDocument{\let\LS@rot\@undefined}
\makeatother

\def \be {\begin{equation}}
\def \ee {\end{equation}}
\def \bee{\begin{equation*}}
\def \eee{\end{equation*}}

\def \bvec {\left( \begin{array}{c}}
	\def \evec {\end{array}\right)}

\def \l1{\bar{1}} 

\newcommand{\nn}{\nonumber \\}

\newcommand{\xfilll}[2][1ex]{%
	\dimen0=#2\advance\dimen0 by #1%
	\leaders\hrule height \dimen0 depth -#1\hfill%
}

\definecolor{deeppink}{rgb}{1.0, 0.08, 0.58}
\definecolor{brilliantrose}{rgb}{1.0, 0.33, 0.64}
\definecolor{darkorange}{rgb}{1.0, 0.55, 0.0}
\definecolor{darkpastelpurple}{rgb}{0.59, 0.44, 0.84}
\definecolor{red}{rgb}{1, 0.0, 0.0}

\DeclareFontFamily{U}{mathb}{\hyphenchar\font45}
\DeclareFontShape{U}{mathb}{m}{n}{
      <5> <6> <7> <8> <9> <10> gen * mathb
      <10.95> mathb10 <12> <14.4> <17.28> <20.74> <24.88> mathb12
      }{}
\DeclareSymbolFont{mathb}{U}{mathb}{m}{n}
\DeclareFontSubstitution{U}{mathb}{m}{n}
\DeclareMathSymbol{\circlearrowleft}       {3}{mathb}{"F6}
\DeclareMathSymbol{\circlearrowright}      {3}{mathb}{"F7}
\DeclareMathSymbol{\curvearrowleftright}   {3}{mathb}{"F2}

\newcommand\inter{\curvearrowleftright}

\newcommand{\intra}{{\mathchoice
    {
        \hspace{-0.7ex}
        {\mathrel{{\ooalign{\hss\raisebox{1.2ex}{
            \rotatebox{180}{$\circlearrowright$} 
            }\hss\cr\raisebox{1.2ex}{
            \rotatebox{180}{$\circlearrowleft$} 
            }}}}
        \hspace{-0.7ex}
        }
    }
    {
        \hspace{-0.7ex}
        {\mathrel{{\ooalign{\hss\raisebox{1.2ex}{
            \rotatebox{180}{$\circlearrowright$} 
            }\hss\cr\raisebox{1.2ex}{
            \rotatebox{180}{$\circlearrowleft$} 
            }}}}
        \hspace{-0.7ex}
        }
    }
    {
        \hspace{-1.5ex}
        {\mathrel{{\ooalign{\hss\raisebox{0.9ex}{
            \rotatebox{180}{
                \scalebox{0.7}{
                    $\circlearrowright$
                }
            } 
            }\hss\cr\raisebox{0.9ex}{
            \rotatebox{180}{
                \scalebox{0.7}{
                    $\circlearrowleft$
                }
            } 
            }}}}
        \hspace{-1.5ex}
        }
    }
    {
        \hspace{-1.3ex}
        {\mathrel{{\ooalign{\hss\raisebox{0.7ex}{
            \rotatebox{180}{
                \scalebox{0.6}{
                    $\circlearrowright$
                }
            } 
            }\hss\cr\raisebox{0.7ex}{
            \rotatebox{180}{
                \scalebox{0.6}{
                    $\circlearrowleft$
                }
            } 
            }}}}
        \hspace{-1.3ex}
        }
    }
}
} 

\newcommand{\nocontentsline}[3]{}
\let\origcontentsline\addcontentsline
\newcommand\stoptoc{\let\addcontentsline\nocontentsline}
\newcommand\resumetoc{\let\addcontentsline\origcontentsline}

\begin{document}

\title{Carrier-mediated spin helices in low-dimensional systems: \\ 
Josephson-interference features and correlation renormalization }

\author{Yung-Yeh Chang}
\affiliation{Institute of Physics, Academia Sinica, Taipei 115201, Taiwan}

\author{Chen-Hsuan Hsu}
\affiliation{Institute of Physics, Academia Sinica, Taipei 115201, Taiwan}
\affiliation{Physics Division, National Center for Theoretical Sciences, Taipei 106319, Taiwan}

\begin{abstract}

Carrier-induced helical magnetism can arise in interacting one-dimensional systems coupled to localized moments. Here, we investigate the coupled-wire sliding-Luttinger-liquid counterparts with and without valley degrees of freedom, using isolated Tomonaga--Luttinger liquids as reference limits. The carrier spin susceptibility mediates a Ruderman--Kittel--Kasuya--Yosida interaction that selects the helical ordering wave vector, while interchannel correlations lock the relative helix phases and stabilize a collective quasi-two-dimensional texture. Using a variational  treatment, we estimate the helix-induced partial gap and analyze the reconstruction of the remaining gapless carrier modes. To probe the spatial texture through transport, we incorporate the helix into an anisotropic superconductor--normal-metal--superconductor Josephson junction motivated by the coupled-channel structure of twisted bilayer WTe$_2$. 
We exploit flux focusing near the junction interfaces to make an applied in-plane field a controlled  probe of the spin helix. 
Relative to  junctions without the helix, the helical exchange field modifies the central-lobe width and relative lobe weights. In the transverse-channel geometry, it also produces field-dependent side-lobe asymmetry and central-peak displacement, which remain negligible in the matched references for the parameters considered.
The reconstructed carrier sector also exhibits modified local-density-of-states exponents, spin-relaxation behavior, and competing correlations, with distinct signatures in isolated and coupled-wire systems. Our results connect collective helical order in coupled-wire systems to carrier-sector reconstruction and identify field-tunable Josephson interferometry as a complementary transport probe of the spatial spin texture.

\end{abstract}

\maketitle

\section{Introduction}  

Twisted moir\'e materials have emerged as a promising platform for realizing one-dimensional (1D) electronic states within otherwise two-dimensional (2D) structures~\cite{San-Jose-TBG-network,Efimkin-2018-TBG-network,moire-DW-Ensslin,LeRoy-moire-DW-exp,Koshino-moireDW-2020,2022-Nature-Pengjie-1DTLL-in-2DMoire,Sanfeng2023:tWTe2,YiMing2024:tWTe2,Duarte:2026PNAS_collapseMoire,Koshino:2026PRB_tWTe2,Hsu:2026,Huynh:2DM2026}.  
A prominent example is twisted bilayer WTe$_2$ (tWTe$_2$), where experiments have revealed arrays of gapless 1D conducting channels with signatures consistent with coupled (Tomonaga--)Luttinger-liquid (TLL) behavior~\cite{2022-Nature-Pengjie-1DTLL-in-2DMoire,Sanfeng2023:tWTe2}. These channels form a 2D array of interacting 1D modes described at low energies as a sliding Luttinger liquid (SLL). On the hole-doped side, the relevant modes do not carry an additional valley degree of freedom~\cite{2022-Nature-Pengjie-1DTLL-in-2DMoire,Sanfeng2023:tWTe2,YiMing2024:tWTe2}. This system therefore provides a natural valleyless counterpart to twisted bilayer graphene (TBG) under bias, where the conducting channels emerging between different local stacking configurations retain valley degrees of freedom~\cite{San-Jose-TBG-network,Efimkin-2018-TBG-network,Koshino-moireDW-2020,moire-DW-Ensslin,LeRoy-moire-DW-exp,PhilipKim-moire-DW,Xu-moireDW-GiantOscil,Carr-DW-hetero,Peeters-PRMater,Fleischmann-NanoLett-2020,Ren-PRB-Metallic-network,Walet-2020,Verbakel-PRB,HCWang-TBG}.

More broadly, systems hosting interacting 1D carrier modes can be classified according to two key characteristics: whether the modes form an isolated 1D channel or a 2D coupled-wire array, and whether an additional valley degree of freedom is present. This leads naturally to four classes: a TLL, a valleyful TLL, an SLL, and a valleyful SLL. Representative realizations include III--V semiconductor quantum wires~\cite{Auslaender:2005Science,Tserkovnyak:2003PRL,Tserkovnyak:2003PRB,Jompol:2009Science}, carbon nanotubes (CNTs)~\cite{Kane:1997PRL_nanotube,Egger:1997PRL_nanotube,Egger:1998EPJ_nanotube,Bockrath:1999Sci_nanotube,Yao:1999Nature,Ishii:2003Science}, tWTe$_2$~\cite{2022-Nature-Pengjie-1DTLL-in-2DMoire,Sanfeng2023:tWTe2,YiMing2024:tWTe2,Lian:2024PRB_tWTe2,StrainWTe2:2024}, 
and the domain-wall network of voltage-biased TBG~\cite{Cenke-PRB-moire-coupledWire,YCChou-PRB-moire-network,castro-TBG-network-PRB,Hsu:2023,Hsu:2026}, respectively. 
Despite their distinct microscopic origins, these systems share closely related bosonic low-energy descriptions.  

Coupling these interacting carriers to localized moments introduces an additional level of collective behavior. The finite-momentum spin susceptibility of the carriers mediates an indirect Ruderman--Kittel--Kasuya--Yosida (RKKY) interaction between the localized moments. In isolated 1D systems, the enhanced spin response of a TLL is known to favor helical magnetic order at low temperatures~\cite{Braunecker-2009,Braunecker-PRL-2009-C13,CHH-CNT-helix}. More recently, this mechanism was extended to the valleyful coupled-wire network of voltage-biased TBG, where interchannel correlations lock the relative helix phases and stabilize a collective quasi-2D texture~\cite{YYC-2Dhelix-2025}. Here, we take the isolated TLL systems as reference limits and treat valleyless and valleyful SLL systems within a common framework, focusing on the static reconstruction of the carrier sector produced by the established helical order. This comparison allows us to distinguish the effects arising in coupled wires from those already present in isolated 1D systems.

These developments motivate a comparison of how dimensionality and valley structure modify the RKKY-selected ordering wave vector, characteristic helix-ordering scale, helix-induced partial gap, and remaining gapless modes. A complementary objective is to identify the spatially rotating texture through experimentally accessible signatures, 
which  
is particularly important because a spin helix can have a small or vanishing spatially averaged magnetization, making its onset difficult to resolve through bulk magnetometry alone. 

Josephson interference provides a complementary approach. In a Josephson weak link, the field-dependent critical current is governed by the phase accumulated across the junction and encodes the spatial distribution of the supercurrent~\cite{BaronePaterno:1982,DynesFulton:1971}. When the helical region forms the weak link, its spatially rotating exchange field can reshape this interference response, consistent with the broader sensitivity of Josephson transport to magnetic exchange fields and nonuniform magnetic textures~\cite{Buzdin:2005RMP,Bergeret:2005RMP,Wu:2015PRB}.
In the present setting, the resulting features provide information about the coupled-channel texture and its orientation relative to the supercurrent. To obtain a field-tunable probe of the texture, we exploit flux focusing near the interfaces between the normal region and the superconducting leads, which converts an applied in-plane field into a nonuniform out-of-plane flux profile~\cite{FlensbergPRB:2017_AnomalousFraunhoferPattern,YCTsai:2026arxiv}. The out-of-plane field scans the Josephson interference pattern, while the in-plane field controls the additional orbital-phase accumulation near the interfaces. Tracking the resulting evolution of the lobe structure therefore provides a systematic way to assess the helical contribution beyond a single distorted interference pattern.
Josephson interferometry therefore offers a transport probe of this spatially rotating,  helical magnetic state complementary to spectroscopic and magnetic resonance measurements.

In this work, we develop a  low-energy framework for all four classes. We use the valleyless 2D coupled-wire system as our primary formulation, since it retains the essential SLL physics. The valleyful 2D system follows by introducing an additional index, whereas the corresponding 1D systems are obtained in the single-channel limit and serve as reference cases.  
Within this framework, we first determine the charge and spin responses, with the spin susceptibility generating the carrier-mediated RKKY interaction and selecting the helical ordering wave vector. We then analyze the formation of the spin helix and its static feedback on the itinerant carriers. Complementing previous renormalization-group analyses, we estimate the helix-induced partial gap within a variational strong-coupling treatment, which is consistent with the corresponding   scaling form.
 
We next examine how the resulting spatial texture can be detected through transport.
We incorporate the static helix into an anisotropic superconductor-normal metal-superconductor (SNS) Josephson junction motivated by the coupled-channel geometry of  tWTe$_2$ and consider channels oriented either parallel or perpendicular to the supercurrent. Relative to the corresponding junctions without the helix formation, the helical exchange field modifies the central-lobe width, redistributes the weight between the central and side lobes, and generates an asymmetry between the positive- and negative-flux side lobes. These features depend strongly on the coupled-channel and helix orientation and evolve systematically with the applied in-plane field. 

Finally, we examine the complementary consequences of the static carrier reconstruction. The helix-induced partial gap and the renormalization of the remaining gapless modes modify the local density of states, spin-relaxation behavior, and hierarchy of competing correlations. The isolated and coupled-wire systems exhibit distinct low-temperature exponents, thereby separating the effects of interchannel correlations and valley structure. Our results establish coupled-wire carrier--localized-spin systems as a broad setting for RKKY-driven collective helical order, whose carrier-sector reconstruction and spatial magnetic texture can be probed through complementary correlation and Josephson-interference signatures.

The remainder of the paper is organized as follows. Section~\ref{sec:model} introduces the four classes of interacting 1D carrier systems and their  bosonized description, together with the corresponding charge and spin responses. Section~\ref{sec:1DKondolattice} develops the carrier-mediated RKKY interaction, the resulting helical order and spin-wave excitations, and the variational treatment of the helix-induced partial gap. Section~\ref{sec:hx-JJ} investigates the Josephson-interference signatures of the helical texture, while Secs.~\ref{sec:Observable} and~\ref{sec:helix_induced_order} examine the resulting carrier-sector observables and competing correlations. Finally, Sec.~\ref{sec:Conclusions} summarizes the main results and discusses their broader implications.
Calculation details are provided in the appendices. Appendix~\ref{app:constraint_lambda} discusses the stability conditions of the SLL. Appendix~\ref{app:LDOS} presents the unfolding method used   in the coupled-wire system. Appendix~\ref{sec:order-parameter} gives the bosonized expressions for the order-parameter operators.

\section{Models for 1D wire and 2D coupled-wire systems}
\label{sec:model} 

We begin by establishing the low-energy models for the itinerant-carrier subsystems of the four classes considered in this work: (1) a 2D coupled-wire system, (2) a valleyful 2D coupled-wire system, (3) a 1D wire, and (4) a valleyful 1D wire. To avoid repeating closely related constructions, we first introduce a unified low-energy description and subsequently specialize it to the individual platforms within the bosonized representation.

For a single conducting channel, we denote the low-energy fermion fields generically by $\psi_{\ell,\Xi}(r)$, where $r$ is the coordinate along the channel and $\ell=+(-)$ labels right- (left-) moving carriers. The collective index $\Xi$ collects the wire index and internal degrees of freedom. The itinerant-carrier Hamiltonian then takes the generic form
\begin{align}
H_{\rm it}
=&
\sum_{\ell,\Xi}
\int dr\,
\psi_{\ell,\Xi}^{\dagger} (r) 
\left(-i\hbar v_F\ell\partial_r\right)
\psi_{\ell,\Xi}(r)
\nn
&+ \frac{1}{2}
\int dr\,dr^{\prime}\,
U(|r-r^{\prime}|)
\rho(r)\rho(r^{\prime}),
\label{eq:H_ee-general}
\end{align}
where the first term describes the linearized carrier dispersion with Fermi velocity $v_F$, while the second describes the screened Coulomb interaction $U$ between carrier densities $\rho $. Here, we use $v_F$ as a common notation for all systems discussed below.

We now specify the collective index $\Xi$: $\Xi=(\sigma,m)$ for a 2D coupled-wire system, $\Xi=(\sigma,\delta,m)$ for a valleyful 2D coupled-wire system, $\Xi=\sigma$ for a 1D wire, and $\Xi=(\sigma,\delta)$ for a valleyful 1D wire. Here, $\sigma\in\{\uparrow,\downarrow\}$ labels spin, $\delta\in\{-,+\}$ labels the two low-energy branches associated with the valley degree of freedom, and $m \in \{ 1,\ldots,N_{\rm ch} \}$ labels the 1D conduction channels, with the number of channels, $N_{\rm ch}$.
As introduced above, these four theoretical systems can be associated with representative material platforms, with their characteristic parameters  summarized in
Table~\ref{Table:system-parameters}.

\begin{table*}[t]
\centering
\caption{Representative realizations and characteristic parameters for the
four classes of interacting carrier systems considered in this work.
Here, $\Delta_a$
is the carrier bandwidth and $J_K$ denotes the coupling to localized moments.}
\label{Table:system-parameters}
\small
\setlength{\tabcolsep}{6pt}
\renewcommand{\arraystretch}{1.25}

\begin{tabular}{c c c c c c}
\hline\hline
Class
&
Carrier geometry
&
Valley
&
Representative system
&
$\Delta_a$
&
$J_K$$^{\rm {\color{blue}a}}$
\\
\hline
1D
&
isolated
&
no
&
GaAs quantum wire
&
$0.23\,\mathrm{eV}$
&
$90\,\mu\mathrm{eV}$
\\
1D with valley (1d+v)
&
isolated
&
yes
&
$^{13}\mathrm{C}$ CNT
&
$2.1\,\mathrm{eV}$
&
$0.6$--$6\,\mu\mathrm{eV}$$^{\rm {\color{blue}b}}$
\\
2D
&
coupled-wire
&
no
&
hole-doped tWTe$_2$
&
$\sim 10 \,\mathrm{meV}$$^{\rm {\color{blue}c}}$
&
$O(0.1\text{--}1)\,\mathrm{meV}$$^{\rm {\color{blue}d}}$
\\
2D with valley (2d+v)
&
coupled-wire
&
yes
&
 TBG
&
$2$--$10\,\mathrm{meV}$$^{\rm {\color{blue}e}}$
&
$\sim1.25\,\mathrm{meV}$$^{\rm {\color{blue}f}}$
\\
\hline\hline
\end{tabular}

\vspace{0.4em}

\noindent
\begin{minipage}[t]{0.98\textwidth}
\raggedright
\footnotesize
$^{\rm {\color{blue}a}}$ Including either exchange or hyperfine couplings.

$^{\rm {\color{blue}b}}$ Estimated from
Refs.~\cite{Pennington:1996RMP,Churchill2009-naturephysics,
Churchill:2009PRL,Fischer2009}.

$^{\rm {\color{blue}c}}$ Estimated from
Ref.~\cite{YiMing2024:tWTe2}.

$^{\rm {\color{blue}d}}$ In our analysis, we adopt $J_K=O(0.1\text{--}1)~\rm{meV}$ as a representative exchange-coupling range, partially motivated by experiments on bulk WTe$_2$~\cite{Fominykh:2026FedopedWTe2,Manna:2026WTe2TK}. 
We remark that this should, however, be regarded only as an order-of-magnitude estimate.

$^{\rm {\color{blue}e}}$ Estimated from
Ref.~\cite{HCWang-TBG}. 

$^{\rm {\color{blue}f}}$ Adopted as a representative effective coupling for magnetic adatoms, motivated by the meV-scale exchange inferred for fluorinated graphene~\cite{F-adatom-graphene}. 

\end{minipage}
\end{table*}

\subsection{Bosonic description}
\label{eq:bosonic-description}

We next formulate the four systems in their bosonized representation. We begin with the valleyless 2D coupled-wire system, which corresponds to the standard SLL~\cite{Emery-sLL-2000,Mukhopadhyay-sLL-PRB-2001,Mukhopadhyay-csLL-PRB,Carpentier-sLL-2001}. 
Its itinerant-carrier Hamiltonian is
\begin{align}
H_{\rm it}^{\rm (2d)}
=
&\frac{1}{N_{\rm ch}}
\sum_{q_{\perp},\nu}
\int\frac{\hbar dr}{2\pi}
\left[
\frac{v_{\nu}(q_{\perp})}{K_{\nu}(q_{\perp})}
\left|\partial_{r}\phi_{\nu,q_{\perp}}(r)\right|^{2}
\right.
\nn
&\hspace{1.5cm}
\left.
+
v_{\nu}(q_{\perp})K_{\nu}(q_{\perp})
\left|\partial_{r}\theta_{\nu,q_{\perp}}(r)\right|^{2}
\right],
\label{2Dcoupled_novalley_Hee}
\end{align}
where $\nu\in\{c,s\}$ labels the charge and spin sectors, and $\phi_{\nu,q_\perp}$ and $\theta_{\nu,q_\perp}$ are canonically conjugate bosonic fields. 
We impose periodic boundary conditions perpendicular to the channels and Fourier transform in this direction, with $q_\perp$ denoting the transverse momentum. 
We retain the spin-sector parameters $v_s$ and $K_s$ as general constants in the analytical derivations below. For numerical evaluations, we set $v_s=v_{\rm ch}^{\rm (2d)}$ and $K_s=1$, where $v_{\rm ch}^{\rm (2d)}$ denotes the velocity of the itinerant carriers.
The short-distance cutoff $a$ is related to the carrier bandwidth $\Delta_a$ by $a=\hbar v_{\rm ch}^{\rm (2d)}/\Delta_a$.
In contrast to the sliding Luther-Emery regime~\cite{YiMing2024:tWTe2}, where a pre-existing spin gap is  included, our analysis concerns the regime in which the carrier spin sector remains gapless before including localized moments in Sec.~\ref{sec:1DKondolattice}. 

We parametrize the   charge-sector interaction by
\begin{align}
K_{c}(q_{\perp})
=
K_{c}
\left[
1+\lambda_{1}\cos(q_{\perp}d)
+\lambda_{2}\cos(2q_{\perp}d)
\right],
\label{eq:tildeK_c}
\end{align}
where $d$ is the channel spacing, and $K_c<1$ corresponds to repulsive interactions in the isolated 1D limit. The decoupled-channel limit is recovered for $\lambda_{1,2}=0$. Stability of Eq.~\eqref{2Dcoupled_novalley_Hee} requires $K_c(q_\perp)>0$ throughout the transverse Brillouin zone; the corresponding constraints for $\lambda_1$ and $\lambda_2$ are discussed in Appendix~\ref{app:constraint_lambda}. In evaluating the correlation functions and observables below, we make an additional approximation $v_c(q_\perp)\approx v_{\rm ch}^{\rm (2d)}$, while retaining the $q_\perp$ dependence  in
$K_c(q_\perp)$.

The valleyful 2D coupled-wire system follows by organizing the bosonic fields into symmetric and antisymmetric combinations, labeled by $P\in\{S\,({\rm symmetric}),\, A\,({\rm antisymmetric})\}$. The Hamiltonian then becomes~\cite{YYC-2Dhelix-2025}
\begin{align}
H_{\rm it}^{\rm (2d+v)}
=
&\frac{1}{N_{\rm ch}}
\sum_{q_{\perp},\nu,P}
\int\frac{\hbar dr}{2\pi}
\left[
\frac{v_{\nu P}(q_{\perp})}{K_{\nu P}(q_{\perp})}
\left|\partial_{r}\phi_{\nu P,q_{\perp}}(r)\right|^{2}
\right.
\nn
&\hspace{1.5cm}
\left.
+
v_{\nu P}(q_{\perp})K_{\nu P}(q_{\perp})
\left|\partial_{r}\theta_{\nu P,q_{\perp}}(r)\right|^{2}
\right], 
\label{eq:2Dcoupledvalley_Hee}
\end{align}
where the charge-symmetric sector is described by 
\begin{align}
K_{cS}(q_{\perp})
=
K_{cS}
\left[
1+\lambda_{1}\cos(q_{\perp}d)
+\lambda_{2}\cos(2q_{\perp}d)
\right],
\label{eq:tildeK_cs}
\end{align}
while the other sectors are assumed to be independent of $q_\perp$, as in the valleyless counterpart.    Thus, the valleyful coupled-wire system is an extension of the SLL framework, with the symmetric and antisymmetric sectors providing the additional low-energy degrees of freedom.

The 1D systems follow directly by taking the decoupled-channel limit of Eqs.~\eqref{2Dcoupled_novalley_Hee} and \eqref{eq:2Dcoupledvalley_Hee}, thereby removing the transverse momentum $q_\perp$. For the valleyless case, this yields the standard spinful TLL Hamiltonian~\cite{Haldane1981a,Voit:1995ROPP,VonDelft1998,Giamarchi2003,Senechal2006},  
\begin{align}
H_{\rm it}^{\rm (1d)}
=
\sum_{\nu}
\int\frac{\hbar dr}{2\pi}
\left[
\frac{v_{\nu}}{K_{\nu}}
\left(\partial_r\phi_{\nu}\right)^2
+
v_{\nu}K_{\nu}
\left(\partial_r\theta_{\nu}\right)^2
\right].
\label{eq:1DHee_novalley}
\end{align}

For a valleyful 1D system, the additional symmetric and antisymmetric sectors are retained, giving
\begin{align}
H_{\rm it}^{\rm (1d+v)}
=
\sum_{\nu,P}
\int\frac{\hbar dr}{2\pi}
\left[
\frac{v_{\nu P}}{K_{\nu P}}
\left(\partial_r\phi_{\nu P}\right)^2
+
v_{\nu P}K_{\nu P}
\left(\partial_r\theta_{\nu P}\right)^2
\right].
\label{eq:1Dvalley_Hee}
\end{align}
For CNTs, the Fermi momentum is measured relative to the two Dirac points, but this does not alter the quadratic bosonic Hamiltonian~\cite{Kane:1997PRL_nanotube,Egger:1997PRL_nanotube,Egger:1998EPJ_nanotube}.

The four systems are described within a common bosonic framework, differing in the presence or absence of transverse coupled-channel correlations and valley degrees of freedom.
In the following, we use the valleyless 2D coupled-wire system as the main example and present the corresponding results for all four systems in summary tables. The remaining cases follow systematically by introducing the valley degree of freedom and/or taking the single-channel limit. 
The underlying physics, however, is not restricted to any particular realization and applies broadly across the four classes of platforms considered here.

The charge and spin responses are determined by the corresponding density--density and spin--spin correlation functions, respectively, and therefore provide direct access to the spatial and temporal correlations of the itinerant carriers.  In isolated TLLs, the charge  and spin correlation functions and their characteristic power-law behavior are well established within bosonization~\cite{Voit:2000AIPConf,Schulz:2000TLL,Gimarchi:2007ChargeCorrelation}. For coupled-wire systems, nonlocal correlations between different wires have also been considered. In particular, Ref.~\cite{KunYang:2017PRB} explicitly evaluated interwire density--density correlations in a valleyless SLL, while Ref.~\cite{Schulz:2020PRB} studied density correlations between interacting quantum wires in an intersecting-wire geometry. 
Here, we extend these considerations to the nonlocal spin response between spatially separated channels of  valleyful SLLs. 
Below we examine the static charge and spin response functions.

\subsection{Charge response}
\label{sec:charge_response}

The charge density operator in channel $m$ is
\begin{align}
\rho_m(r)
=
\sum_{\sigma}
\psi_{\sigma,m}^{\dagger}(r)
\psi_{\sigma,m}(r),
\end{align}
where $\psi_{\sigma,m}(r)=\sum_{\ell}\psi_{\ell\sigma,m}(r)$ for the valleyless 2D coupled-wire system considered here. The fermion bilinears can be separated into forward-scattering ($\ell^\prime=\ell$) and backscattering ($\ell^\prime\neq\ell$) contributions. We focus on the latter, which generate the singular finite-momentum response of interest, and omit the less singular $q\sim0$ contribution.

The static charge response between channels separated by $n$ channel spacings is determined by the retarded density-density correlation function
\begin{align}
\chi^{c}_{n}(r)
=
(-i)
\lim_{\epsilon\to0^{+}}
\int_{0}^{\infty}dt\,
e^{-\epsilon t}
\left\langle
\left[
\rho_{n}(r,t),
\rho_{0}(0,0)
\right]
\right\rangle_{\rm it},
\label{eq:chi_n}
\end{align}
where $\langle \cdots \rangle_{\rm it}$ denotes the average taken with respect to the itinerant-carrier subsystem. Here, the superscript $c$ denotes the charge response and distinguishes it from the spin response below. Different backscattering processes are characterized by their momentum transfer $2Q$, allowing the susceptibility to be decomposed as
\begin{align}
\chi^{c}_{n}(r)
=
\sum_Q
\chi^{c}_{n,2Q}(r).
\end{align}
In the low-temperature regime $\xi_T  \gg |r|$, where $\xi_T = \hbar v/(k_BT)$ is the thermal length and $v$ denotes the relevant itinerant-carrier velocity, each contribution exhibits the characteristic power-law behavior
\begin{align}
\chi^{c}_{n,2Q}(r)
\sim
\cos(2Qr)
\left|
\frac{a}{r}
\right|^{2g^0_{n}-1},
\label{eq:chi_c_r}
\end{align}
with the corresponding correlation exponent $g^0_n = (\bar{\Delta}_{\phi_c,n} + K_s)/2$ and the dimensionless quantities, 
\begin{subequations}
\label{eq:Delta_bar_phi_cS}
    \begin{align}
    \bar{\Delta}_{\phi_c,n} & = \int^\pi_{-\pi} \frac{d(q_\perp d)}{2\pi} {K}_c(q_\perp)\cos(nq_\perp d),\\
    \bar{\Delta}_{\theta_c,n} & = \int^\pi_{-\pi} \frac{d(q_\perp d)}{2\pi}\frac{\cos(nq_\perp d)}{{K}_c(q_\perp)}.
\end{align}
\end{subequations}

The resulting correlation exponents are listed in Table~\ref{tab:g}. More generally, at finite temperature the susceptibility obeys the scaling form
\begin{align}
\chi^{c}_{n,2Q}(r,T)
\sim
\cos(2Qr)
\left(
\frac{a}{\xi_T}
\right)^{2g^0_n-1}
\mathcal{F}_{g^0_n}
\left(
\frac{\pi|r|}{\xi_T}
\right) . 
\label{eq:chi_c_T}
\end{align}
The crossover function $\mathcal{F}_{g}(x)$ is given by 
\begin{align}
\mathcal{F}_{g}(x) = 
    \frac{_2 F_1[1/2,g;1;-\sinh^{-2}(x)]}{\sinh^{2g}(x)},
    \label{eq:hypergeometric}
\end{align}
where $_2 F_1[a,b;c;z]$ denotes the hypergeometric function. Substituting Eq.~\eqref{eq:hypergeometric} into Eq.~\eqref{eq:chi_c_T} recovers Eq.~\eqref{eq:chi_c_r} in the limit $x\ll1$. 

For valleyful systems, the finite-momentum contributions can be further classified as intrabranch ($\delta^\prime=\delta$) or interbranch ($\delta^\prime\neq\delta$) processes, labeled by $\mathcal{B}\in\{\intra,\inter\}$. The corresponding correlation exponents are denoted by $g^0_{\mathcal{B},n}$. For the 1D systems, the corresponding results follow from the single-channel limit $n=0$, in which $\bar{\Delta}_{\phi_c,0}=K_c$ for the valleyless case and $\bar{\Delta}_{\phi_{cS},0}=K_{cS}$ for the valleyful case. The correlation exponents for all four platforms are summarized in Table~\ref{tab:g}.

From an experimental perspective, electron energy-loss spectroscopy (EELS) provides access to the dynamical charge susceptibility $\chi^c(\bm q,\omega)$ and therefore offers a natural probe of the frequency- and momentum-dependent counterpart of the density correlations considered here~\cite{AbbamonteEELS:2017SciPost,AbbamonteEELS:2025AnnuRev}. The finite-momentum power laws derived above determine the corresponding low-energy scaling behavior and can thus provide signatures of the underlying TLL or SLL correlations.
EELS has been widely applied to low-dimensional systems~\cite{Fink:1998PRL_scSeparation,Nagao:2006PRL,Nagao:2008Nanotechnology,KrambergerEELS_CNT:2008PRL,HageEELS_CNT:2017PRL}, where characteristic signatures of 1D correlations have been observed. Examples include features associated with spin--charge separation in quasi-1D Sr$_2$CuO$_3$~\cite{Fink:1998PRL_scSeparation} and the linewidth broadening of 1D plasmon excitations, with electronic-correlation effects proposed as a mechanism in Au atomic-wire systems~\cite{Nagao:2006PRL,Nagao:2008Nanotechnology}. For the coupled-wire systems considered here, the additional dependence on transverse separation and momentum provides a route to probing interchannel density correlations beyond those of an isolated TLL.

\begin{table}[t]
\caption{A summary of   the correlation exponents and their explicit forms for the charge and spin density correlation functions between two wires at $n$th-neighbor separation in 2D coupled-wire systems. The definition of $\bar{\Delta}_{\phi_c,n}$ is given in Eq.~\eqref{eq:Delta_bar_phi_cS}, while $\bar{\Delta}_{\phi_{cS},n}$ is defined analogously, with $K_c(q_\perp)$ replaced by $K_{cS}(q_\perp)$.  
}
    \label{tab:g}
    \centering
    \setlength{\tabcolsep}{6pt} 
    \renewcommand{\arraystretch}{1.7} 
    \begin{tabular}{c c c}
        \hline\hline
         \begin{tabular}{@{}c@{}} 
         2D
         \end{tabular}
         & Exponent & 
         Explicit form   \\ 
         \hline
         ${\rm CDW}$  & $2g_{n}^{0}$ & $\bar{\Delta}_{\phi_{c},n}+K_{s}$   \\
         \hline
       ${\rm SDW}_{z}$ & $2g_{n}^{z} = 2g_{n}^{0}$ & $\bar{\Delta}_{\phi_{c},n}+K_{s} $ 
       \\
       \hline
       ${\rm SDW}_{x}$, ${\rm SDW}_{y}$ & $2g_{n}^{x} = 2g_{n}^{y}$ & $\bar{\Delta}_{\phi_{c},n}+K_{s}^{-1}$  
       \\
         \hline\hline
          \begin{tabular}{@{}c@{}}
          2D with 
          valley 
         \end{tabular}   &  Exponent  & Explicit form    \\ 
         \hline
       ${\rm CDW}_{\intra}$ & $2g_{\intra, n}^{0}$ & 
       \begin{tabular}{@{}c@{}}
       $\begin{aligned}
        &\tfrac{1}{2}\Big( \bar{\Delta}_{\phi_{cS},n}+K_{cA}\nn
       & \quad +K_{sS}+K_{sA} \Big)
       \end{aligned}$
       \end{tabular}
       \\
         \hline
         ${\rm CDW}_{\inter}$ & $2g_{\inter, n}^{0}$ & 
           \begin{tabular}{@{}c@{}}
       $\begin{aligned}
        &\tfrac{1}{2}\Big( \bar{\Delta}_{\phi_{cS},n}+K_{cA}^{-1}\nn
       & \quad +K_{sS}+K_{sA}^{-1} \Big)
       \end{aligned}$
       \end{tabular}
        \\
       \hline
        ${\rm SDW}_{\intra, z}$ & $2g_{\intra, n}^{z} = 2g_{\intra, n}^{0}$ & 
        \begin{tabular}{@{}c@{}}
       $\begin{aligned}
        &\tfrac{1}{2}\Big( \bar{\Delta}_{\phi_{cS},n}+K_{cA}\nn
       & \quad +K_{sS}+K_{sA} \Big)
       \end{aligned}$
       \end{tabular}
         \\
       \hline
       ${\rm SDW}_{\intra, x/y}$ & $2g_{\intra, n}^{x}=2g_{\intra , n}^{y}$ & 
       \begin{tabular}{@{}c@{}}
       $\begin{aligned}
        &\tfrac{1}{2}\Big( \bar{\Delta}_{\phi_{cS},n}+K_{cA}\nn
       & \quad +K_{sS}^{-1}+K_{sA}^{-1} \Big)
       \end{aligned}$
       \end{tabular}
       \\
       \hline
        ${\rm SDW}_{\inter, z}$ & $2g_{\inter, n}^{z} = 2g_{\inter, n}^{0}$ & 
         \begin{tabular}{@{}c@{}}
       $\begin{aligned}
        &\tfrac{1}{2}\Big( \bar{\Delta}_{\phi_{cS},n}+K_{cA}^{-1}\nn
       & \quad +K_{sS}+K_{sA}^{-1}
       \Big)
       \end{aligned}$
       \end{tabular}
         \\
       \hline
        ${\rm SDW}_{\inter, x/y}$ & $2g_{\inter, n}^{x}=2g_{\inter , n}^{y}$  &  
         \begin{tabular}{@{}c@{}}
       $\begin{aligned}
        &\tfrac{1}{2}\Big( \bar{\Delta}_{\phi_{cS},n}+K_{cA}^{-1}\nn
       & \quad +K_{sS}^{-1}+K_{sA}
       \Big)
       \end{aligned}$
       \end{tabular}
          \\
       \hline
       \hline
    \end{tabular}
\end{table}

\subsection{Spin response}

We next consider the spin response. The spin-density operator in channel $m$ is
\begin{align}
s_m^\mu(r)
=
\frac{1}{2}
\sum_{\sigma,\sigma^\prime}
\psi_{\sigma,m}^{\dagger}(r)
\sigma^\mu_{\sigma\sigma^\prime}
\psi_{\sigma^\prime,m}(r).
\end{align}
In the static limit, the spin response characterized by the diagonal susceptibility between channels separated by $n$ channel spacings is
\begin{align}
\chi_{n}^{s,\mu}(r)
=
(-i)
\lim_{\epsilon\to0^{+}}
\int_{0}^{\infty}dt\,
e^{-\epsilon t}
\left\langle
\left[
s_n^\mu(r,t),
s_0^\mu(0,0)
\right]
\right\rangle_{\rm it}.
\label{eq:chi_s_n}
\end{align}

The analysis proceeds analogously to that for the charge response. We retain the finite-momentum backscattering contributions and decompose them according to their momentum transfer $2Q$. In the low-temperature regime $\xi_T \gg |r|$, each contribution exhibits the power-law form
\begin{align}
\chi_{n,2Q}^{s,\mu}(r)
\sim
\cos(2Qr)
\left|
\frac{a}{r}
\right|^{2g_{n}^{\mu}-1},
\label{eq:chi_s_r}
\end{align}
where $g_n^\mu$ is the corresponding spin-correlation exponent. At finite temperature, the same power law crosses over at the thermal length $\xi_T$, analogous to the charge response discussed above. For valleyful systems, the contributions are further separated into intrabranch and interbranch processes, with exponents $g_{\mathcal{B},n}^{\mu}$, where $\mathcal{B}\in\{\intra,\inter\}$. The  correlation exponents for all four platforms are summarized in Table~\ref{tab:g}; the 1D results follow from the single-channel limit $n=0$.

Nitrogen-vacancy (NV) centers provide a promising route for probing such nonlocal spin responses at the nanoscale~\cite{Peter_NVcenter:2013,Rondin_NV:2014ROPP,SchirhaglNV:2014AnnuRev,DegenNV:2017RMP,Elio:2020PRBNV}. In a source-probe geometry, a magnetized tip generates a localized magnetic field that induces a spin polarization in the sample, while a spatially separated NV-center probe detects the resulting stray magnetic field. Within linear-response theory, this setup allows the nonlocal spin susceptibility to be inferred from the applied field profile and the induced magnetic signal. The separation $r$ in Eq.~\eqref{eq:chi_s_r} then corresponds to the source-probe separation, while the spatial resolution is controlled by the spatial profiles and stand-off distances of the source and probe.

The finite-momentum backscattering correlation exponents summarized in Table~\ref{tab:g} determine both the dominant momentum channel and the spatial range of the carrier-mediated RKKY interaction. We now use the corresponding susceptibilities to determine the magnetic texture selected by coupling the carriers to localized moments.

\section{Hybrid itinerant-carrier--localized-spin systems }
\label{sec:1DKondolattice}

\subsection{RKKY-induced helical magnetism}
\label{sec:RKKY} 

We now couple the itinerant-carrier systems introduced above to localized moments through an isotropic exchange interaction,
\begin{align}
H
&=
H_{\rm it}+H_{\rm K},
\nn
H_{\rm K}
&=\frac{J_K a}{N_{\perp}} 
\sum_{m}\int dr
\bm{s}_m(r)\cdot\bm{S}_m(r),
\label{eq:Kondo_H}
\end{align}
where $\bm{S}_m(r)$ is the spin density of a localized moment of magnitude $S$ at position $r$ in channel $m$. Here, $N_{\perp}$ denotes the number of localized moments within the transverse extent of a gapless carrier mode. For an isolated 1D system, the channel index $m$ is not needed. In the parameter regime considered here, the Kondo temperature is sufficiently low that Kondo screening can be neglected, and the dominant interaction between the localized moments is the carrier-mediated RKKY interaction. This regime has been studied in isolated 1D systems~\cite{Braunecker-PRL-2009-C13,Braunecker-2009,CHH-CNT-helix,2018-CHH-nuclear-2DTI,Hsu:2021} and was recently extended to a valleyful 2D coupled-wire system~\cite{YYC-2Dhelix-2025}. Here, we treat the valleyless and valleyful 1D and 2D cases within a common framework, as summarized in Table~\ref{Table:system-parameters}.

The RKKY interaction in the continuum limit reads 
\begin{align}
H_{\rm R}
=
\sum_{m,n}\sum_{\mu}
\int
\frac{dr\,dr^\prime}{N_{\perp}^{2}}\,
J_{n}^{\mu}(|r-r^\prime|)
S^{\mu}_{m+n}(r)S_m^{\mu}(r^\prime),
\label{eq:RKKY}
\end{align}
where $
J_n^\mu(r)
\propto
J_K^2\chi_n^{s,\mu}(r) $
is determined by the static spin susceptibility of the itinerant-carrier subsystem defined in Eq.~\eqref{eq:chi_s_n}. The RKKY interaction therefore inherits the interaction dependence of the carrier spin response and, particularly in gate-tunable moir\'e platforms, can be controlled by modifying the electronic properties.

The dominant finite-momentum backscattering contributions  give rise to pronounced extrema of $\chi_n^{s,\mu}$ at characteristic momenta  $2Q$. The resulting RKKY interaction favors helical ordering of the localized moments at the corresponding wave vector. This mechanism gives rise to spin helices or spirals in isolated 1D systems~\cite{Braunecker-PRL-2009-C13,Braunecker-2009, CHH-CNT-helix,2018-CHH-nuclear-2DTI,Hsu:2021} and to phase-coherent helical order across the channels in 2D coupled-wire systems~\cite{YYC-2Dhelix-2025}. For the latter, the ordered magnetization can be written as
\begin{align}
&\bm{m}_{{\rm hx},m}(r,T)
=
m_{\rm hx}(T)
\frac{N_{\perp}S}{a}
\bm{n}(r),
\nn 
{\rm and} \quad & \bm{n}(r) =
\hat{x}\cos(2Qr)
+
\hat{y}\sin(2Qr),
\label{eq:helix}
\end{align}
where $0\leq m_{\rm hx}(T)\leq1$ is the normalized helix magnetization.
The ordering wave vector $2Q$ is selected by the dominant extremum of the spin susceptibility and is therefore platform dependent.
For the coupled-wire parameter regime considered here, the dominant transverse component of the RKKY interaction occurs at $q_\perp=0$. The corresponding energy is minimized when the helices on different channels share the same phase, producing the phase-locked configuration in Eq.~\eqref{eq:helix}, with $\bm{n}(r)$  independent of $m$.

Expanding about the ordered state by performing a local spin rotation and a Holstein-Primakoff transformation yields the quadratic spin-wave Hamiltonian $H_{\rm mag}$ and the magnon dispersion. For the nanoscale systems considered here,  the relevant magnon branch is characterized by the energy scale
\begin{align}
\hbar\omega_{0}
\equiv
\frac{S\left|J^x_{q_\perp=0}(2Q)\right|}{2N_{\perp}},
\label{eq:omega_0}
\end{align}
set by the RKKY scale, with $J^\mu_{q_\perp} (q)$ being the Fourier component of $J^\mu_{n} (r)$. 
The characteristic helix-ordering scale\footnote{Throughout this work, $T_{\rm hx}$ denotes the characteristic finite-size helix-ordering scale determined from the thermal reduction of $m_{\rm hx}(T)$,
rather than a thermodynamic transition temperature of an infinite strictly 1D or spin-rotation-invariant 2D system.}
$T_{\rm hx}$ is determined self-consistently from the reduction of the ordered magnetization by thermally excited magnons~\cite{Braunecker-2009,Braunecker-PRL-2009-C13,Meng-strip-EuroPhys}. In the low-temperature limit,  numerical fitting yields  $m_{\rm hx}(T)=1-(T/T_{\rm hx})^{2 \alpha^{\rm }}$  with $\alpha$ being a nonuniversal fractional exponent specific to the 2D coupled-wire system.  

The helix-ordering scale also depends on the concentration   of the localized moments. If the random local-moment occupation is represented by a homogeneous reduction of the effective local-spin density by a factor $p_s < 1$, it reduces the RKKY interaction and hence the helix-ordering scale. The precise dependence of $T_{\rm hx}$ on $p_s$ is generally nonuniversal, since dilution can also modify the RKKY coupling and the magnon spectrum. A related power-law dependence on the isotope concentration was discussed for isolated CNTs~\cite{CHH-CNT-helix}.

For 2D coupled-wire systems, the spin helix maps onto a ferromagnetic configuration in the locally rotated frame. Their spin-wave theories therefore share the same single-branch structure, with platform dependence entering primarily through the microscopic parameters and RKKY couplings. By contrast, the antiferromagnetic configuration relevant to CNTs gives rise to two magnon branches and can be separately treated.

The ordered moments act back on the itinerant carriers as an effective rotating Zeeman field,
\begin{align}
H_{\rm hx}
=
B_{\rm hx}
\sum_m
\int dr\,
\bm{n}(r)\cdot\bm{s}_m(r),
\label{eq:H_hx}
\end{align}
where $B_{\rm hx}\equiv J_Km_{\rm hx}S$. For the valleyless 2D coupled-wire system, bosonization gives the sine-Gordon term, 
\begin{align}
H_{\rm hx}^{\rm (2d)}
=
\frac{B_{\rm hx}}{2\pi a}
\sum_m
\int dr\,
\cos\left[2\Phi_m^{+}(r)\right],
\label{eq:Hhx-modified}
\end{align}
where
\begin{subequations}
\label{Eq:new-basis}
\begin{align}
\Phi_m^\eta
&=
\frac{1}{\sqrt{2}}
\left(
\phi_{c,m}
+
\eta\theta_{s,m}
\right),
\\
\Theta_m^\eta
&=
\frac{1}{\sqrt{2}}
\left(
\theta_{c,m}
+
\eta\phi_{s,m}
\right),
\end{align}
\end{subequations}
with $\eta\in\{+,-\}$. The fields $\Phi_m^\eta$ and $\Theta_m^\eta$ form a new dual bosonic basis and obey the standard bosonic commutation relations. The valleyless 1D expression follows by dropping the channel index, while the corresponding valleyful transformations are given in Refs.~\cite{CHH-CNT-helix,YYC-2Dhelix-2025}.
In the parameter regime of interest, the sine-Gordon term is RG relevant. 
It gaps part of the itinerant-carrier spectrum and reconstructs the remaining gapless modes, providing the static feedback of the ordered localized moments on the carrier sector.
In the following subsection, we go beyond the previous RG analysis and determine the helix-induced gap  using a variational approach.
 
\subsection{Variational treatment of the helix-induced gap} 
\label{Sec:helix-induced gap}

From the renormalization-group perspective, the sine-Gordon term in Eq.~\eqref{eq:Hhx-modified} flows toward strong coupling when it is relevant. We therefore examine the strong-coupling regime and determine the resulting helix-induced gaps. To this end, we employ the variational approach~\cite{Giamarchi2003} and introduce the variational free-energy functional,
\begin{align}
F_{\rm var}[S_{0}]
\equiv
F_{0}
+
\frac{k_{\rm B}T}{\hbar}
\left\langle S-S_{0}\right\rangle_{0},
\label{eq:F_var}
\end{align}
where $S_{0}$ is the trial action to be introduced below, $F_{0}$ is the corresponding free energy, and $\langle\cdots\rangle_{0}$ denotes an expectation value evaluated with respect to $S_{0}$.

Taking the 2D system described by Eq.~\eqref{eq:Hhx-modified} as an example, the sine-Gordon interaction involves only the $\Phi_m^{+}$ fields, while the other sectors remain gapless. We therefore retain the $\Phi_m^+$ sector when evaluating the helix-induced gap:
\begin{subequations}
\begin{align}
S_{\rm it} ^{\rm (2d)} 
&=
\frac{1}{N_{\rm ch}}
\int dr d\tau
\sum_{q_\perp}
\frac{\hbar}{2\pi\tilde{K}^{\rm (2d)}(q_\perp)}
\Bigg[
\frac{1}{\tilde{v}_{\rm ch}^{\rm (2d)}}
\left|\nabla_{\tau}\Phi_{q_\perp}^{+}\right|^{2}
\nn
&\hspace{4.5cm}
+
\tilde{v}_{\rm ch}^{\rm (2d)}
\left|\nabla_{r}\Phi_{q_\perp}^{+}\right|^{2}
\Bigg],
\label{eq:See_modified}
\\
S_{\rm hx} ^{\rm (2d)} 
&=
\frac{B_{\rm hx}}{2\pi a}
\sum_{m}
\int dr d\tau\,
\cos\left(2\Phi_{m}^{+}\right) ,
\label{eq:Shx_modified}
\end{align}
\end{subequations}
Here, we adopt a decoupled-sector approximation in the basis introduced in Eq.~\eqref{Eq:new-basis}, neglecting the quadratic gradient terms that mix the $\pm$ sectors. This reduces the gap calculation to a sine-Gordon theory for $\Phi_m^+$, while the remaining sector is described by a separate gapless Gaussian theory. 
Within this approximation, the effective interaction parameter is given by
\begin{align}
\tilde{K}^{\rm (2d)}(q_\perp)
&\equiv
\sqrt{\frac{2[K_c(q_\perp)]^2}{[K_c(q_\perp)]^2+1}}.
\label{eq:tilde-v-K}
\end{align}
The corresponding effective velocity is estimated as $\tilde v_{\rm ch}^{\rm (2d)}\approx v_{\rm ch}^{\rm (2d)}\sqrt{(1+K_c^{-2})/2}$, using the additional approximation $K_c(q_\perp)\approx K_c$ only in the velocity expression.
For the valleyful coupled-wire system, the effective interaction parameter within the same decoupled-sector approximation is given by
\begin{align}
\tilde K^{\rm (2d+v)}(q_\perp)
&=
\frac{2K_{cS}(q_\perp)}{\sqrt{1+3[K_{cS}(q_\perp)]^2}}.
\label{eq:tilde-v-K-valley}
\end{align}
The corresponding effective velocity is estimated as $\widetilde v_{\rm ch}^{\rm (2d+v)}\approx v_{\rm ch}^{\rm (2d+v)}\sqrt{(3+K_{cS}^{-2}) / 4 }$, using the additional approximation $K_{cS}(q_\perp)\approx K_{cS}$  in the velocity expression.
 
Without loss of generality, we take $B_{\rm hx}>0$ and expand the field about the saddle point $\Phi_{m,0}^{+}$, which minimizes Eq.~\eqref{eq:Shx_modified}, as
\begin{align}
\Phi_{m}^{+}(r,\tau)
=
\Phi_{m,0}^{+}
+
\varphi_{m}^{+}(r,\tau),
\end{align}
where $\varphi_{m}^{+}$ describes fluctuations about the saddle-point configuration.
Since the saddle-point configuration is uniform, the shift leaves the quadratic terms unchanged. Expanding the cosine term to quadratic order in $\varphi_{m}^{+}$ generates a positive mass term proportional to $(\varphi_{m}^{+} )^2$. 

Within the variational approach, the action is approximated by a quadratic trial action for the fluctuation fields $\varphi_{m}^{+}$.
Fourier transforming the index $m$ to the transverse momentum gives  
\begin{align}
S_{0} ^{\rm (2d)}
=
\frac{k_B T}{2\hbar L N_{\rm ch}}
\sum_{\bm{q},q_\perp}
\left[\varphi_{q_\perp}^{+}(\bm{q})\right]^{\dagger}
\left[G^{+}(\bm{q},q_\perp)\right]^{-1}
\varphi_{q_\perp}^{+}(\bm{q}),
\end{align}
where  $G^{+}(\bm{q},q_\perp)$ is the trial Green's function. Since $S^{\rm (2d)}_0$ is fully specified by $G^{+}$, the optimal trial action is obtained by minimizing $F_{\rm var}$ with respect to the trial Green's function carrying $\bm{q} \equiv (\omega_n, \, q)$ and $q_\perp$,
\begin{align}
\frac{\partial F_{\rm var}}
{\partial G^{+}(\bm{q},q_\perp)}
=
0.
\end{align}

To proceed, we evaluate the variational free energy in Eq.~\eqref{eq:F_var} with the above trial Green's function and obtain  
\begin{align}
F_{\rm var}
=&
-\frac{k_BT}{2}
\sum_{\bm q,q_\perp}
\ln\left[
G^{+}(\bm{q},q_\perp)
\right]
\nn
&+
\sum_{\bm{q},q_\perp}
\frac{k_B T}{2\pi \tilde{K}^{\rm (2d)}(q_\perp)}
\left[
\tilde{v}^{\rm (2d)}_{\rm ch}q^{2}
+
\frac{\omega_{n}^{2}}{\tilde{v}_{\rm ch}^{\rm (2d)}}
\right]
G^{+}(\bm{q},q_\perp)
\nn
&-
\frac{B_{\rm hx}LN_{\rm ch}}{2\pi a}
\exp\left[
-\frac{2 k_B T}{\hbar LN_{\rm ch}}
\sum_{\bm{q},q_\perp}
G^{+}(\bm{q},q_\perp)
\right],
\label{eq:F_var_final}
\end{align}
with $\tilde{K}^{\rm (2d)}(q_\perp)$ given in Eq.~\eqref{eq:tilde-v-K}. Minimizing this expression with respect to the trial Green's function yields the self-consistent equation, 
\begin{align}
\left[G^{+}(\bm{q},q_\perp)\right]^{-1}
=& 
\frac{1}{\pi\tilde{K}^{\rm (2d)}(q_\perp)}
\left(
\tilde{v}_{\rm ch}^{\rm (2d)}q^{2}
+
\frac{\omega_{n}^{2}}{\tilde{v}_{\rm ch}^{\rm (2d)}}
\right) 
\nn
&+ 
\frac{2B_{\rm hx}}{\pi\hbar a}
\exp\left[
-\frac{2 k_B T}{\hbar LN_{\rm ch}}
\sum_{\bm{q},q_\perp}
G^{+}(\bm{q},q_\perp)
\right] ,
\label{eq:selfconsistent_G}
\end{align}
where the second term is the self-energy generated by the helix-induced mass.

We adopt the trial Green's function,  
\begin{align}
G^{+}(\bm{q},q_\perp)
=
\frac{\pi\tilde{K}^{\rm (2d)}(q_\perp)\tilde{v}_{\rm ch}^{\rm (2d)}}
{\omega_{n}^{2}+\big[\tilde{v}_{\rm ch}^{\rm (2d)}\big]^{2}q^{2}+\left[\Delta_{g}(q_\perp)/\hbar\right]^{2}},
\label{eq:massive_G_ansatz}
\end{align}
where the mass parameter $\Delta_g(q_\perp)$ is identified with the helix-induced  gap of the itinerant carriers for each transverse mode $q_\perp$. Within the present variational approximation, the sine-Gordon interaction generates a  self-energy independent of $q$ and $q_\perp$ [see Eq.~\eqref{eq:selfconsistent_G}]. 
The gap therefore depends only on $q_\perp$ through $\tilde{K}^{\rm (2d)}(q_\perp)$ and remains independent of $q$, as reflected in   Eq.~\eqref{eq:selfconsistent_Deltag} below.

Within this parametrization, the problem of  determining the optimal Green's function reduces to solving self-consistently for $\Delta_g$. Specifically, substituting Eq.~\eqref{eq:massive_G_ansatz} into Eq.~\eqref{eq:selfconsistent_G} yields 
\begin{align}
[& \Delta_{g}(q_\perp)]^{2}
=
\frac{2\hbar\tilde{v}_{\rm ch}^{\rm (2d)}B_{\rm hx}\tilde{K}^{\rm (2d)}(q_\perp)}{a}
\nn
&\quad\times
\exp\left\{
-\frac{2k_B T}{\hbar L N_{\rm ch}}
\sum_{\bm{q},q^\prime_\perp}
\frac{\pi\tilde{K}^{\rm (2d)}(q_\perp^\prime)\tilde{v}_{\rm ch}^{\rm (2d)}}
{\omega_{n}^{2}+[\tilde{v}_{\rm ch}^{\rm (2d)}]^{2}q^{2}
+\left[\Delta_{g}(q_\perp^\prime)/\hbar\right]^{2}}
\right\}.
\label{eq:selfconsistent_Deltag}
\end{align}

To extract a single characteristic gap scale, we approximate the transverse dependence of the interacting sector by replacing $\tilde K^{\rm (2d)}(q_\perp)$ with its transverse-momentum average $\bar K_+^{\rm  (2d)}$, defined by $\bar{K}^{\rm (2d)}_{\pm} \equiv \bar{K}^{\rm (2d)}_{\pm , n = 0}$, where
\begin{align}
\bar{K}^{\rm (2d)}_{\pm , n}
\equiv
\int_{-\pi}^{\pi}
\frac{d(q_\perp d)}{2\pi}\,
\cos(nq_\perp d)\,
[\tilde{K}^{\rm (2d)}(q_\perp)]^{\pm 1}.
\label{eq:Kbar-n}
\end{align}
The transverse averaging at this step is used only to estimate a single gap scale; the transverse dependence of the effective interaction parameter is retained when evaluating the correlation exponents below. Equation~\eqref{eq:Kbar-n} characterizes the correlations between the newly defined bosonic fields on channels separated by $n$ transverse  spacings. 

The resulting variational propagator describes the common low-energy mass scale of the phase-locked channels, 
\begin{align}
\Delta_{g}^{2}
&\approx
\frac{2\hbar\tilde{v}_{\rm ch}^{\rm (2d)}B_{\rm hx}\bar{K}^{\rm (2d)}_+}{a}
\nn
&\quad\times
\exp\left\{
-\frac{2 k_B T }{\hbar L}
\sum_{\bm{q}}
\frac{\pi\bar{K}^{\rm (2d)}_+ \tilde{v}_{\rm ch}^{\rm (2d)}}{\omega_{n}^{2}+[\tilde{v}_{\rm ch}^{\rm (2d)}]^{2}q^{2}
+\left(\Delta_{g}/\hbar\right)^{2}}
\right\}.
\label{eq:self-consistent_Delta_g}
\end{align}
Replacing the momentum and Matsubara-frequency sums in Eq.~\eqref{eq:self-consistent_Delta_g} by integrals over continuous variables and introducing the ultraviolet cutoff $\tilde{\Delta}_a$, we obtain the helix-induced partial gap in the spectrum of the itinerant carriers,
\begin{align}
\Delta_g
=
\tilde{\Delta}_a
\left(
\frac{2B_{\rm hx}\bar{K}^{\rm (2d)}_+}{\tilde{\Delta}_a}
\right)^{1/(2-\bar{K}^{\rm (2d)}_+)},
\end{align}
for sufficiently low temperatures.

The variational result exhibits the strong-coupling power-law form characteristic of a relevant sine-Gordon perturbation, as in previous RG treatments~\cite{YYC-2Dhelix-2025}. Here, the exponent is evaluated using the effective parameters of the decoupled-sector approximation. The same variational procedure can be applied to the valleyful coupled-wire system and to the corresponding isolated 1D limits.

\section{Josephson-interference features of the spin helix}
\label{sec:hx-JJ}

Having established the formation, stability, and electronic consequences of the spin helix, we now examine how the same ordered texture can be detected through phase-coherent transport. 
To determine how the carrier-induced helix modifies an experimentally measurable interference pattern, we consider a 2D SNS Josephson junction in which the normal region contains a static helical exchange field motivated by the interacting analysis above.

\subsection{Josephson junction with a helix weak link}
\label{sec:JJmodel}

We consider a 2D SNS Josephson junction comprising a normal region that forms the weak link between two conventional $s$-wave superconducting leads, modeled as
\begin{align}
H_{\rm J}
=
H_{\rm N}
+H_{{\rm S},1}
+H_{{\rm S},2}
+H_{{\rm NS},1}
+H_{{\rm NS},2}.
\label{eq:H_J}
\end{align}
Here, $H_{\rm N}$ describes the normal region in which the helix can form, $H_{{\rm S},1}$ and $H_{{\rm S},2}$ describe the two $s$-wave superconducting leads, and $H_{{\rm NS},1}$ and $H_{{\rm NS},2}$ describe spin-conserving tunneling between the normal region and the respective superconducting leads. We denote the superconducting phase difference between the two leads by $\varphi$.

Motivated particularly by the coupled-channel structure of tWTe$_2$, we discretize the normal region on an anisotropic rectangular lattice, with the conduction channels either along the junction axis (taken as $x$) or perpendicular to it ($y$).  
Thus, the rectangular lattice should be understood as a coarse-grained representation of the coupled-channel system rather than as the microscopic atomic lattice of the material. 
Using the spinor $c_{\bm j}=(c_{\bm j\uparrow},c_{\bm j\downarrow})^T$ at
$\bm r_{\bm j}=[(j_x-1)a_0,(j_y-1)b_0]$, the normal-region term is
\begin{align}
H_{\rm N}
=&
\sum_{\bm j}
\Bigg\{
c_{\bm j}^{\dagger}
\left[
-\mu_0 \sigma^0
+( \bm h_{\bm j}^{\rm hx} +  {\bm h}^{\rm ext} ) \cdot\bm\sigma
\right]
c_{\bm j}
\nonumber\\
&
-
\Big(
t_x e^{i\vartheta_{\bm j,x}} c_{\bm j+\hat{x}}^{\dagger} 
 c_{\bm j}
+t_y e^{i\vartheta_{\bm j,y}}
c_{\bm j+\hat{y}}^{\dagger} 
 c_{\bm j}
+\mathrm{H.c.}\Big) \Bigg\}.
\label{eq:JJ-HN}
\end{align} 
Here, $\mu_0$ denotes the effective chemical potential,  $\bm h^{\rm ext}=g\mu_B\bm B/2$ describes the Zeeman coupling to the external magnetic field $\bm B = (B_{\rm in},0,B_z)$, $\sigma^0$ is a $2\times2$ identity matrix, and $\sigma^\mu$ denotes the corresponding Pauli matrix acting in spin space.  We consider an in-plane field $B_{\rm in}$, whose sign distinguishes fields parallel and antiparallel to the supercurrent direction, together with an out-of-plane field $B_z$. The orbital effect of $B_z$ is incorporated through the Peierls phases $\vartheta_{{\bm j},x}$ and $\vartheta_{{\bm j},y}$.

The additional term specific to the present problem is the static helical exchange field,
\begin{align}
\bm h_{\bm j}^{\rm hx}
= (B_{\rm hx}/2 )
&\big[
{\bm e}_1 \cos\left( {\bm Q}_{\rm hx} \cdot {\bm a}_{{\rm hx},{\bm j}} + \phi_{{\rm hx},{\bm j}}  \right)  
\nn
& \quad +
{\bm e}_2 \sin\left( {\bm Q}_{\rm hx} \cdot {\bm a}_{{\rm hx},{\bm j}} + \phi_{{\rm hx},{\bm j}} \right)  
\big].
\label{eq:JJ-helix}
\end{align}
Here, $B_{\rm hx}$ and ${\bm Q}_{\rm hx}$ specify the amplitude and wave vector of the static helical exchange field. In the interacting theory, the corresponding texture is selected by the RKKY interaction, with $|{\bm Q}_{\rm hx}|=2k_F$ for the valleyless channels considered above. In the effective transport model, their numerical values are treated as prescribed parameters, as specified below.
When the coupled channels are parallel to the junction axis, and hence to the current direction, the anisotropy is characterized by $t_x>t_y$ and $a_0<b_0$. In this case, the helix propagates along the channels, with ${\bm Q}_{\rm hx}\parallel\hat{x}$ and
${\bm a}_{{\rm hx},{\bm j}}=(j_x-1)a_0\hat{x}$, while the helix phase may vary between different channels as $\phi_{{\rm hx},{\bm j}}=\phi_{\rm hx}(j_y)$. For channels oriented perpendicular to the current direction, the anisotropy is reversed, with $t_x<t_y$ and $a_0>b_0$. The helix then propagates along the $y$ direction, such that ${\bm Q}_{\rm hx}\parallel\hat{y}$ and ${\bm a}_{{\rm hx},{\bm j}}=(j_y-1)b_0\hat{y}$, while $\phi_{{\rm hx},{\bm j}}=\phi_{\rm hx}(j_x)$. In Eq.~\eqref{eq:JJ-helix}, we retain the channel-dependent phase $\phi_{{\rm hx},{\bm j}}$ to allow for a general transverse phase configuration. For the phase-coherent helix considered as the default case, the same phase is assigned to all participating channels.

The unit vectors $\bm e_1$ and $\bm e_2$ are mutually orthogonal and span the spin-rotation plane.  The parametrization in Eq.~\eqref{eq:JJ-helix} ensures $|\bm h_{\bm j}^{\rm hx}|=B_{\rm hx}/2$ at every lattice site. 
The parametrization also allows different spin-rotation planes, including the following representative configurations: 
\begin{align}
(\bm e_1,\bm e_2)=
\begin{cases}
(\hat{\bm x},\hat{\bm y}), & xy\text{ configuration},\\
(\hat{\bm y},\hat{\bm z}), & yz\text{ configuration},\\
(\hat{\bm z},\hat{\bm x}), & zx\text{ configuration}.
\end{cases}
\label{eq:helix-configurations}
\end{align}
The characteristic features of the Josephson interference pattern remain qualitatively robust across these configurations, including when random channel-dependent phases $\phi_{{\rm hx},{\bm j}}$ are introduced.
 
We use flux focusing to make the applied in-plane field $B_{\rm in}$ a controlled probe of the helical texture. The motivation is to examine how an established helix modifies the response to a tunable, spatially nonuniform magnetic-field profile, rather than relying only on the shape of an interference pattern at a single field. Owing to the Meissner response of the superconducting leads, the applied field lines bend near the two NS interfaces, generating additional out-of-plane components with opposite signs on the two sides of the junction~\cite{FlensbergPRB:2017_AnomalousFraunhoferPattern,YCTsai:2026arxiv}. Varying the magnitude and sign of $B_{\rm in}$ therefore controls the additional orbital phases accumulated along the trajectories contributing to the Josephson current.

We represent this effect by a dipole-like profile, with induced out-of-plane fields $\pm f_B B_{\rm in}$ confined to strips of length $L_{\rm dip}$ near the two NS interfaces. Here, the dimensionless parameter $0\leq f_B\leq1$ specifies the strength of the induced field in our model. The additional orbital coupling is incorporated through the Peierls phases of the hopping terms. At each fixed $B_{\rm in}$, sweeping $B_z$ produces an interference pattern; varying $B_{\rm in}$ then reveals how that pattern responds to the controlled change in the interface field profile. Comparison with the corresponding $B_{\rm hx}=0$ junction separates the contribution of the helical exchange field from the response of the same anisotropic geometry and flux-dipole profile alone.

The Josephson current is obtained from
\begin{align}
I_s(\varphi)=\frac{2e}{\hbar}\frac{\partial F_{\rm J}}{\partial\varphi},
\end{align}
with $F_{\rm J}$ the free energy. 
Following the tunneling formulation of Refs.~\cite{Flensberg:2016PRB,FlensbergPRB:2017_AnomalousFraunhoferPattern}, we evaluate $F_{\rm J}$ perturbatively in the NS tunneling amplitude,  $\gamma_{\rm NS} $, retaining the leading nonvanishing contribution at fourth order. After integrating out the fermionic degrees of freedom in the superconducting leads, this contribution can be expressed in terms of the  Green's functions of the normal region connecting the two NS interfaces:
\begin{align}
\mathbb G^{RL}(i\omega_n)
=
\left\langle N_R\right|
\left(i\hbar\omega_n\mathbb I-H_{\rm N}\right)^{-1}
\left|N_L\right\rangle , 
\end{align}
with $\omega_n$ the fermionic Matsubara frequency and $\mathbb{I}$ the identity matrix. Here, $N_L$ and $N_R$ denote the left and right interface slices, respectively.
The interface Green's functions are evaluated using the recursive Green's-function method under open  boundary conditions~\cite{KTLawPRB:2018AJE,FurusakiPhysicaB:1994DirtyJJ,AsanoPRB:2002,diez2023symmetry}. The critical current for the interference pattern is then obtained as $I_c(B_z)=  \max_{\varphi} |I_s(\varphi,B_z)|$.

Although the carrier system responsible for forming the helix is interacting, the junction calculation addresses the phase-coherent transport response of an already established texture rather than its many-body formation. The interacting analysis motivates the formation of the spatially rotating exchange field, while the effective junction model treats the exchange amplitude $B_{\rm hx}$ and ordering wave vector ${\bm Q}_{\rm hx}$ as phenomenological inputs. This separation allows us to isolate the roles of the spin texture, interchannel propagation, and orbital interference without requiring a microscopic fit of all junction parameters to a particular material realization. 

Throughout the analysis, we fix the area of the normal region to $L\times W=210\times840\,{\rm nm}^2$, where $L$ and $W$ denote the junction dimensions along and transverse to the current direction, respectively. The spatial extent of the flux-dipole region is fixed at $L_{\rm dip}=30\,{\rm nm}$. To model the anisotropic coupled-channel geometry, we consider two orientations of the conducting channels relative to the current direction.
When the conducting channels are parallel to the current direction, we take $a_0=3\,{\rm nm}$ as the longitudinal discretization length in Eq.~\eqref{eq:JJ-HN}, while $b_0=10\,{\rm nm}$ represents the effective interchannel separation. We use the intrachannel hopping $t_x\equiv t$ as the energy unit and set the weaker interchannel hopping to $t_y=t/8$ (here $t$ denotes the hopping amplitude, not time).  In this configuration, the helix propagates along the supercurrent direction, with $\bm{Q}_{\rm hx}\parallel\hat{x}$. 
For conducting channels oriented perpendicular to the current direction, we exchange the longitudinal and transverse parameters, taking $a_0=10\,{\rm nm}$ and $b_0=3\,{\rm nm}$, together with $t_x=t/8$ and $t_y\equiv t$. The helix then propagates along the transverse direction, $\bm{Q}_{\rm hx}\parallel\hat{y}$.  The remaining parameters specifying the helix configuration are chosen consistently with the corresponding orientation.

Unless otherwise specified, we use the following representative parameters: overall energy scale $t=10\,{\rm meV}$, chemical potential $\mu_0/t=-1.7$, Land\'e factor $g=2$, superconducting pairing amplitude $\Delta/t=0.2$, effective NS tunneling strength $\gamma_{\rm NS}/t=0.12$, thermal energy $k_BT/t=0.002$, flux-focusing factor $f_B=0.2$, and helix-induced exchange energy $B_{\rm hx}/t=0.9$. This parameter set defines a finite-exchange benchmark for examining how the coupled-channel orientation and applied in-plane field modify the Josephson interference pattern, rather than a quantitative fit to the material estimates in Table~\ref{Table:system-parameters}. In particular, the exchange amplitude is chosen to resolve the effect of an established texture on the lobe structure within the finite junction considered here.

We choose a helical wavelength of $12\,{\rm nm}$ along the conducting-channel direction, corresponding to four longitudinal lattice spacings per period.  The wavelength is prescribed as an input to the transport model rather than determined self-consistently from the RKKY susceptibility of its auxiliary tight-binding band. For the phase-coherent configuration, we set $\phi_{{\rm hx},{\bm j}}=0$ on every channel. The comparisons below retain the same junction parameters when the helix is removed, thereby isolating the changes associated with the prescribed helical exchange field.

\begin{figure}[t]
    \centering
    \includegraphics[width=0.48\textwidth]{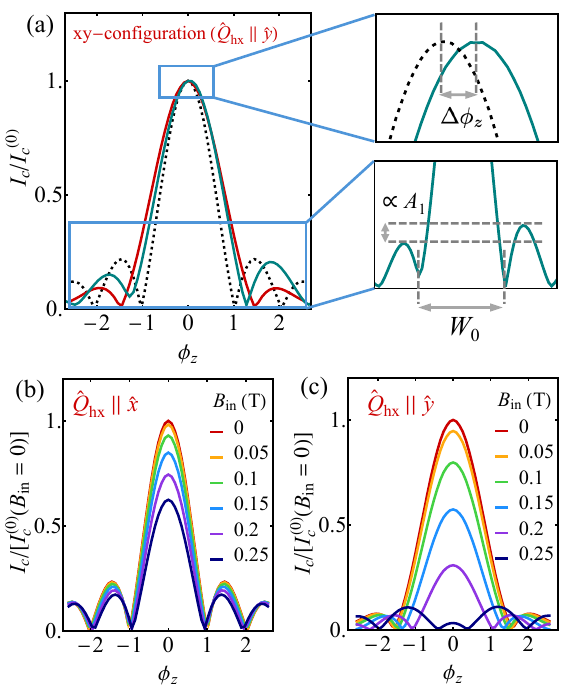}
    \caption{(a) Josephson interference patterns for an SNS junction with three different normal-region states: conventional (black dashed), phase-coherent spin-helix state (red), and spin-helix state with random phase offsets (green); see the main text for their definitions. In all panels, we focus on the helix rotating within the $xy$ plane [see Eq.~\eqref{eq:helix-configurations}]. The normalized critical current $I_c/I_c^{(0)}$ is plotted as a function of the dimensionless magnetic flux $\phi_z$. The insets show enlarged views illustrating the central-peak displacement $\Delta\phi_z$, the side-lobe asymmetry $A_1$, and $W_0$.
    Panels (b) and (c) show the interference patterns for different values of $B_{\rm in}$ with $\hat{\bm Q}_{\rm hx}\parallel\hat{x}$ and $\hat{\bm Q}_{\rm hx}\parallel\hat{y}$,  respectively. 
    For each channel orientation, the critical current is scaled by the dipole-off, zero-flux value, $I_{c }^{(0)} (B_{\rm in} = 0)$. 
    }
    \label{fig:pattern}
\end{figure}

\subsection{Observable features in interference patterns}

In this section, we examine how the helical spin texture and magnetic-field profile modify the Josephson interference pattern of the helix-based SNS junction.  Unless otherwise stated, the results presented in this section focus on the $xy$ helix configuration.

Figure~\ref{fig:pattern} illustrates the Josephson interference patterns of the 2D helix-based SNS junction and highlights the characteristic modifications induced by the helical texture and magnetic-field profile. In the absence of both the static helical field and the flux-dipole contribution, the junction exhibits the conventional Fraunhofer pattern,
\begin{align}
I_c(\phi_z)
\propto
\left|
\frac{\sin(\pi\phi_z)}{\pi\phi_z}
\right|,
\end{align}
as shown by the dashed curve in Fig.~\ref{fig:pattern}(a). Here,
$\phi_z\equiv B_zLW/\Phi_0$ is the dimensionless external magnetic flux through the junction and $\Phi_0=h/2e$ is the superconducting flux quantum.
When the helical exchange field and the flux-dipole contribution are included, the interference pattern generally departs from the conventional Fraunhofer form and develops several characteristic features, as illustrated in Fig.~\ref{fig:pattern}. In particular, the relative weights of the central and side lobes are redistributed, while the positions of the interference minima can shift, thereby modifying the lobe widths. The side lobes can also become asymmetric between positive and negative flux.

To quantify these features, we introduce four measures. First, we characterize  the width, $W_p$, of the $p$th lobe by the flux interval between its two neighboring interference minima, 
 where $p=0$ labels the central lobe and $p\in\{\pm1,\pm2,\ldots\}$ labels the side lobes.
The central lobe width, $W_0$, indicated in the inset of Fig.~\ref{fig:pattern}(a) is particularly relevant for our analysis. 

Second, to characterize the relative weight of the central and side lobes, we define the central-to-$p$th-side-lobe ratio,
\begin{align}
R_p
\equiv
\frac{I_c^{(0)}}
{\left[I_c^{(p)}+I_c^{(-p)}\right]/2},
\qquad p>0,
\label{eq:R_p}
\end{align}
where the average of the positive- and negative-flux side-lobe peaks accounts for a possible asymmetry between them. Here, $I_c^{(0)}$ denotes the peak critical current of the central lobe, while $I_c^{(p)}$ and $I_c^{(-p)}$ denote the peak critical currents of the $p$th side lobes at positive and negative flux, respectively.
Third, the side-lobe asymmetry is quantified by the normalized difference,
\begin{align}
A_p
\equiv
\frac{I_c^{(p)}-I_c^{(-p)}}
{\left[I_c^{(p)}+I_c^{(-p)}\right]/2}
\times 100\%,
\qquad p>0,
\label{eq:A_p}
\end{align}
as illustrated in the inset of Fig.~\ref{fig:pattern}(a). 
Finally, we denote the displacement of the central peak from zero flux by $\Delta\phi_z$, as illustrated in the inset of Fig.~\ref{fig:pattern}(a).
It is expressed as a percentage of one flux quantum; thus, $\Delta\phi_z=100\%$ corresponds to a shift by $\Phi_0$.

We first examine how the interference pattern evolves with increasing in-plane field $B_{\rm in}$ for the two orientations of the coupled channels and the associated helix propagation direction. Figures~\ref{fig:pattern}(b) and~\ref{fig:pattern}(c) show representative results for $\hat{\bm Q}_{\rm hx}\parallel\hat{x}$ and $\hat{\bm Q}_{\rm hx}\parallel\hat{y}$, respectively. In both cases, increasing $B_{\rm in}$ drives the interference pattern away from the conventional Fraunhofer form, but the evolution depends strongly on the orientation of the conduction channels relative to the supercurrent. For $\hat{\bm Q}_{\rm hx}\parallel\hat{x}$, where the channels and helix propagation direction are parallel to the current, the overall lobe structure remains comparatively robust over the field range considered, with only a moderate suppression of the central peak. By contrast, for $\hat{\bm Q}_{\rm hx}\parallel\hat{y}$, the central lobe is unusually broad at low field and its peak is progressively suppressed as $B_{\rm in}$ increases, accompanied by a pronounced redistribution of the side-lobe intensities.
The interference patterns can also develop an imbalance between the positive- and negative-flux side lobes and a nonzero displacement of the central peak from zero flux. 

\begin{figure}[t]
    \centering
    \includegraphics[width=0.48\textwidth]{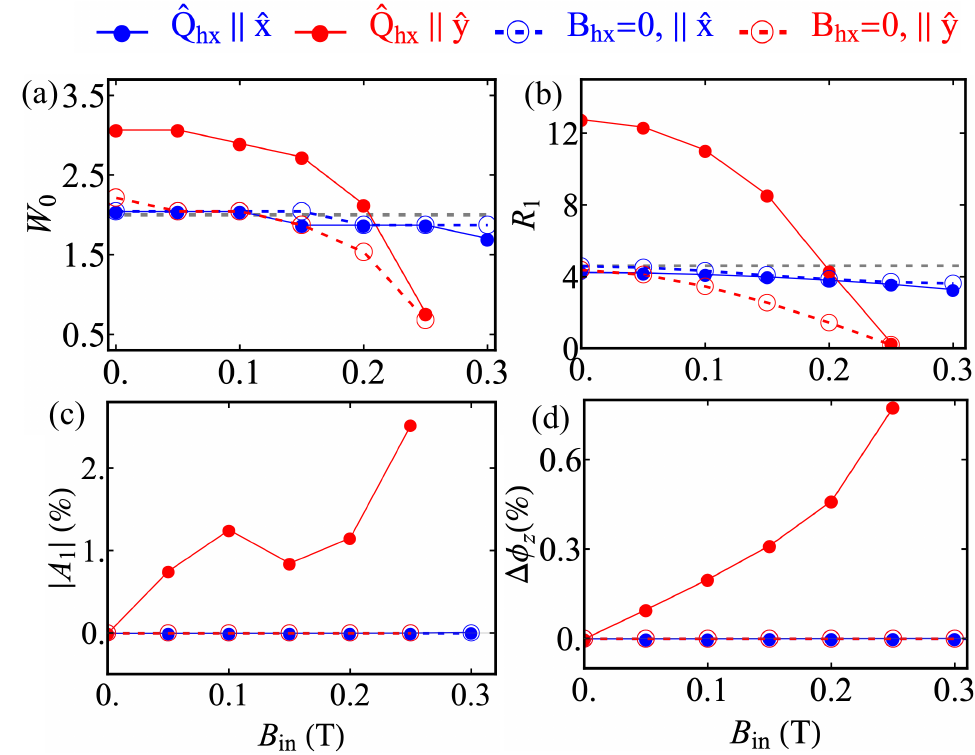}
    \caption{ 
    Characteristic measures of the Josephson interference patterns for phase-coherent spin helices in the two coupled-channel orientations (solid dots with solid curves). Panels (a)--(d) show the central-lobe width $W_0$, the central-to-first-side-lobe ratio $R_1$ [Eq.~\eqref{eq:R_p}], the magnitude of the first-side-lobe asymmetry $|A_1|$ [Eq.~\eqref{eq:A_p}], and the central-peak displacement $\Delta\phi_z$, respectively, as functions of the in-plane field $B_{\rm in}$. The gray dashed lines in panels (a) and (b) indicate the ideal Fraunhofer values $W_0=2$ and $R_1\simeq 4.6$, respectively. The corresponding no-helix results (open circles with dashed lines), obtained by setting $B_{\rm hx}=0$ while retaining the same anisotropic geometry and flux-dipole profile, are shown for comparison. Some higher-$B_{\rm in}$ data for $\hat{\bm Q}_{\rm hx}\parallel\hat{\bm y}$ are omitted because the corresponding interference patterns become strongly distorted, so that the measures used here are no longer well defined.
    All other parameter values are the same as those used in Fig.~\ref{fig:pattern}.    
    }
   \label{fig:patt_features}
\end{figure}

Figure~\ref{fig:patt_features} summarizes the four measures for each channel orientation together with matched junction references with no helix formation, allowing the helical contribution to be distinguished from the response of the anisotropic geometry and flux-dipole profile. 
These features evolve continuously with $B_{\rm in}$ and differ between the two coupled-channel orientations. The in-plane field therefore provides a useful external control parameter.

We first consider the width $W_0$ of the central lobe, as shown in Fig.~\ref{fig:patt_features}(a). For an ideal Fraunhofer pattern, the adjacent interference minima occur at $\phi_z=\pm1$, giving $W_0=2$. A pronounced deviation appears for $\hat{\bm Q}_{\rm hx}\parallel\hat{\bm y}$, where the central lobe remains substantially broader than this ideal value, although it narrows at higher $B_{\rm in}$. By comparison, the matched $B_{\rm hx}=0$ reference, which retains the same anisotropic hopping and flux-dipole profile, has a width near the ideal value at low fields that decreases below $W_0=2$ with increasing $B_{\rm in}$. 
The helix therefore broadens the central lobe relative to the matched reference over the common field range shown, distinguishing its contribution from the background response of the anisotropic junction geometry and flux focusing.
By contrast, for $\hat{\bm Q}_{\rm hx}\parallel\hat{\bm x}$, $W_0$ remains close to the conventional value and differs only weakly from the corresponding no-helix reference. These comparisons reveal a strongly orientation-dependent helical modification of the central-lobe width.

We next examine the relative weight of the central lobe, characterized by $R_1$; see Fig.~\ref{fig:patt_features}(b). Since $R_1$ compares the central-lobe peak with the average height of the first side lobes, a smaller $R_1$ indicates a stronger relative suppression of the central lobe. Consistent with Figs.~\ref{fig:pattern}(b) and~\ref{fig:pattern}(c), $R_1$ decreases with increasing $B_{\rm in}$ for both channel orientations, with a larger reduction for $\hat{\bm Q}_{\rm hx}\parallel\hat{\bm y}$ than for $\hat{\bm Q}_{\rm hx}\parallel\hat{\bm x}$ over the common field range. The matched $B_{\rm hx}=0$ references further separate the helix-induced contribution from that of the anisotropic junction geometry and flux focusing. For $\hat{\bm Q}_{\rm hx}\parallel\hat{\bm x}$, the helix produces only a modest change relative to the corresponding junction without helix. 
By contrast, for the transverse-channel geometry with $\hat{\bm Q}_{\rm hx}\parallel\hat{\bm y}$, the helix substantially increases $R_1$ above the matched no-helix reference. Overall, the relative peak weight $R_1$ decreases with increasing $B_{\rm in}$, and eventually drops below the conventional value. These comparisons demonstrate an orientation-dependent helical modification of the relative lobe weights.

The side-lobe asymmetry provides a complementary characterization of the interference response. A finite $A_1$ indicates unequal peak critical currents of the first side lobes at positive and negative flux, $I_c^{(1)}\neq I_c^{(-1)}$. 
As shown in Fig.~\ref{fig:patt_features}(c), $|A_1|$ becomes finite for the transverse-channel geometry with $\hat{\bm Q}_{\rm hx}\parallel\hat{\bm y}$ in the presence of the helix and exhibits a pronounced field dependence, reaching its largest value at the highest $B_{\rm in}$ shown.
In contrast, the asymmetry remains negligible for the parallel-channel geometry with $\hat{\bm Q}_{\rm hx}\parallel\hat{\bm x}$ and for both matched no-helix references. These comparisons reveal an orientation-dependent helical contribution to the side-lobe imbalance beyond the response of the anisotropic junction and flux-dipole profile alone.

Finally, Fig.~\ref{fig:patt_features}(d) shows the displacement $\Delta\phi_z$ of the central peak. Similar to $|A_1|$, a finite $\Delta\phi_z$ develops with increasing $B_{\rm in}$ only for $\hat{\bm Q}_{\rm hx}\parallel\hat{\bm y}$ in the presence of the helix. The shift remains negligible for the parallel-channel geometry and for both matched no-helix references. Thus, for the parameters considered, the central-peak displacement provides another orientation-dependent signature of the helix that is not reproduced by the anisotropic junction geometry and flux focusing alone.

Taken together, $W_p$, $R_p$, $A_p$, and $\Delta\phi_z$ provide complementary information about the interference response of the helix-based junction. Comparing their systematic field dependence with the matched no-helix references allows the helical contribution to be resolved for each coupled-channel orientation. While the lobe widths and relative weights also vary with field in the absence of the helix, finite side-lobe asymmetry and central-peak displacement emerge only in the transverse-channel junction with the helix for the parameters considered. Their combined evolution with $B_{\rm in}$ therefore helps distinguish the helical contribution from the background response of the anisotropic junction and flux focusing.
  
To isolate the role of flux focusing in this field-tuning strategy, we repeat the calculation with $f_B=0$, retaining the helical exchange field and the direct Zeeman coupling to the applied magnetic field. For the phase-coherent configurations considered above, the interference patterns are symmetric, with vanishing side-lobe asymmetry and central-peak displacement, while $W_0$ and $R_1$ become nearly independent of $B_{\rm in}$ over the field range examined. For  $\hat{\bm Q}_{\rm hx}\parallel\hat{\bm y}$, however, $W_0$ and $R_1$ still deviate from the ideal Fraunhofer values. Thus, flux focusing is not required for a nonstandard lobe structure to emerge. Rather, we exploit its orbital contribution to make $B_{\rm in}$ an effective control parameter for tracking how the helix modifies the interference response. The dipole-off comparison confirms that the pronounced field dependence seen above arises primarily from the tunable interface field profile, rather than from the direct Zeeman coupling. At larger $B_{\rm in}$, the latter may become increasingly important.

We further examine the effects of reversing the helicity, ${\bm Q}_{\rm hx}\to-{\bm Q}_{\rm hx}$, and the in-plane field, $B_{\rm in} \to - B_{\rm in}$. Both the reversal operations transform the  entire interference pattern according to $I_c(\phi_z)\to I_c(-\phi_z)$. As a result, $A_1$ and $\Delta \phi_z$ reverse their signs under both operations, whereas $W_0$ and $R_1$ remain unchanged.  
The reversal of an applied in-plane field thus provides an additional operational knob on the field-dependent interference response.

To examine the effect of losing interchannel helix-phase locking, we assign each channel an independent phase offset $\phi_{{\rm hx},{\bm j}}$ drawn uniformly from $[0,2\pi)$ in Eq.~\eqref{eq:JJ-helix}. We consider independent realizations for each coupled-channel orientation. In the realizations examined, random interchannel phase offsets can enhance $|A_1|$ relative to the phase-coherent reference for both $\hat{\bm Q}_{\rm hx}\parallel\hat{\bm x}$ and $\hat{\bm Q}_{\rm hx}\parallel\hat{\bm y}$, while the magnitude and sign of $A_1$ depend on the specific randomness realization.  
These results demonstrate the sensitivity of the interference pattern to the relative helix phases across different channels.

As a remark, the above helix-induced interference features, such as side-lobe asymmetry, qualitatively resemble features found when flux focusing coexists with onsite disorder in junctions without helix~\cite{FlensbergPRB:2017_AnomalousFraunhoferPattern,YCTsai:2026arxiv}.
Moreover, the interplay of spin-orbit coupling, magnetic-field orientation, and junction crystallographic orientation has been shown to strongly affect the critical currents in  Josephson junctions~\cite{Alex:2022PRB}.
A distorted interference pattern alone therefore does not uniquely identify its microscopic origin. In our system, these helix-induced features become more pronounced as the helical exchange-field amplitude $B_{\rm hx}$ increases, with all other parameters held fixed.  The helical contribution, however, is tied to the ordering scale $T_{\rm hx}$, whereas onsite disorder does not itself disappear at this scale. 
Suppression of the additional changes in $W_p$, $R_p$, $A_p$, and $\Delta\phi_z$ upon increasing the temperature through $T_{\rm hx}$, relative to the background response without the helix, would therefore provide further evidence for a contribution from the spin helix rather than from disorder alone.

\begin{table*}[t]
\centering
\caption{Power-law exponents governing the bulk and boundary LDOS,  and the spin relaxation rate in the isolated 1D systems at different temperature regimes.
For the valleyful system, the spin-relaxation exponent refers to the dominant interbranch transverse-spin
channel.}
\label{Table:summary-1D}
\small
\setlength{\tabcolsep}{7pt}
\renewcommand{\arraystretch}{1.35}
\begin{tabular}{c c c}
\hline\hline
Observable
&
1D
&
1D with valley
\\
\hline

\begin{tabular}{@{}c@{}}
Bulk LDOS\\[-0.1em]
$\beta_{\rm bulk}$ for $T>T_{\rm hx}$\\
$\widetilde{\beta}_{\rm bulk}$ for $T\ll T_{\rm hx}$
\end{tabular}
&
\begin{tabular}{@{}l@{}}
$\displaystyle
\beta^{\rm (1d)}_{\rm bulk}
=
\frac{1}{4}
\left(
K_c+K_c^{-1}+K_s+K_s^{-1}
\right)-1$
\\[0.6em]
$\displaystyle
\widetilde{\beta}^{\rm (1d)}_{\rm bulk}
=
\frac{1}{2}
\left[
\bar K^{\rm (1d)}
+
\left(\bar K^{\rm (1d)}\right)^{-1}
\right]-1$
\end{tabular}
&
\begin{tabular}{@{}l@{}}
$\begin{aligned}
\beta^{\rm (1d+v)}_{\rm bulk}
={}&
\frac{1}{8}
\Big(
K_{cS}+K_{cS}^{-1}
+K_{cA}+K_{cA}^{-1}
\\[-0.2em]
&\qquad
+K_{sS}+K_{sS}^{-1}
+K_{sA}+K_{sA}^{-1}
\Big)-1
\end{aligned}$
\\[0.6em]
$\displaystyle
\widetilde{\beta}^{\rm (1d+v)}_{\rm bulk}
=
\frac{1}{2}
\left[
\bar K^{\rm (1d+v)}
+
\left(\bar K^{\rm (1d+v)}\right)^{-1}
\right]-1$
\end{tabular}
\\
\hline

\begin{tabular}{@{}c@{}}
Boundary LDOS\\[-0.1em]
$\beta_{\rm bdry}$ for $T>T_{\rm hx}$\\
$\widetilde{\beta}_{\rm bdry}$ for $T\ll T_{\rm hx}$
\end{tabular}
&
\begin{tabular}{@{}l@{}}
$\displaystyle
\beta^{\rm (1d)}_{\rm bdry}
=
\frac{1}{2}
\left(
K_c^{-1}+K_s^{-1}
\right)-1$
\\[0.6em]
$\displaystyle
\widetilde{\beta}^{\rm (1d)}_{\rm bdry}
=
1/\bar K^{\rm (1d)}
-1$
\end{tabular}
&
\begin{tabular}{@{}l@{}}
$\displaystyle
\beta^{\rm (1d+v)}_{\rm bdry}
=
\frac{1}{4}
\left(
K_{cS}^{-1}
+K_{cA}^{-1}
+K_{sS}^{-1}
+K_{sA}^{-1}
\right)-1$
\\[0.6em]
$\displaystyle
\widetilde{\beta}^{\rm (1d+v)}_{\rm bdry}
=
1/ 
\bar K^{\rm (1d+v)}
 -1$
\end{tabular}
\\
\hline

\begin{tabular}{@{}c@{}}
Spin-relaxation$^{\rm {\color{blue}a}}$: $1/(T_1 T)$ \\[-0.1em]
$ {2g^x-2}$ for $T>T_{\rm hx}$\\
$ {2\widetilde g^x-2}$ for $T\ll T_{\rm hx}$ 
\end{tabular}
&
\begin{tabular}{@{}l@{}}
$\displaystyle
2g^{x\,{\rm (1d)}} - 2
=
 \left(
K_c+K_s^{-1}
\right) - 2$
\\[0.6em]
$\displaystyle
2\widetilde g^{x\,{\rm (1d)}} -2
=
1$
\end{tabular}
&
\begin{tabular}{@{}l@{}}
$\displaystyle
2 g_{\inter}^{x\,{\rm (1d+v)}} - 2
=
\frac{1}{2}
\left(
K_{cS}+K_{cA}^{-1}
+K_{sS}^{-1}+K_{sA}
\right)
-2
$
\\[0.6em]
$\displaystyle
2 \widetilde g_{\inter}^{x\,{\rm (1d+v)}} - 2
=
1
$
\end{tabular}
\\
\hline\hline
\end{tabular}

\vspace{0.4em}

\noindent
\begin{minipage}[t]{0.98\textwidth}
\raggedright
\footnotesize
$^{\rm {\color{blue}a}}$  
The exponents of the low-temperature normalized spin-relaxation $1/(T_1 T)$ include the phenomenological depletion factor $1-m_{\rm hx}(T)$.    The 1D values are conditional estimates discussed in Sec.~\ref{sec:relaxation_rate}.

\end{minipage}
\end{table*}

\begin{table*}[t]
\centering
\caption{Power-law exponents governing the bulk and boundary LDOS,  and the spin relaxation rate in the 2D coupled-wire systems at different temperature regimes.
For the valleyful system, the spin-relaxation exponent refers to the dominant interbranch transverse-spin
channel.}
\label{Table:summary-2D}
\small
\setlength{\tabcolsep}{6pt}
\renewcommand{\arraystretch}{1.35}
\begin{tabular}{c c c}
\hline\hline
Observable
&
2D
&
2D with valley
\\
\hline

\begin{tabular}{@{}c@{}}
Bulk LDOS\\[-0.1em]
$\beta_{\rm bulk}$ for $T>T_{\rm hx}$\\
$\widetilde{\beta}_{\rm bulk}$ for $T\ll T_{\rm hx}$
\end{tabular}
&
\begin{tabular}{@{}l@{}}
$\begin{aligned}
\beta^{\rm (2d)}_{\rm bulk}
={}&
\frac{1}{4}
\Big(
K_c
+\bar\Delta_{\theta_c,0}
+K_s+K_s^{-1}
\Big)-1
\end{aligned}$
\\[0.6em]
$\displaystyle
\widetilde{\beta}^{\rm (2d)}_{\rm bulk}
=
\frac{1}{2}
\left(
\bar K_+^{\rm (2d)}
+\bar K_-^{\rm (2d)}
\right)-1$
\end{tabular}
&
\begin{tabular}{@{}l@{}}
$\begin{aligned}
\beta^{\rm (2d+v)}_{\rm bulk}
={}&
\frac{1}{8}
\Big(
K_{cS}
+\bar\Delta_{\theta_{cS},0}
\\[-0.2em]
&\qquad
+K_{cA}+K_{cA}^{-1}
+K_{sS}+K_{sS}^{-1}
\\[-0.2em]
&\qquad
+K_{sA}+K_{sA}^{-1}
\Big)-1
\end{aligned}$
\\[0.6em]
$\displaystyle
\widetilde{\beta}^{\rm (2d+v)}_{\rm bulk}
=
\frac{1}{2}
\left(
\bar K_+^{\rm (2d+v)}
+\bar K_-^{\rm (2d+v)}
\right)-1$
\end{tabular}
\\
\hline

\begin{tabular}{@{}c@{}}
Boundary LDOS\\[-0.1em]
$\beta_{\rm bdry}$ for $T>T_{\rm hx}$\\
$\widetilde{\beta}_{\rm bdry}$ for $T\ll T_{\rm hx}$
\end{tabular}
&
\begin{tabular}{@{}l@{}}
$\displaystyle
\beta^{\rm (2d)}_{\rm bdry}
=
\frac{1}{2}
\left(
\bar\Delta_{\theta_c, 0}
+K_s^{-1}
\right)-1$
\\[0.6em]
$\displaystyle
\widetilde{\beta}^{\rm (2d)}_{\rm bdry}
=
\bar K_-^{\rm (2d)}-1$
\end{tabular}
&
\begin{tabular}{@{}l@{}}
$\begin{aligned}
\beta^{\rm (2d+v)}_{\rm bdry}
={}&
\frac{1}{4}
\Big(
\bar\Delta_{\theta_{cS}, 0}
+K_{cA}^{-1}
+K_{sS}^{-1}
+K_{sA}^{-1}
\Big)-1
\end{aligned}$
\\[0.6em]
$\displaystyle
\widetilde{\beta}^{\rm (2d+v)}_{\rm bdry}
=
\bar K_-^{\rm (2d+v)}-1$
\end{tabular}
\\
\hline

\begin{tabular}{@{}c@{}}
Spin-relaxation$^{\rm {\color{blue}a}}$: 
$1/(T_1 T)$  \\[-0.1em]
$ {2g^x-2}$ for $T>T_{\rm hx}$\\
$ {2\widetilde g^x-2}$ for $T\ll T_{\rm hx}$ 
\end{tabular}
&
\begin{tabular}{@{}l@{}}
$\displaystyle
2 g^{x\,{\rm (2d)}} - 2
=
\left(
K_c
+K_s^{-1}
\right) - 2$
\\[0.6em]
$\displaystyle
2 \widetilde g^{x\,{\rm (2d)}} - 2
=
2 \bar K_+^{\rm (2d)}
+ 2\alpha^{\rm (2d)}
-2
$
\end{tabular}
&
\begin{tabular}{@{}l@{}}
$\begin{aligned}
2g_{\inter}^{x\,{\rm (2d+v)}} - 2
={}&
\frac{1}{2}
\Big(
K_{cS}
+K_{cA}^{-1}
+K_{sS}^{-1}
+K_{sA}
\Big) - 2
\end{aligned}$
\\[0.6em]
$\displaystyle
2 \widetilde g_{\inter}^{x\,{\rm (2d+v)}} - 2
=
2\bar K_+^{\rm (2d+v)}
+2\alpha^{\rm (2d+v)} 
-2$
\end{tabular}
\\
\hline\hline
\end{tabular}

\vspace{0.4em}

\noindent
\begin{minipage}[t]{0.98\textwidth}
\raggedright
\footnotesize
$^{\rm {\color{blue}a}}$  
The exponents of the low-temperature normalized spin-relaxation rate $1/(T_1T)$ include the phenomenological depletion factor $1-m_{\rm hx}(T)$.  

\end{minipage}
\end{table*}

\section{Helix-modified power laws of physical observables}

\label{sec:Observable}

In this section, we discuss several experimentally accessible features arising in the itinerant-carrier sector, with particular emphasis on the scaling exponents governing the various power-law behaviors. For completeness, we also summarize the corresponding exponents when the helix is absent, allowing a direct comparison between the regimes before and after the formation of spin-helix order. 
The low-temperature correlation exponents presented below are evaluated within the same decoupled-sector approximation used in Sec.~\ref{Sec:helix-induced gap}.

\subsection{Local density of states}

\label{sec:LDOS}

We first examine the local density of states (LDOS). We formulate the LDOS in the two temperature regimes, $T>T_{\rm hx}$ and $T\ll T_{\rm hx}$, and distinguish between positions in the interior of a conducting channel, referred to as the ``bulk,'' and at the boundary  of a channel, referred to as the ``boundary.'' As before, we begin with the valleyless 2D coupled-wire system, while the results for the remaining platforms follow straightforwardly.

The LDOS can be computed via the equal-position  correlation function~\cite{FisherGlazman1997,Eggert2007Bosonization,Bena:2012PRB,Jia:2022HLL}, which we generalize for coupled wires, 
\begin{align}
    \rho_{{\rm dos},m} (\omega,T) = & \frac{1}{\pi \hbar}{\rm Re}\left[ \int^\infty_0 dt e^{i\omega t}\right. \nn
   &  \qquad    \times \sum_{\sigma}  \langle \big\{\psi_{ \sigma,m}(r,t),\, \psi_{  \sigma,m}^\dagger(r,0)  \big\}\rangle \Bigg].
   \label{eq:rho-dos}
\end{align}
For the bulk LDOS, the fermion field $\psi_{\sigma,m}(r,t)$ is evaluated sufficiently far from the channel boundaries.
By contrast, the boundary LDOS is determined by the corresponding boundary fermion field $\psi^{\rm bdry}_{\sigma,m}(t)$, whose explicit form is given in Appendix~\ref{app:LDOS}.

The LDOS takes a generic form, 
\begin{align}
\rho_{{\rm dos},m} (\omega,T) & \propto T^{\beta_m} \cosh \left(\frac{\hbar \omega}{2k_B T}\right) \nn
 & \quad \times \left| \Gamma \left( \frac{i\hbar \omega }{2\pi k_B T} + \frac{1+\beta_m}{2}\right) \right|^2 ,
    \label{eq:rho_E_T}
\end{align}
characterized, in general, by an exponent $\beta_m$ that depends on the temperature regime, probe location, and material platform. 
We remark that, under the periodic boundary conditions imposed in the direction transverse to the channels,  the LDOS exponent is independent of the channel index $m$.
The proportionality factor in Eq.~\eqref{eq:rho_E_T} is nonuniversal and need not be identical in the two temperature regimes. We therefore focus on the scaling exponents and the associated energy and temperature dependence, rather than on a quantitative comparison of the absolute LDOS amplitudes across $T_{\rm hx}$. 

For the representative 2D coupled-wire system without valley, the bulk LDOS exponent above $T_{\rm hx}$ is
\begin{align}
\beta_{\rm bulk}^{\rm (2d)} = \frac{1}{4}\Bigg(K_c+\bar{\Delta}_{\theta_{c},0}+K_{s} +K_{s}^{-1}\Bigg)-1, 
\label{eq:beta_bulkL_2dNoValley}
\end{align}
where we have used the fact $\bar{\Delta}_{\phi_c,n=0} = K_c$ with 
 $\bar{\Delta}_{\phi_c,n}$ given in Eq.~\eqref{eq:Delta_bar_phi_cS}. 
Deep in the spin-helix phase, the helix-induced interaction gaps out half of the low-energy modes, such that the bulk LDOS is governed by the remaining gapless sector. The corresponding exponent becomes
\begin{align}
\widetilde{\beta}_{\rm bulk}^{\rm (2d)}
=
\frac{1}{2} \left( \bar{K}^{\rm (2d)}_{+} + \bar{K}^{\rm (2d)}_{-} \right) -1,
\label{eq:beta_tilde-2dNoValley}
\end{align}
with $\bar{K}^{\rm (2d)}_{\pm}$ defined in Eq.~\eqref{eq:Kbar-n}.

We next consider the boundary LDOS for parallel channels terminating at the same longitudinal coordinate, $r=0$. As derived in Appendix~\ref{app:LDOS}, the exponent above $T_{\rm hx}$  is
\begin{align}
\beta_{\rm bdry}^{\rm (2d)}
=
\frac{1}{2}
\left(
\bar{\Delta}_{\theta_{c},0} + K_{s}^{-1} \right) -1.
\label{eq:beta_bdry-2dNoValley}
\end{align}
For $T\ll T_{\rm hx}$, the gapped modes are suppressed at low energies, leaving
\begin{align}
\widetilde{\beta}_{\rm bdry}^{\rm (2d)}
=
\bar{K}^{\rm (2d)}_{-}
-1.
\label{eq:beta_tilde_bdry-2dNoValley}
\end{align}
The corresponding bulk and boundary exponents for the remaining platforms can be obtained using the same procedure as above. 
The scaling parameters associated with the bulk and boundary behaviors of the platforms considered are summarized in  Tables~\ref{Table:summary-1D}--\ref{Table:summary-2D}.

\begin{figure}[t]
    \centering
    \includegraphics[width=0.48\textwidth]{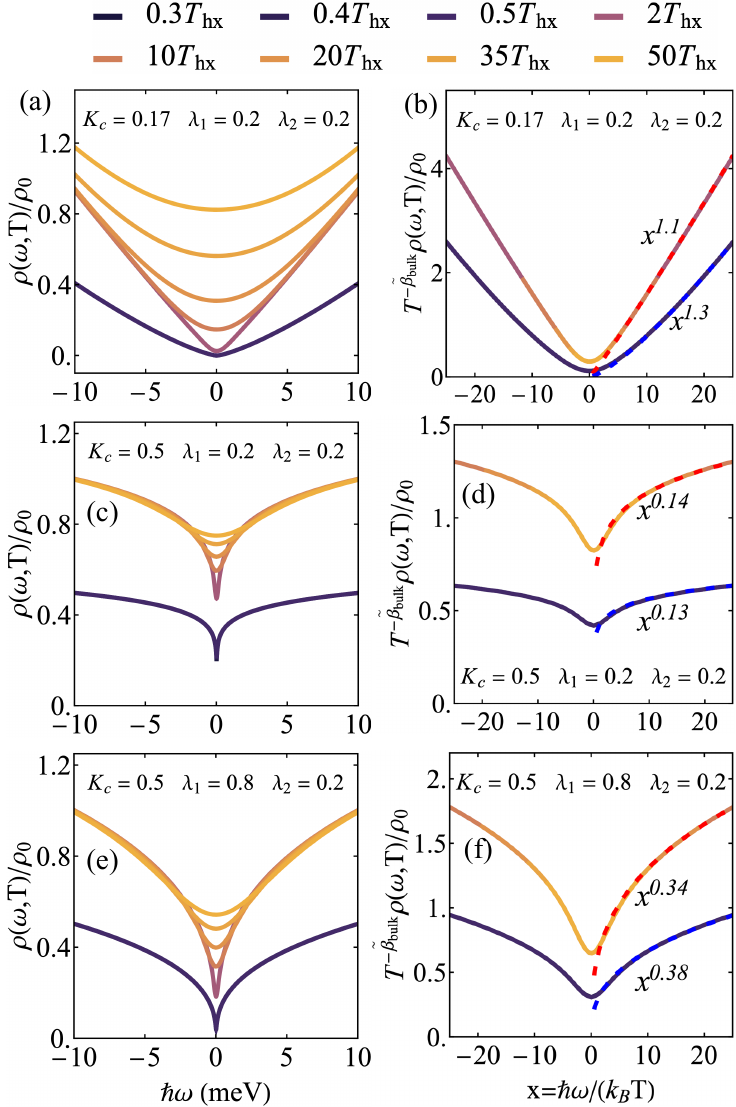}
    \caption{Illustration of the bulk-LDOS scaling forms (left column) and the corresponding finite-temperature scaling collapses (right column) for different interaction parameters $K_c$, $\lambda_1$, and $\lambda_2$. The two sets of curves use the exponents $\beta_{\rm bulk}^{\rm (2d)}$ for $T>T_{\rm hx}$ and $\widetilde{\beta}_{\rm bulk}^{\rm (2d)}$ for $T\ll T_{\rm hx}$. Panels (a,c,e) show $\rho(\omega,T)/\rho_0$, while panels (b,d,f) show the rescaled curves. The red and blue dashed lines indicate the large-$|\hbar\omega|/(k_BT)$ power laws. The low-energy expressions require $\max(|\hbar\omega|,k_BT)\ll\Delta_a$ and, in the helix phase, additionally $\max(|\hbar\omega|,k_BT)\ll\Delta_g(T)$. The displayed curves include formal extrapolations beyond these validity windows and are intended to illustrate the scaling behavior, not the complete LDOS near the gap or ultraviolet cutoff. The reference $\rho_0$ is defined as the average of the plotted $T>T_{\rm hx}$ scaling curves evaluated at $\hbar\omega=10\,{\rm meV}$ and serves only to set the vertical scale.  We take $J_K=1.0\,{\rm meV}$ and $N_{\rm ch}=20$, yielding estimated helix-ordering scales of $T_{\rm hx}\approx1\,{\rm K}$ and $T_{\rm hx}\approx0.2\,{\rm K}$ for $K_c=0.17$ and $K_c=0.5$, respectively. The color scale is set by $T/T_{\rm hx}$.}
    \label{fig:ldos}
\end{figure}

Taking the valleyless 2D coupled-wire system as a representative example, Fig.~\ref{fig:ldos} illustrates the finite-temperature scaling forms obtained from Eq.~\eqref{eq:rho_E_T} using the bulk exponents in Table~\ref{Table:summary-2D}. The two temperature regimes are described by the same functional form but different exponents, $\beta_{\rm bulk}^{\rm (2d)}$ for $T>T_{\rm hx}$ and $\widetilde{\beta}_{\rm bulk}^{\rm (2d)}$ for $T\ll T_{\rm hx}$. The dashed curves in Figs.~\ref{fig:ldos}(b,d,f) display the corresponding large-$|\hbar\omega|/(k_BT)$ power laws. In the helix phase, the gapped branches no longer contribute to the asymptotic low-energy LDOS, and the remaining gapless sector exhibits a modified scaling exponent. This change in the energy and temperature dependence provides a spectroscopic signature of the reconstructed carrier sector.

The low-temperature expression is applicable only when $\max(|\hbar\omega|,k_BT)\ll\min[\Delta_g(T),\Delta_a]$. Its asymptotic power law therefore requires $k_BT\ll|\hbar\omega|\ll\min[\Delta_g(T),\Delta_a]$. Figure~\ref{fig:ldos} displays the two scaling forms over an extended energy range, including their formal continuation outside these controlled windows. It does not describe the spectral crossover near the helix-induced gap or the temperature crossover near $T_{\rm hx}$. Since the relative microscopic prefactors of the two regimes are not determined here, the comparison concerns the scaling exponents rather than the absolute amount of spectral weight removed by the helix.

The exponents also reflect the additional tunability provided by interactions in the 2D coupled-wire systems. Both intra- and interchannel interactions can modify the magnitudes of $\beta_{\rm bulk}^{\rm (2d)}$ and $\widetilde{\beta}_{\rm bulk}^{\rm (2d)}$, 
thereby tuning the low-energy power-law dependence of the LDOS in each temperature regime. 
For the representative parameter set with $\lambda_1=\lambda_2=0.2$, weakening the intrachannel repulsion by increasing $K_{c}$ from $0.17$ (motivated by the value reported in Ref.~\cite{2022-Nature-Pengjie-1DTLL-in-2DMoire}) to $0.5$ reduces both exponents, as seen by comparing Figs.~\ref{fig:ldos}(b) and (d). The interchannel interactions provide an additional tuning degree of freedom. At fixed $K_{c}=0.5$ and $\lambda_2=0.2$, increasing $\lambda_1$ from $0.2$ to $0.8$ modifies $\widetilde{\beta}_{\rm bulk}^{\rm (2d)}$ and $\beta_{\rm bulk}^{\rm (2d)}$ again, as shown in Figs.~\ref{fig:ldos}(d) and (f).  
Although   varying the intrachannel interaction could also be possible   for a purely 1D system, it lacks the additional tuning freedom provided by interchannel interactions, a qualitative distinction between the isolated quantum wires or nanotubes  and the coupled-wire regime in tWTe$_2$ or TBG.

We now discuss possible experimental probes of the LDOS considered above. Scanning tunneling microscopy and spectroscopy (STM/STS) provide a natural means of probing the distinct bulk and boundary power laws of the LDOS. 
More generally, STM imaging has demonstrated that exchange coupling between itinerant electrons and localized moments can imprint a spatially varying magnetic texture on the itinerant-electron LDOS~\cite{HanaguriNatComm:2020-GdRuSiSTM}.
Accessing gated moiré coupled-wire devices, including tWTe$_2$ and TBG, however, generally requires electrostatic gating, for which a top gate can obstruct direct access of the STM tip to the sample surface. This limitation may be circumvented by using atomically thin layers that remain sufficiently transparent to tunneling, as demonstrated in related van der Waals STM/STS geometries~\cite{Zhizhan:2021NatComm,Yanyu:2022NatPhys}.
Alternatively, recent developments in cross-sectional STM/STS provide a complementary route by accessing a 2D device through its cross section, thereby enabling direct measurements of the boundary tunneling spectrum even in the presence of overlying device components~\cite{YPChiu:2026Nature,Weber:2026NatureNews}. Applied to a dual-gated TBG device, such a geometry could in principle probe the domain-wall modes from the side while preserving electrostatic control, providing a possible means of extracting the boundary LDOS exponent during \textit{in situ} tuning of the interlayer bias.


\subsection{Spin relaxation rate}
\label{sec:relaxation_rate}

The spin relaxation time $T_1$ provides a sensitive probe of low-energy spin fluctuations and therefore offers a means of detecting the formation of spin-helix order. The experimentally relevant quantity is the relaxation rate $1/T_1$, which acquires distinct power-law temperature dependences above and below $T_{\rm hx}$  as the low-energy spin correlations are reorganized. For the valleyful 2D coupled-wire system, this behavior was analyzed previously~\cite{YYC-2Dhelix-2025}. Here, we extend the analysis to the broader class of 1D and 2D systems with and without valley degrees of freedom, allowing a direct comparison of their characteristic relaxation exponents.

We start with the general expression. 
The relaxation rate can be expressed in terms of the local retarded transverse spin susceptibility of the itinerant carriers~\cite{slichter2013principles}, 
\begin{align}
    \frac{1}{T_{1}T}=-J_{K}^{2}a^{2}\frac{k_{B}}{\hbar^{3}}\lim_{\omega\to0}\frac{{\rm Im}\left[\chi_{\perp}^{R}(r=0,\omega)\right]}{\omega}\;,
    \label{eq:spin-relaxation-rate}
\end{align}
with  the dynamical transverse  spin susceptibility,  $\chi_{\perp}^{R}$, which we  compute by retaining the most relevant contributions.

When the spin helix is not developed, all the boson modes are gapless. The dynamical transverse susceptibility for the 2D valleyless coupled wires for $T>T_{{\rm hx}}$ is 
\begin{align}
    \chi^{R }_{\perp}(r=0,\omega) \propto & -\frac{2 \Delta_a \sin\left(\pi g^{x{\rm }}\right)}{\pi^2 \hbar v_{\rm ch}^2}\left(\frac{2\pi k_B T}{\Delta_a}\right)^{2g^{x{\rm }}-1}\nn
    & \times B \left(-\frac{i\hbar\omega}{2\pi k_B T}+g^{x{\rm  }};1-2g^{x{\rm }}\right),
\end{align}
where $B(x,y)$ denotes the Euler $\beta$-function.
In the zero-frequency limit, $1/(T_1 T)$ exhibits a  power-law temperature dependence with a nonuniversal, interaction-dependent fractional exponent:
\begin{align}
        \frac{1}{T_{1}T} \propto & \frac{J_{K}^{2}}{2\pi^{4}k_{B}\hbar}\sin\big (\pi g^{x{\rm }}\big ) \left(\frac{2\pi k_{B}}{\Delta_a}\right)^{2g^{x{\rm }}}T^{2g^{x{\rm }}-2}\nn
    & \times  \frac{\Gamma\left(1-2g^{x{\rm }}\right)\Gamma\left(g^{x{\rm }}\right)}{\Gamma\left(1-g^{x{\rm }}\right)} \left[\Psi(1-g^{x{\rm }})-\Psi (g^{x{\rm }})\right] ,  
     \label{eq:spinrelaxation-above-Thx}
\end{align}
where $\Psi(z)\equiv\left[1/\Gamma(z)\right]d\Gamma(z)/dz$ denotes the digamma function and the parameters $g^{x }$ for  the platforms are given by 
\begin{subequations}
\label{Eq:g_x-all-platform}
  \begin{align}
    g^{x{\rm (2d)}} \equiv & 
    g^{x{\rm (2d)}}_{ 0}
     = \frac{1}{2}(K_c + K^{-1}_s) ,  
    \\[0.6em]
   g^{x{\rm (2d+v)}}_\inter = & \tfrac{1}{4} \Big(
    K_{cS}+K_{cA}^{-1}  +K_{sS}^{-1}+K_{sA}\Big), 
    \\[0.6em]
g^{x{\rm (1d)}} = & \tfrac{1}{2}\left(K_c + K_s^{-1}\right),
\\[0.6em]
g^{x{\rm (1d+v)}}_\inter  = &  \tfrac{1}{4}
\Big(K_{cS}+K_{cA}^{-1} +K_{sS}^{-1}+K_{sA}\Big) .
\end{align} 
 \end{subequations}
Although the intrabranch and interbranch contributions have the same scaling exponent and hence the same power-law temperature dependence for $T>T_{\rm hx}$ under the assumption $K_{\nu P\neq cS}(q_\perp)=1$, we list only the interbranch exponent for the valleyful systems here, as the intrabranch contributions are subdominant~\cite{YYC-2Dhelix-2025} and thus not listed.

Once the spin helix develops, we evaluate the transverse spin susceptibility using the helix-adapted bosonic basis. At energies below the helix-induced gap, the remaining gapless modes determine the leading carrier contribution, with the corresponding relaxation power law proportional to $T^{2\bar K-2}$. To estimate the relaxation response of the ordered local-moment system, we additionally adopt a phenomenological spin-wave-depletion weighting. Specifically, we use $1-m_{\rm hx}(T)$ as a measure of the thermally fluctuating local-spin contribution, rather than interpreting it as a literal fraction of independently disordered moments. Within this approximation, the relaxation response takes the form
\begin{align}
\frac{1}{T_1T}
\propto
[1-m_{\rm hx}(T)]T^{2\bar K-2}  ,
    \label{eq:spinrelax-higher-Thx}
\end{align}  
with the parameter $\bar{K}$ given by 
\begin{subequations}
\label{Eq:K-bar_x-all-platform}
\begin{align}
\bar{K}_{+}^{\rm (2d)} = & \int_{-\pi}^\pi
\frac{d(q_\perp d)}{2\pi}\widetilde{K}^{\rm (2d)}(q_\perp), 
\\[0.3em]
\bar{K}_{+}^{\rm (2d+v)} = & \int_{-\pi}^\pi
\frac{d(q_\perp d)}{2\pi}\widetilde{K}^{\rm (2d+v)}(q_\perp), 
\\[0.3em]
\bar{K}^{\rm (1d)} = & \sqrt{
  \frac{v_c K_c+ {v_s}/{K_s}} {v_sK_s+ {v_c}/{K_c}}
},
\\[0.3em]
\bar{K}^{\rm (1d+v)} = & \sqrt{
  \frac{\sum_P \left(v_{cP}K_{cP}+ {v_{sP}} / {K_{sP}}\right)}{\sum_{P'}\left(v_{sP'}K_{sP'}+ {v_{cP'}}/{K_{cP'}}\right)}
}.
\end{align}
 \end{subequations}
In the above, $\widetilde{K}^{\rm (2d)}(q_\perp)$ is given by Eq.~\eqref{eq:tilde-v-K};  the valleyful effective parameter $\widetilde K^{\rm (2d+v)}(q_\perp)$ is given in Eq.~\eqref{eq:tilde-v-K-valley}.
 
The temperature dependence can be summarized as 
\begin{subnumcases}
{\frac{1}{T_1 T } \propto}
T^{2g^{x}-2}, & $T>T_{{\rm hx}}$, 
\label{eq:pwr-spinrelaxT-1} \\
T^{2\widetilde{g}^x-2} ,  
 & $T \ll T_{{\rm hx}},$ 
 \label{eq:pwr-spinrelaxT-2} 
\end{subnumcases}
as summarized in Tables~\ref{Table:summary-1D}--\ref{Table:summary-2D}.
Here, we denote   $\widetilde{g}^x \equiv \bar{K}_{}^{\rm } + \alpha^{\rm }$,  with  $\bar{K}$ given in Eq.~\eqref{Eq:K-bar_x-all-platform} and $\alpha^{\rm }$  a nonuniversal fractional exponent specific to the 2D coupled-wire system, as introduced in Sec.~\ref{sec:RKKY}. For example, Ref.~\cite{YYC-2Dhelix-2025} extracted $\alpha$ from a numerical fit over the temperature range $0.01\leq T/T_{\rm hx} \leq 0.1$.

The isolated 1D systems provide useful reference limits for this comparison. Previous finite-size spin-wave analyses obtain a generalized Bloch law for the helix magnetization, with an exponent determined by the transverse carrier susceptibility entering the RKKY interaction~\cite{Braunecker-2009,CHH-CNT-helix}. If the RKKY interaction and the relaxation channel are governed by the same effective exponent $\bar K$, this law gives $1-m_{\rm hx}(T)\propto T^{3-2\bar K}$. Combining it with the depletion-weighted relaxation estimate then yields $1/(T_1T)\propto T$ over the temperature range where these approximations apply.  For the coupled-wire systems, the numerically determined exponents $\alpha^{\rm (2d)}$ and $\alpha^{\rm (2d+v)}$ enter the relaxation response through $\widetilde g^x=\bar K+\alpha$, retaining the dependence on the coupled-channel magnetic fluctuations.

Spin-resonance techniques provide a natural probe of spin dynamics and the spin relaxation rate~\cite{Book-Resistive-NMR}. For the low-dimensional and nanoscale platforms considered here, conventional inductive detection may have reduced sensitivity because of the relatively small number of localized moments participating in the resonance. This limitation can be overcome when a sufficiently large ensemble of samples is available. For example, conventional NMR has been performed on $^{13}\mathrm{C}$-enriched carbon nanotubes using macroscopic collections of nanotube samples, thereby providing sufficient signal for detection~\cite{Single-CNT-NMR-PRL}. Such an ensemble-based approach, however, is less suitable for individually gated moir{\'e} devices.

Resistively detected spin resonance provides an attractive alternative for these systems~\cite{Klitzing:1991PRL,Resistive-NMR-Klitzing,Hirayama:2009SST,muraki:2012,Lyon:2017PRL}. Rather than detecting the inductive signal generated directly by the resonating localized moments, this approach measures the accompanying change in electrical transport. A change in the spin polarization of the localized moments can modify the carrier spectrum or spin-dependent scattering and thereby produce a measurable electrical response. This mechanism is particularly favorable for nanoscale systems: even when the number of localized moments is small, their coupling to a narrow conducting channel can generate a detectable transport signature. The spin relaxation time $T_1$ can then be extracted from the recovery of the electrical signal following the resonance excitation~\cite{Klitzing:1991PRL,muraki-science-2006}.

\begin{table*}[t]
\caption{Scaling exponents governing the leading algebraic contributions to the connected order-parameter correlations of the itinerant-carrier subsystem in the spin-helix phase, $T\ll T_{\rm hx}$, for the four systems considered in this work. 
For valleyful systems,  intrabranch and interbranch contributions are distinguished where applicable.
The CDW, SDW$_z$, and TS$_{x/y}$ correlations are omitted because their bosonized operators contain dual fields of the pinned sectors and therefore have exponentially decaying correlations. 
}
\centering
\setlength{\tabcolsep}{10pt}
\renewcommand{\arraystretch}{1.7} 
\begin{tabular}{c c c c c}
\hline\hline
Platform
&
\multicolumn{2}{c}{SDW$_{x/y}$}
&
\multicolumn{2}{c}{SS/TS$_z$}
\\
\hline

2D without valley
&
\multicolumn{2}{c}{$2\bar{K}_{+}^{\rm (2d)}$}
&
\multicolumn{2}{c}{$2\bar{K}_{-}^{\rm (2d)}$}
\\
\hline

2D with valley
&
$(\intra)\quad
\bar{K}_{+}^{\rm (2d+v)}
+
\bar{K}_{-}^{\rm (2d+v)}$
&
$(\inter)\quad
2\bar{K}_{+}^{\rm (2d+v)}$
&
$(\intra)\quad
\bar{K}_{+}^{\rm (2d+v)}
+
\bar{K}_{-}^{\rm (2d+v)}$
&
$(\inter)\quad
2\bar{K}_{-}^{\rm (2d+v)}$
\\
\hline

1D without valley
&
\multicolumn{2}{c}{$2\bar{K}^{\rm (1d)}$}
&
\multicolumn{2}{c}{$\displaystyle
{2}/{\bar{K}^{\rm (1d)}}$}
\\
\hline

1D with valley
&
$(\intra)\quad
\displaystyle
\bar{K}^{\rm (1d+v)}
+
{1}/{\bar{K}^{\rm (1d+v)}}$
&
$(\inter)\quad
2\bar{K}^{\rm (1d+v)}$
&
$(\intra)\quad
\displaystyle
\bar{K}^{\rm (1d+v)}
+
{1}/{\bar{K}^{\rm (1d+v)}}$
&
$(\inter)\quad
\displaystyle
{2}/{\bar{K}^{\rm (1d+v)}}$
\\

\hline\hline
\end{tabular}
\label{tab:exponents_below_Thx}
\end{table*}

\section{Competing correlations in the spin-helix phase} 

\label{sec:helix_induced_order}

As discussed in Sec.~\ref{sec:1DKondolattice}, the formation of the spin helix strongly renormalizes selected bosonic sectors, which  can substantially modify the scaling exponents and therefore the relative strength of competing ordering tendencies. In this section, we examine the correlation functions of density-wave and superconducting operators and determine their corresponding scaling exponents. This allows us to identify how the spin helix  reshapes the dominant ordering tendencies. 

We begin by defining the order-parameter operators for the valleyless coupled-wire system. As in the response-function analysis, we focus on the backscattering contributions. The density-wave operators can be written as
\begin{align}
\mathcal{O}_{{\rm dw}, m}
=
\sum_{\sigma,\sigma^{\prime}}
\psi_{R\sigma,m}^{\dagger}
\left(\mathbf{M}_{{\rm dw}} \right)_{\sigma\sigma^{\prime}}
\psi_{L\sigma^{\prime},m}.
\label{eq:O_dw_short}
\end{align}
For the CDW, $\mathcal{O}_{{\rm dw},m}=\mathcal{O}_{{\rm cdw},m}$, we take $\mathbf{M}_{\rm dw}=\sigma^{0}\equiv\mathbb{I}$, while for the $\mu$ component of the SDW, $\mathcal{O}_{{\rm dw},m}=\mathcal{O}_{{\rm sdw},m}^{\mu}$, we take $\mathbf{M}_{\rm dw}=\sigma^{\mu}$. The corresponding explicit bosonized forms are given in Eq.~\eqref{eq:B_dw_2d} of Appendix~\ref{sec:order-parameter}.

 We consider both conventional zero-center-of-mass-momentum pairing channels and, for valleyful systems, the finite-momentum interbranch pairing channels introduced below. The corresponding operators of the former are defined as
\begin{align}
\mathcal{O}_{{\rm sc},m}
=
\sum_{\sigma,\sigma^{\prime}}
\sigma
\psi_{R\sigma,m}^{\dagger}
\left(\mathbf{M}_{\rm sc}\right)_{\sigma\sigma^{\prime}}
\psi_{L\bar{\sigma}^{\prime},m}^{\dagger}.
\label{eq:O_sc_short}
\end{align}
Here,  $\bar{\sigma}$ denotes the spin opposite to $\sigma$. 
For spin-singlet    superconductivity (SS), $\mathcal{O}_{{\rm sc},m}=\mathcal{O}_{{\rm ss},m}$, we take $\mathbf{M}_{\rm sc}=\sigma^0$, whereas for the $\mu$ component of spin-triplet superconductivity (TS), $\mathcal{O}_{{\rm sc},m}=\mathcal{O}_{{\rm ts},m}^{\mu}$, we take $\mathbf{M}_{\rm sc}=\sigma^{\mu}$. The explicit bosonized forms of the  pairing operators are given in Eq.~\eqref{eq:B_SC_2d} of Appendix~\ref{sec:order-parameter}.

For the valleyful 1D and 2D systems, both intrabranch and interbranch processes can be considered for the density-wave and  pairing channels. The extension of the density-wave operators is straightforward. Here, we introduce the pairing operators, for which the distinction between the two processes becomes particularly important. The intrabranch pairing operator is
\begin{align}
\mathcal{O}_{{\rm sc},\delta,m}^{\intra}
=
\sum_{\sigma,\sigma^{\prime}}
\sigma
\psi_{R\delta\sigma,m}^{\dagger}
\left(\mathbf{M}_{\rm sc}\right)_{\sigma\sigma^{\prime}}
\psi_{L\delta\bar{\sigma}^{\prime},m}^{\dagger},
\end{align}
shown in Eq.~\eqref{eq:B_SC} of Appendix~\ref{sec:order-parameter}.
Defining $\bar{\delta}$ as the branch opposite to $\delta$, the interbranch pairing operator is
\begin{align}
\mathcal{O}_{{\rm sc},\delta,m}^{\inter}
=
\sum_{\sigma,\sigma^\prime}
\sigma
\psi_{R\delta\sigma,m}^{\dagger}
\left(\mathbf{M}_{\rm sc}\right)_{\sigma\sigma^\prime}
\psi_{L\bar{\delta}\bar{\sigma}^{\prime},m}^{\dagger},
\end{align}
whose bosonized form follows from Eq.~\eqref{eq:B_SC}, as discussed in Appendix~\ref{sec:order-parameter}. For both cases, $\mathbf{M}_{\rm sc}=\sigma^0$ and $\mathbf{M}_{\rm sc}=\sigma^\mu$ correspond to the spin-singlet and spin-triplet components, respectively. In contrast to the zero-momentum pairing channel, the interbranch pairing considered here carries a finite center-of-mass momentum and therefore corresponds to a spatially modulated superconducting correlation.

Having introduced the candidate order-parameter operators, we now examine their correlation functions in the spin-helix phase. 
We consider connected correlations, with the static expectation-value contribution induced by the established helix subtracted. The exponents below characterize the leading algebraic contributions from the remaining gapless modes. 
As discussed in Sec.~\ref{sec:RKKY}, the helix-induced interaction gaps part of the itinerant-carrier spectrum and renormalizes the remaining gapless modes. Incorporating this effect, we obtain the corresponding scaling exponents for the four systems considered here, as summarized in Table~\ref{tab:exponents_below_Thx}. These results describe the low-temperature regime $T\ll T_{\rm hx}$, where the spin helix is well developed. As seen from the table, the scaling exponents of the 1D and 2D valleyless systems are related through the replacement $\bar{K}_{\pm}^{{\rm (2d)}} \leftrightarrow \left[\bar{K}^{{\rm (1d)}}\right]^{\pm}$. An analogous correspondence holds for the valleyful systems.
 
The ordering tendencies are determined by comparing the scaling exponents of the corresponding correlation functions: a smaller exponent implies a more slowly decaying correlation, or a stronger order-parameter susceptibility, and hence a stronger tendency toward the corresponding order among the correlation channels considered here.

Within the interaction model adopted here, one can show that the transverse SDW$_{x/y}$ correlation has the smallest scaling exponent throughout the stable repulsive region of the $(\lambda_1,\lambda_2)$ parameter space.
It therefore represents the dominant ordering tendency. Likewise, for a valleyful SLL, one finds
$\bar{K}_{-}^{\mathrm{(2d+v)}}-\bar{K}_{+}^{\mathrm{(2d+v)}}>0$
throughout the repulsive regime $0<K_{cS}<1$ within the stable valleyful-SLL region. Consequently, the interbranch SDW$_{x/y}$ correlation has the smallest scaling exponent and is therefore the dominant ordering tendency. Physically, the enhanced interbranch SDW$_{x/y}$ tendency can be traced to the spin-helix state, whose formation is driven by the dominant interbranch contribution to the transverse spin susceptibility. Because the corresponding intrabranch contribution remains subleading, the low-temperature transverse spin correlations in the helical regime are governed primarily by interbranch processes, thereby favoring the interbranch SDW$_{x/y}$ channel.   
For the isolated 1D valleyful and valleyless systems, the ordering tendency follows directly by setting $\lambda_1=\lambda_2=0$, resulting in the same dominant ordering tendency as their coupled-wire counterparts.

As a remark, these ordering tendencies are obtained within the carrier sector reconstructed by an already established spin helix. Therefore, coexistence of the helix with any additional order requires the reconstructed carrier spin susceptibility to continue to support the RKKY interaction responsible for the helix. 
This situation is analogous to RKKY-induced helical order in proximity-coupled superconducting wires, where the opening of a pairing gap modifies and suppresses the $2k_F$ spin susceptibility, while a residual RKKY interaction can nevertheless stabilize the helix self-consistently~\cite{Klinovaja:2013,Braunecker:2013}. In the present case, an additional ordering gap in the remaining carrier modes may similarly weaken the RKKY interaction.
Determining whether such an instability emerges, its characteristic temperature scale, and its coexistence with the helix requires a self-consistent treatment beyond the correlation exponents obtained here. Accordingly,  Table~\ref{tab:exponents_below_Thx} establishes the hierarchy of secondary correlations within an already developed spin-helix phase, rather than an additional thermodynamic transition.

\section{Discussion and Conclusions}
\label{sec:Conclusions}

Several general conclusions emerge from comparing the four classes considered here. Across both TLL and SLL systems, with and without valley degrees of freedom, the carrier-mediated RKKY interaction can stabilize helical order of the localized moments. The resulting static helical field partially gaps the itinerant-carrier spectrum and renormalizes the remaining gapless modes. Dimensionality and valley structure do not alter this basic mechanism, but they determine the ordering wave vector, characteristic helix-ordering scale, helix-induced gap, and the scaling exponents governing the low-energy responses.
The corresponding LDOS and spin-relaxation exponents for the isolated
and coupled-wire systems are summarized in
Tables~\ref{Table:summary-1D} and~\ref{Table:summary-2D}, respectively.

The main distinction between the TLL and SLL cases lies in the collective nature of the magnetic order. In an isolated TLL, the helix is stabilized within a single channel, whereas in an SLL, the RKKY interaction couples localized moments on different wires and locks their helical phases. The resulting quasi-2D helix is therefore more robust than a collection of independent 1D helices. For the demonstrated valleyful case in Ref.~\cite{YYC-2Dhelix-2025}, this collective enhancement raises $T_{\rm hx}$ by more than an order of magnitude relative to the corresponding isolated-wire estimate. The valley degree of freedom instead changes the dominant backscattering processes and the combinations of carrier modes affected by the helical field, leading to distinct ordering wave vectors, gaps, and low-energy exponents.

The reconstructed carrier sector provides several possible signatures of the ordered state. In particular, the helix-induced partial gap and the modified power-law behavior of the remaining gapless modes can be probed through tunneling and spin-sensitive measurements. These observables establish the feedback of the static magnetic order on the itinerant carriers, but they do not directly resolve its spatially rotating structure. 
Dynamical coupling between the reconstructed carriers and fluctuations of the ordered helix may further modify the correlation functions and transport response, but lies beyond the static treatment considered here. 
The small or vanishing spatially averaged magnetization also makes the helical order difficult to identify through conventional bulk magnetometry.

Our Josephson-junction results provide a complementary transport probe of the helical texture through its orientation-dependent interference response. Since changing the channel orientation also interchanges the hopping and lattice anisotropies in our coarse-grained model, we identify the helical contribution by comparison with a matched $B_{\rm hx}=0$ junction for each orientation. For phase-coherent helices with channels and helix propagation perpendicular to the current, the central-lobe width $W_0$ substantially exceeds both the ideal Fraunhofer value and the matched reference. The relative peak weight $R_1$ decreases with increasing $B_{\rm in}$ but remains above the reference over the common field range shown. Both modifications are much weaker for parallel channels. Finite side-lobe asymmetry $|A_1|$ and central-peak displacement $\Delta\phi_z$ also develop with increasing field in the perpendicular geometry, while both remain negligible for parallel channels and in the matched no-helix references for the parameters considered.
Flux focusing makes $B_{\rm in}$ a tunable  probe through the interface field profile.  Tracking the combined field and orientation dependence of $W_0$, $R_p$, $A_p$, and $\Delta\phi_z$ against matched references therefore provides more information about the helical contribution than a single distorted interference pattern.

While similar features can arise from disorder or an inhomogeneous current distribution, a systematic weakening of the additional changes in $W_0$, $R_p$, $A_p$, and $\Delta\phi_z$ near $T_{\rm hx}$, relative to the nonhelical background, would support their association with the helix. Correlating these transport changes with the helix-induced gap or modified power-law responses would provide further evidence for carrier-mediated helical order. More broadly, the coupled-channel geometry considered here is not restricted to moiré systems. Parallel arrays of 1D conducting channels can also be engineered by etching semiconductor quantum wells~\cite{Sato:2019PRB}, providing an alternative route to coupled-wire Josephson junctions.

The 2D moir\'e platforms offer an additional advantage through their electrical tunability. In TBG, experimentally adjustable parameters such as the twist angle, interlayer bias, carrier density, and distance to a metallic screening gate can modify the carrier velocity, bandwidth, spatial profile of the 1D conduction modes, and effective interaction strengths~\cite{Efimkin-2018-TBG-network,HCWang-TBG}. Since the RKKY coupling is determined by the itinerant-carrier spin susceptibility, these controls provide indirect means of tuning the ordering wave vector, $T_{\rm hx}$, the effective helical field $B_{\rm hx}$, and the helix-induced gap $\Delta_g$. They also modify the bulk and boundary LDOS exponents and the spin-relaxation behavior. For TBG, the effective interaction parameters can be estimated from the domain-wall density profiles obtained from the continuum Bistritzer-MacDonald model in the presence of an interlayer bias~\cite{Bistritzer-PNAS-2011,HCWang-TBG}. Such control is generally less direct in conventional 1D platforms such as CNTs and GaAs quantum wires. 

The range of potential realizations extends beyond the four representative platforms studied explicitly. Twisted anisotropic homobilayers have been proposed to exhibit a ``moir\'e collapse,'' in which a 2D moir\'e pattern is transformed into an effectively 1D crystal near a new class of magic angles~\cite{Duarte:2026PNAS_collapseMoire}. Related 1D structures have also been discussed in twisted trilayer graphene~\cite{Nakatsuji:2023PRX_TTG}, graphene nanoslides~\cite{Christophe:PRB2026_grapheneslides}, and strained twisted bilayer 1T$^\prime$-WTe$_2$~\cite{StrainWTe2:2024}. Quasi-1D electronic structures further occur in van der Waals materials such as the layered magnet CrSBr~\cite{Quasi1DCrSBr:2023}. These developments suggest additional settings in which interacting 1D carrier modes coexist with localized magnetic moments or can be coupled to them through magnetic proximity, implanted moments, or hyperfine interactions. Studies of honeycomb networks of 1D conducting channels, incorporating scattering at the junctions, have predicted flat bands and enhanced superconductivity~\cite{Jongjun:2020PRL}, as well as non-Fermi-liquid transport at finite temperatures~\cite{Jongjun:2021PRL}. These findings further motivate examining how the combined effect of TLL correlations within the channels and scattering at the junctions modifies the carrier spectrum and the spin susceptibility that mediates the RKKY interaction.

Our results connect carrier-mediated helical order with complementary signatures of carrier-sector reconstruction and spatial magnetic texture. In coupled-wire systems, tunneling and spin-sensitive measurements probe the reconstructed low-energy modes, while field-tunable Josephson interferometry probes the orientation-dependent transport response of the helix.

\begin{acknowledgments}

We acknowledge interesting discussions with S.-Y.~Chen, Y.-P.~Chiu, T.~Hanaguri, D.~Hsieh, Y.~Kato, C.-T.~Ke, S.-F.~Lee, A.~Matos-Abiague, M.~Oshikawa, J.~Puebla, K.~Shiozaki, T.~Tohyama, K.~Totsuka,  Y.-C.~Tsai,  J.~I.-J. Wang, and N.-C.~Yeh. This work was supported financially by the National Science and Technology Council (NSTC), Taiwan (Grant No.~NSTC-114-2112-M-001-057 and Grant No.~115-2112-M-001-024), and Academia Sinica (AS), Taiwan  (AS-iMATE-114-12). 
Y.-Y.C. acknowledges the financial support from The 2023 Postdoctoral Scholar Program of Academia Sinica, Taiwan. 

\end{acknowledgments}

\section*{Data Availability }
The data that support the findings of this work are openly available at \url{https://doi.org/10.5281/zenodo.22872659}, Ref.~\cite{zenodo_data}.


\appendix

\section{Stability constraints on $\lambda_1$ and $\lambda_2$}
\label{app:constraint_lambda}

In this section, we derive the constraints on $\lambda_1$ and $\lambda_2$ required for the stability of the SLL. The stability requires 
\begin{align}
f(q_{\perp})\equiv1+\lambda_{1}\cos\left(q_{\perp}d\right)+\lambda_{2}\cos\left(2q_{\perp}d\right)>0
\end{align}
for all $q_{\perp}d\in[-\pi,\pi]$. For convenience, we define $y=\cos\left(q_{\perp}d\right)$, so that $y\in[-1,1]$, and $f(y)=2\lambda_{2}y^{2}+\lambda_{1}y+1-\lambda_{2}$.
For $\lambda_2 \neq 0$, the stationary point of the quadratic polynomial occurs at $y=y_{0}\equiv-\frac{\lambda_1}{4\lambda_2}$, with
\begin{align}
f(y_{0})=\frac{-\lambda_{1}^{2}-8\lambda_{2}^{2}+8\lambda_{2}}{8\lambda_{2}}\;.
\end{align}
In the following, we divide the discussion into two cases: $\lambda_{2}>0$ and $\lambda_{2}<0$:
\begin{itemize}
    \item For $\lambda_{2}>0$: If $|y_0|\leq1$, $f(y_{0})$ corresponds to the global minimum of $f(y)$. Incorporating $\lambda_{2}>0$ and $|y_0|\leq1$, the constraint on $\lambda_{1}$ and $\lambda_{2}$ for this case is given by
    \begin{align}
    {\rm (i):}~\frac{\lambda_{1}^{2}}{2}+4\left(\lambda_{2}-\frac{1}{2}\right)^{2}<1
    ~~\cap~~\left|-\frac{\lambda_1}{4\lambda_2}\right|\leq1
    ~~\cap~~\lambda_2>0\,.
    \label{eq:constraint_i}
    \end{align}
    If $|y_0|>1$, ${\rm min}[f(y)]$ occurs at $y=\pm1$ for $\lambda_{1}\lessgtr0$, with $f(y=\pm1)=\lambda_{2}-|\lambda_{1}|+1$, since $y_0$ lies outside the domain $y\in[-1,1]$. We arrive at the second constraint on $\lambda_{1}$ and $\lambda_{2}$, given by
    \begin{align}
    {\rm (ii):}~\lambda_{2}-|\lambda_{1}|+1>0
    ~~\cap~~\left|-\frac{\lambda_1}{4\lambda_2}\right|>1
    ~~\cap~~\lambda_{2}>0\,.
    \label{eq:constraint_ii}
    \end{align}

    \item For $\lambda_{2}<0$: The minimum of $f(y)$ always occurs at $y=\pm1$ for $\lambda_{1}\lessgtr0$, regardless of the value of $y_{0}$. Therefore, we obtain the third constraint, expressed as
    \begin{align}
    {\rm (iii)}:~\lambda_{2}-|\lambda_{1}|+1>0~~\cap~~\lambda_{2}<0\,.
    \label{eq:constraint_iii}
    \end{align}
\end{itemize}
For $\lambda_2=0$, positivity over $y\in[-1,1]$ requires $|\lambda_1|<1$. 

In summary, the constraint on $\lambda_{1}$ and $\lambda_{2}$ that ensures $f(q_\perp)>0$ for all $q_{\perp}d\in[-\pi,\pi]$ is given by the union of conditions (i)--(iii), as specified in Eqs.~\eqref{eq:constraint_i} to \eqref{eq:constraint_iii}, together with the condition from $\lambda_2 = 0$.

\section{Unfolding method and the local density of states near the boundary}
\label{app:LDOS}

In this section, we present details of the evaluation of the LDOS of the 2D coupled-wire system near the system boundary, using the unfolding method~\cite{Eggert:1992PRB,Fabrizio:1995PRB,Giamarchi2003}. 
We focus on the valleyless SLL, which can be extended straightforwardly to the valleyful SLL.
For a semi-infinite system, the bosonic fields are first rescaled to cast the interacting Hamiltonian into a free-boson form. The boundary relation between the left- and right-moving chiral fields is then used to extend the theory onto the whole system. We first work in the channel basis and then impose the periodic boundary condition along the direction perpendicular to the channels.

To proceed, we begin by expressing the bosonized Hamiltonian $H^{\rm (2d)}_{\rm it}$ of the 2D valleyless coupled-wire system [Eq.~\eqref{2Dcoupled_novalley_Hee}] in the real-space channel basis:
\begin{align}
H^{\rm (2d)}_{{\rm it}}
=
\sum_{\nu}\int_{0}^{\infty}\frac{\hbar dr}{2\pi}
\big[&
(\partial_{r}\boldsymbol{\phi}_{\nu})^{T}\mathbf{V}_{\phi_{\nu}}(\partial_{r}\boldsymbol{\phi}_{\nu})
\nn
&  +
(\partial_{r}\boldsymbol{\theta}_{\nu})^{T}\mathbf{V}_{\theta_{\nu}}(\partial_{r}\boldsymbol{\theta}_{\nu})
\big].
\end{align}
Here, $\boldsymbol{\phi}_{\nu}$ and $\boldsymbol{\theta}_{\nu}$ are $N_{{\rm ch}}$-component vectors in the real-space channel basis, 
\begin{align}
\boldsymbol{\phi}_{\nu}
&=
\left(
\phi_{\nu,1},
\phi_{\nu,2},
\dots,
\phi_{\nu,N_{\rm ch}}
\right)^T, \nn
\boldsymbol{\theta}_{\nu}
&=
\left(
\theta_{\nu,1},
\theta_{\nu,2},
\dots,
\theta_{\nu,N_{\rm ch}}
\right)^T.
\label{eq:boson_channel_basis}
\end{align}
Here, $\mathbf V_{\phi_\nu}$ and $\mathbf V_{\theta_\nu}$ are symmetric positive-definite velocity matrices, which reduce to $v_{\nu}/K_{\nu}$ and $v_{\nu}K_{\nu}$, respectively, in the single-channel limit. We define the corresponding Luttinger-liquid matrix $\boldsymbol{\mathcal K}_{\nu}$ through
\begin{equation}
\boldsymbol{\mathcal{K}}_{\nu}
\mathbf{V}_{\phi_{\nu}}
\boldsymbol{\mathcal{K}}_{\nu}
=
\mathbf{V}_{\theta_{\nu}}.
\label{eq:lutt-matrix-definition}
\end{equation}
In the single-wire limit, $\boldsymbol{\mathcal K}_{\nu}$ reduces to the conventional TLL parameters $K_{\nu}$, and Eq.~\eqref{eq:lutt-matrix-definition} becomes
a trivial equality.

To implement the unfolding procedure, we transform the coupled-wire Hamiltonian to the normal-mode basis and simultaneously rescale the bosonic fields:
\begin{align}
\widetilde{\boldsymbol{\phi}}_{\nu}
=
\mathbf{O}_{\nu}^{T}
\bm{\mathcal{K}}_{\nu}^{-1/2}
\boldsymbol{\phi}_{\nu},
&
\widetilde{\boldsymbol{\theta}}_{\nu}
=
\mathbf{O}_{\nu}^{T}
\bm{\mathcal{K}}_{\nu}^{1/2}
\boldsymbol{\theta}_{\nu},
\end{align}
where $\widetilde{\boldsymbol{\phi}}_{\nu}$ and $\widetilde{\boldsymbol{\theta}}_{\nu}$ are the normal-mode bosons, and $\mathbf O_{\nu}$ is an orthogonal matrix that diagonalizes the velocity matrix $\mathbf{U}_{\nu}$,
\begin{align}
\mathbf{O}_{\nu}^{T}\mathbf{U}_{\nu}\mathbf{O}_{\nu}
=
{\rm diag}
\left(
v_{\nu,1},
\ldots,
v_{\nu,N_{{\rm ch}}}
\right),
\end{align}
with $N_{\rm ch}$ distinct normal-mode velocities, and
\begin{align}
\mathbf{U}_{\nu}
=
\boldsymbol{\mathcal{K}}_{\nu}^{1/2}
\mathbf{V}_{\phi_{\nu}}
\boldsymbol{\mathcal{K}}_{\nu}^{1/2}
=
\boldsymbol{\mathcal{K}}_{\nu}^{-1/2}
\mathbf{V}_{\theta_{\nu}}
\boldsymbol{\mathcal{K}}_{\nu}^{-1/2}.
\end{align}

We next impose the open-boundary condition. We assume that spin-flipping and single-particle interchannel hopping are absent, while density correlations across different channels are retained. Under these assumptions, the open-boundary condition can be imposed separately on the fermion field of each channel. In the bosonized description, the vanishing of the full fermion fields at the boundary fixes $\phi_{\nu,m}$ to time-independent values,
\begin{align}
\boldsymbol{\phi}_{\nu}(r=0,\tau)
=
\boldsymbol{\phi}_{\nu,0},
\label{eq:phi_pinning}
\end{align}
where $\boldsymbol{\phi}_{\nu,0}$ is a constant vector independent of time. In the normal-mode basis, the corresponding boundary condition becomes
\begin{align}
\widetilde{\boldsymbol{\phi}}_{\nu}(r=0,\tau)
=
\mathbf O_{\nu}^{T}
\boldsymbol{\mathcal K}_{\nu}^{-1/2}
\boldsymbol{\phi}_{\nu,0}
\equiv
\widetilde{\boldsymbol{\phi}}_{\nu,0}.
\label{eq:tilde_phi_pinning}
\end{align}

For the dual boson fields of the $\nu$ sector in the normal-mode basis, we introduce the corresponding right-/left-moving chiral fields as
\begin{align}
\widetilde{\bm{\phi}}_{\nu}^{R/L}
=
\widetilde{\boldsymbol{\theta}}_{\nu}
\mp
\widetilde{\boldsymbol{\phi}}_{\nu},
\end{align}
where $\widetilde{\bm{\phi}}_{\nu}^{R/L}$ are the right/left-moving chiral fields associated with the $\nu$ sector. The boundary condition of Eq.~\eqref{eq:tilde_phi_pinning} can be further extended to an arbitrary position, given by
\begin{align}
\widetilde{\bm{\phi}}_{\nu}^{L}(r,t)
-
\widetilde{\bm{\phi}}_{\nu}^{R}(-r,t)
=
2\widetilde{\boldsymbol{\phi}}_{\nu,0}.
\label{eq:unfolding}
\end{align}
Equation~\eqref{eq:unfolding} allows us to introduce a set of free chiral fields where one half-space is described by $\widetilde{\bm{\phi}}_{\nu}^{R}$ and the other by $\widetilde{\bm{\phi}}_{\nu}^{L}$, defined as follows,
\begin{subequations}
\label{eq:new-chiral-field}
\begin{align}
\widetilde{\bm{\phi}}_{\nu}^{R}(r,t)
&\to
\widetilde{\bm{\phi}}_{\nu}^{R,e}(r,t),
\\
\widetilde{\bm{\phi}}_{\nu}^{L}(r,t)
&\to
\widetilde{\bm{\phi}}_{\nu}^{R,e}(-r,t).
\end{align}
\end{subequations}
This indicates that the system on the half interval can be unfolded onto the whole interval and represented entirely in terms of a single set of free chiral fields $\widetilde{\bm{\phi}}_{\nu}^{R,e}$ defined on the entire system. The constant $\widetilde{\boldsymbol{\phi}}_{\nu,0}$ contributes only an overall phase and does not affect the scaling properties.

Using  Eqs.~\eqref{eq:phi_pinning}--\eqref{eq:new-chiral-field}, we express the fermion-field operator on the boundary as
\begin{align}
\psi^{\rm bdry}_{\ell\sigma,m}
&\propto
\exp\left[
\frac{i}{\sqrt{2}}
\sum_{\nu}
\sum_{p=1}^{N_{\rm ch}}
\eta_{\nu}^{\ell\sigma}
\left(
\boldsymbol{\mathcal{K}}_{\nu}^{-1/2}\mathbf{O}_{\nu}
\right)_{mp}
\widetilde{\phi}_{\nu,p}^{R}
\right]
\nonumber  \\
&=
\exp\Bigg[
\frac{i}{\sqrt{2}}
\sum_{\nu}
\sum_{p=1}^{N_{{\rm ch}}}
\eta_{\nu}^{\ell\sigma}
\left(
\boldsymbol{\mathcal{K}}_{\nu}^{-1/2}\mathbf{O}_{\nu}
\right)_{mp}
\left(
\widetilde{\theta}_{\nu,p}^{e}
-
\widetilde{\phi}_{\nu,p}^{e}
\right)
\Bigg],
\end{align}
where $\eta_{\nu}^{\ell\sigma}$ is a numerical prefactor, with $(\eta_{\nu}^{\ell\sigma})^2 = 1$. The equal-position boundary-fermion correlation function in the zero-temperature limit can be computed as
\begin{widetext}
    \begin{align}
\left\langle
\psi_{\ell\sigma,m}^{{\rm bdry}}(\tau)
\left[
\psi_{\ell\sigma,m}^{{\rm bdry}}(0)
\right]^{\dagger}
\right\rangle
\propto{}&
\prod_{p=1}^{N_{\rm ch}}
\prod_{\nu}
\exp\Bigg\{
-\frac{1}{4}
\left[
\left(
\boldsymbol{\mathcal{K}}_{\nu}^{-1/2}
\mathbf{O}_{\nu}
\right)_{mp}
\right]^{2}
\left\langle
\left[
\widetilde{\theta}_{\nu,p}^{e}(\tau)
-
\widetilde{\theta}_{\nu,p}^{e}(0)
\right]^{2}
\right\rangle
\Bigg\}
\nonumber \\
&\times
\exp\Bigg\{
-\frac{1}{4}
\left[
\left(
\boldsymbol{\mathcal{K}}_{\nu}^{-1/2}
\mathbf{O}_{\nu}
\right)_{mp}
\right]^{2}
\left\langle
\left[
\widetilde{\phi}_{\nu,p}^{e}(\tau)
-
\widetilde{\phi}_{\nu,p}^{e}(0)
\right]^{2}
\right\rangle
\Bigg\}
\nonumber \\
\propto{}&
|\tau|^{-\Delta_{{\rm bdry},m}}
\end{align}
\end{widetext}
with exponent
\begin{align}
\Delta_{{\rm bdry},m}
&  =
\frac{1}{2}
\sum_{\nu}
\sum_{p=1}^{N_{\rm ch}}
\left[
\left(
\boldsymbol{\mathcal{K}}_{\nu}^{-1/2}
\mathbf{O}_{\nu}
\right)_{mp}
\right]^{2} \nn
& =
\frac{1}{2}
\sum_{\nu}
\left(
\boldsymbol{\mathcal{K}}_{\nu}^{-1}
\right)_{mm}.
\end{align}
Imposing a periodic boundary condition yields the exponent
\begin{align}
\Delta_{{\rm bdry}}
=
\frac{1}{2}
\left(
\bar{\Delta}_{\theta_{c}, 0}
+
K_{s}^{-1}
\right),
\end{align}
independent of $m$. The boundary LDOS exponent $\beta_{\rm bdry}$ can be readily obtained, as shown in Eq.~\eqref{eq:beta_bdry-2dNoValley}.

To compute the boundary LDOS in the $T\ll T_{\rm hx}$ regime, we represent the fermion field $\psi_{\ell\sigma,m}$ in terms of the transformed boson basis $\Phi_{m}^{\eta}$ and $\Theta_{m}^{\eta}$ defined in Eq.~\eqref{Eq:new-basis}. In this basis, the fermion field reads
\begin{align}
\psi_{\ell\sigma,m}(r)
=
\frac{U_{\ell\sigma,m}}{\sqrt{2\pi a}}
e^{i\ell k_{F}r}
\exp\left[
i\left(
\Theta_{m}^{-\ell\sigma}
-
\ell\Phi_{m}^{-\ell\sigma}
\right)
\right],
\label{eq:psi_2d_newbasis}
\end{align}
where $U_{\ell\sigma,m}$ denotes the Klein factor. The boundary LDOS can be analyzed using the same procedure as in the $T>T_{\rm hx}$ regime, but retaining only the fermionic modes that remain gapless. To extract the universal power-law exponent, we evaluate the equal-position correlation function of the boundary fermion fields in the zero-temperature limit, yielding
\begin{align}
\left\langle
\psi_{\ell\sigma,m}^{\rm bdry}(\tau)
\left[
\psi_{\ell\sigma,m}^{\rm bdry}(0)
\right]^{\dagger}
\right\rangle
\propto
|\tau|^{-\widetilde{\Delta}_{\rm bdry}},
\end{align}
where
\begin{align}
\widetilde{\Delta}_{\rm bdry}
=
\bar{K}_-^{\rm (2d)}.
\end{align}
This leads to the corresponding low-temperature boundary-LDOS exponent shown in Eq.~\eqref{eq:beta_tilde_bdry-2dNoValley}.

\section{Bosonized forms of order parameters}
\label{sec:order-parameter}

In this section, we present the bosonized representations of the order-parameter operators associated with the density wave and   pairing channels for both the valleyless and valleyful coupled-wire systems. These expressions are used in Sec.~\ref{sec:helix_induced_order} to analyze the competing correlations in the spin-helix phase.

We begin with the valleyless 2D coupled-wire system. Using Eq.~\eqref{eq:psi_2d_newbasis}, the density-wave order parameters defined in Eq.~\eqref{eq:O_dw_short} can be written  as
\begin{subequations}
\label{eq:B_dw_2d}
    \begin{align}
    \mathcal{O}_{{\rm cdw},m}	(r)\propto ~  & e^{i\left(\Theta_{m}^{+}-\Theta_{m}^{-}+\Phi_{m}^{+}+\Phi_{m}^{-}\right)(r)} \nn
    & \qquad +e^{i\left(-\Theta_{m}^{+}+\Theta_{m}^{-}+\Phi_{m}^{+}+\Phi_{m}^{-}\right)(r)}, \\
\mathcal{O}_{{\rm sdw},m}^{z}	(r) \propto ~  &e^{i\left(\Theta_{m}^{+}-\Theta_{m}^{-}+\Phi_{m}^{+}+\Phi_{m}^{-}\right)(r)}
\nn
    & \qquad -e^{i\left(-\Theta_{m}^{+}+\Theta_{m}^{-}+\Phi_{m}^{+}+\Phi_{m}^{-}\right)(r)},\\
\mathcal{O}_{{\rm sdw},m}^{x}	(r)\propto ~ & e^{2i\Phi_{m}^{+}(r)}+e^{2i\Phi_{m}^{-}(r)},\\
\mathcal{O}_{{\rm sdw},m}^{y}	(r) \propto ~ & e^{2i\Phi_{m}^{+}(r)}-e^{2i\Phi_{m}^{-}(r)}.
\end{align}
\end{subequations}
Using Eq.~\eqref{eq:O_sc_short}, the superconducting pairing operators in the spin-singlet and spin-triplet channels can be written as
\begin{subequations}
\label{eq:B_SC_2d}
    \begin{align}
    \mathcal{O}_{{\rm ss},m}	(r)\propto ~  &  e^{-2i\Theta_{m}^{-}(r)}-e^{-2i\Theta_{m}^{+}(r)}, \\
\mathcal{O}_{{\rm ts},m}^{z}	(r)\propto ~ &  e^{-2i\Theta_{m}^{-}(r)}+e^{-2i\Theta_{m}^{+}(r)},\\
\mathcal{O}_{{\rm ts},m}^{x}	(r) \propto ~ & e^{i\left(-\Theta_{m}^{-}-\Theta_{m}^{+}+\Phi_{m}^{-}  -\Phi_{m}^{+}\right)(r)}
\nn
& \qquad -e^{i\left(-\Theta_{m}^{+}-\Theta_{m}^{-}+\Phi_{m}^{+}-\Phi_{m}^{-}\right)(r)},\\
\mathcal{O}_{{\rm ts},m}^{y}	(r)  \propto ~ &  e^{i\left(-\Theta_{m}^{-}-\Theta_{m}^{+}+\Phi_{m}^{-}
  -\Phi_{m}^{+}\right)(r)} \nn
 & \qquad +e^{i\left(-\Theta_{m}^{+}-\Theta_{m}^{-}+\Phi_{m}^{+}-\Phi_{m}^{-}\right)(r)}.
\end{align}
\end{subequations}
Because our primary objective is to extract the scaling exponents of the various correlation functions, Eqs.~\eqref{eq:B_dw_2d} and \eqref{eq:B_SC_2d} retain only the bosonic operators governing their spatial and temporal behavior. Klein factors and the oscillatory phases associated with the Fermi wave vectors, which do not affect the scaling exponents, are suppressed. The corresponding expressions for an isolated valleyless 1D system can be obtained  by suppressing the channel indices $m$.

To obtain the corresponding order-parameter operators for the valleyful coupled-wire system, we employ the bosonic basis introduced in Refs.~\cite{CHH-CNT-helix,YYC-2Dhelix-2025}, denoted by $\Phi^\eta_{\delta,m}$ and $\Theta^\eta_{\delta,m}$. In this basis, the fermion field is expressed as
\begin{align}
    \psi_{\ell\delta\sigma,m}(r)&=\frac{U_{\ell\delta\sigma,m}}{\sqrt{2\pi a}}\,e^{i\ell k_{F,\delta}r}\exp\left\{ i\left[\Theta_{\ell\delta,m}^{-\ell\sigma}(r)-\ell\Phi_{\ell\delta,m}^{-\ell\sigma}(r)\right]\right\} .
\end{align}
The resulting bosonized forms of the intrabranch  components of the density-wave order parameters are given by
\begin{subequations}
\label{eq:B_dw}
    \begin{align}
\mathcal{O}_{{\rm cdw},\delta,m}^{\intra}(r)
\propto  ~ & 
e^{i\left(-\Theta_{\delta,m}^{-}+\Theta_{\bar{\delta},m}^{+}+\Phi_{\delta,m}^{-}+\Phi_{\bar{\delta},m}^{+}\right)(r)}
\nn
& \qquad+
e^{i\left(-\Theta_{\delta,m}^{+}+\Theta_{\bar{\delta},m}^{-}+\Phi_{\delta,m}^{+}+\Phi_{\bar{\delta},m}^{-}\right)(r)},
\\
\mathcal{O}_{{\rm sdw},\delta,m}^{z,\intra}(r)
\propto ~ &
e^{i\left(-\Theta_{\delta,m}^{-}+\Theta_{\bar{\delta},m}^{+}+\Phi_{\delta,m}^{-}+\Phi_{\bar{\delta},m}^{+}\right)(r)}
\nn
& \qquad-
e^{i\left(-\Theta_{\delta,m}^{+}+\Theta_{\bar{\delta},m}^{-}+\Phi_{\delta,m}^{+}+\Phi_{\bar{\delta},m}^{-}\right)(r)},
 \\
\mathcal{O}_{{\rm sdw},\delta,m}^{x\,\intra}(r)
\propto  ~ &
e^{i\left(-\Theta_{\delta,m}^{-}+\Theta_{\bar{\delta},m}^{-}+\Phi_{\delta,m}^{-}+\Phi_{\bar{\delta},m}^{-}\right)(r)}
\nn
& \qquad+
e^{i\left(-\Theta_{\delta,m}^{+}+\Theta_{\bar{\delta},m}^{+}+\Phi_{\delta,m}^{+}+\Phi_{\bar{\delta},m}^{+}\right)(r)},
 \\
\mathcal{O}_{{\rm sdw},\delta,m}^{y\,\intra}(r)
\propto ~ & 
e^{i\left(-\Theta_{\delta,m}^{-}+\Theta_{\bar{\delta},m}^{-}+\Phi_{\delta,m}^{-}+\Phi_{\bar{\delta},m}^{-}\right)(r)} \nn
& \qquad -
e^{i\left(-\Theta_{\delta,m}^{+}+\Theta_{\bar{\delta},m}^{+}+\Phi_{\delta,m}^{+}+\Phi_{\bar{\delta},m}^{+}\right)(r)}.
\end{align}
\end{subequations}
The interbranch counterparts can be readily obtained by replacing $\Phi^\eta_{\bar{\delta},m}$ and $\Theta^\eta_{\bar{\delta},m}$ in Eq.~\eqref{eq:B_dw} by $\Phi^\eta_{\delta,m}$ and $\Theta^\eta_{\delta,m}$, respectively. The corresponding zero-center-of-mass-momentum intrabranch components of the superconducting pairing operators read
\begin{subequations}
\label{eq:B_SC}
   \begin{align}
\mathcal{O}_{\mathrm{ss},\delta,m}^{\intra}(r)
\propto & 
e^{i\left(-\Theta_{\delta,m}^{-}-\Theta_{\bar{\delta},m}^{-}+\Phi_{\delta,m}^{-}-\Phi_{\bar{\delta},m}^{-}\right)(r)}
\nn
& \qquad   -
e^{i\left(-\Theta_{\delta,m}^{+}-\Theta_{\bar{\delta},m}^{+}+\Phi_{\delta,m}^{+}-\Phi_{\bar{\delta},m}^{+}\right)(r)},
 \\
\mathcal{O}_{\mathrm{ts},\delta,m}^{z\,\intra}(r)
\propto & 
e^{i\left(-\Theta_{\delta,m}^{-}-\Theta_{\bar{\delta},m}^{-}+\Phi_{\delta,m}^{-}-\Phi_{\bar{\delta},m}^{-}\right)(r)} \nn
& \qquad +
e^{i\left(-\Theta_{\delta,m}^{+}-\Theta_{\bar{\delta},m}^{+}+\Phi_{\delta,m}^{+}-\Phi_{\bar{\delta},m}^{+}\right)(r)},
 \\
\mathcal{O}_{\mathrm{ts},\delta,m}^{x\,\intra}(r)
\propto & 
e^{i\left(-\Theta_{\delta,m}^{-}-\Theta_{\bar{\delta},m}^{+}+\Phi_{\delta,m}^{-}-\Phi_{\bar{\delta},m}^{+}\right)(r)} \nn
& \qquad -
e^{i\left(-\Theta_{\delta,m}^{+}-\Theta_{\bar{\delta},m}^{-}+\Phi_{\delta,m}^{+}-\Phi_{\bar{\delta},m}^{-}\right)(r)},
 \\
\mathcal{O}_{\mathrm{ts},\delta,m}^{y\,\intra}(r)
\propto & 
e^{i\left(-\Theta_{\delta,m}^{-}-\Theta_{\bar{\delta},m}^{+}+\Phi_{\delta,m}^{-}-\Phi_{\bar{\delta},m}^{+}\right)(r)} \nn
& \qquad +
e^{i\left(-\Theta_{\delta,m}^{+}-\Theta_{\bar{\delta},m}^{-}+\Phi_{\delta,m}^{+}-\Phi_{\bar{\delta},m}^{-}\right)(r)}.
\end{align} 
\end{subequations}
Similar to the bosonized density-wave operators, the interbranch components of the superconducting pairing operators can be obtained by replacing $\Phi^\eta_{\bar{\delta}}$ and $\Theta^\eta_{\bar{\delta}}$ in Eq.~\eqref{eq:B_SC} by $\Phi^\eta_{\delta}$ and $\Theta^\eta_{\delta}$, respectively. As before, we retain only the bosonic operators relevant to the scaling behavior in Eqs.~\eqref{eq:B_dw} and \eqref{eq:B_SC}.

\bibliographystyle{apsrevown}
\bibliography{Ref_LowDimHelix.bib}

@article{VonDelft1998,
author = {{von Delft}, Jan and Schoeller, Herbert},
doi = {10.1002/(sici)1521-3889(199811)7:4<225::aid-andp225>3.0.co;2-l},
journal = {Ann. Phys.},
number = {4},
pages = {225--305},
title = {{Bosonization for beginners - Refermionization for experts}},
volume = {7},
year = {1998}
}

@article{Duarte:2026PNAS_collapseMoire,
author = {Duarte J. P. de Sousa  and Seungjun Lee  and Francisco Guinea  and Tony Low },
title = {{Moiré collapse and Luttinger liquids in twisted anisotropic homobilayers}},
journal = {Proc. Natl. Acad. Sci.},
volume = {123},
number = {11},
pages = {e2527371123},
year = {2026},
doi = {10.1073/pnas.2527371123},
URL = {https://www.pnas.org/doi/abs/10.1073/pnas.2527371123}
}

@article{Christophe:PRB2026_grapheneslides,
  title = {Gate-tunable resonances and one-dimensional channel in a graphene nanoslide},
  author = {De Beule, Christophe and Liu, Ming-Hao and Partoens, Bart and Covaci, Lucian},
  journal = {Phys. Rev. B},
  volume = {113},
  issue = {20},
  pages = {L201403},
  numpages = {6},
  year = {2026},
  month = {May},
  publisher = {American Physical Society},
  doi = {10.1103/g6fn-xnf3},
  url = {https://link.aps.org/doi/10.1103/g6fn-xnf3}
}

@Article{Hsu:2021,
  author  = {Chen-Hsuan Hsu and Peter Stano and Jelena Klinovaja and Daniel Loss},
  journal = {{Semicond. Sci. Technol.}},
  title   = {{Helical liquids in semiconductors}},
  year    = {2021},
  pages   = {123003},
  volume  = {36},
  doi     = {10.1088/1361-6641/ac2c27},
}

@article{HanaguriNatComm:2020-GdRuSiSTM,
author = {Yasui, Yuuki and Butler, Christopher J. and Khanh, Nguyen Duy and Hayami, Satoru and Nomoto, Takuya and Hanaguri, Tetsuo and Motome, Yukitoshi and Arita, Ryotaro and Arima, Taka-hisa and Tokura, Yoshinori and Seki, Shinichiro},
doi = {10.1038/s41467-020-19751-4},
journal = {Nat. Commun.},
number = {1},
pages = {5925},
pmid = {33230104},
publisher = {Springer US},
title = {{Imaging the coupling between itinerant electrons and localised moments in the centrosymmetric skyrmion magnet GdRu2Si2}},
url = {http://dx.doi.org/10.1038/s41467-020-19751-4},
volume = {11},
year = {2020}
}

@Inbook{FisherGlazman1997,
author="Fisher, Matthew P. A.
and Glazman, Leonid I.",
editor="Sohn, Lydia L.
and Kouwenhoven, Leo P.
and Sch{\"o}n, Gerd",
title="Transport in a One-Dimensional {Luttinger} Liquid",
bookTitle="Mesoscopic Electron Transport",
year="1997",
publisher="Springer Netherlands",
address="Dordrecht",
pages="331--373",
isbn="978-94-015-8839-3",
doi="10.1007/978-94-015-8839-3_9",
url="https://doi.org/10.1007/978-94-015-8839-3_9"
}

@Article{Hsu:2023,
  author    = {Hsu, Chen-Hsuan and Loss, Daniel and Klinovaja, Jelena},
  journal   = {Phys. Rev. B},
  title     = {{General scattering and electronic states in a quantum-wire network of moir{\'e} systems}},
  year      = {2023},
  month     = {Sep},
  pages     = {L121409},
  volume    = {108},
  doi       = {10.1103/PhysRevB.108.L121409},
  issue     = {12},
  numpages  = {9},
  publisher = {American Physical Society},
}

@article{HCWang-TBG,
author = {Wang, Hao-Chien and Hsu, Chen-Hsuan},
doi = {10.1088/2053-1583/ad3b11},
journal = {2D Mater.},
pages = {035007},
title = {{Electrically tunable correlated domain wall network in twisted bilayer graphene}},
url = {http://arxiv.org/abs/2311.14384},
volume = {11},
year = {2024}
}

@article{CHH-CNT-helix,
author = {Hsu, Chen-Hsuan and Stano, Peter and Klinovaja, Jelena and Loss, Daniel},
doi = {10.1103/PhysRevB.92.235435},
journal = {Phys. Rev. B},
number = {23},
pages = {235435},
title = {{Antiferromagnetic nuclear spin helix and topological superconductivity in $^{13}\mathrm{C}$ nanotubes}},
volume = {92},
year = {2015}
}

@article{Voit:1995ROPP,
doi = {10.1088/0034-4885/58/9/002},
url = {https://doi.org/10.1088/0034-4885/58/9/002},
year = {1995},
month = {sep},
publisher = {},
volume = {58},
number = {9},
pages = {977},
author = {J Voit},
title = {{One-dimensional Fermi liquids}},
journal = {Rep. Prog. Phys.}
}

@article{Meng-strip-EuroPhys,
author = {Meng, Tobias and Stano, Peter and Klinovaja, Jelena and Loss, Daniel},
doi = {10.1140/epjb/e2014-50445-1},
journal = {Eur. Phys. J. B},
number = {9},
pages = {203},
title = {{Helical nuclear spin order in a strip of stripes in the quantum Hall regime}},
volume = {87},
year = {2014}
}

@article{Haldane1981a,
author = {Haldane, FDM},
journal = {J. Phys. C Solid State Phys.},
pages = {2585--2609},
title = {{`Luttinger liquid theory' of one-dimensional quantum fluids. I. Properties of the Luttinger model and their extension to the general 1D interacting spinless Fermi gas}},
volume = {14},
year = {1981},
doi = {10.1088/0022-3719/14/19/010},
url = {https://doi.org/10.1088/0022-3719/14/19/010}
}

@incollection{Senechal2006,
  author    = {S{\'e}n{\'e}chal, David},
  title     = {An Introduction to Bosonization},
  booktitle = {Theoretical Methods for Strongly Correlated Electrons},
  editor    = {S{\'e}n{\'e}chal, David and Tremblay, Andr{\'e}-Marie and Bourbonnais, Claude},
  series    = {CRM Series in Mathematical Physics},
  publisher = {Springer},
  address   = {New York},
  year      = {2004},
  pages     = {139--186}
}

@book{Giamarchi2003,
author = {Giamarchi, Thierry},
doi = {10.1093/acprof:oso/9780198525004.001.0001},
month = {dec},
publisher = {Oxford University Press},
title = {{Quantum Physics in One Dimension}},
year = {2003}
}

@article{Peter_NVcenter:2013,
  title = {Local spin susceptibilities of low-dimensional electron systems},
  author = {Stano, Peter and Klinovaja, Jelena and Yacoby, Amir and Loss, Daniel},
  journal = {Phys. Rev. B},
  volume = {88},
  issue = {4},
  pages = {045441},
  numpages = {10},
  year = {2013},
  month = {Jul},
  publisher = {American Physical Society},
  doi = {10.1103/PhysRevB.88.045441},
  url = {https://link.aps.org/doi/10.1103/PhysRevB.88.045441}
}

@article{StrainWTe2:2024,
  title = {Strain-dependent one-dimensional confinement channels in twisted bilayer {$1{T}^{\ensuremath{'}}\text{\ensuremath{-}}{\mathrm{WTe}}_{2}$}},
  author = {Magorrian, S. J. and Hine, N. D. M.},
  journal = {Phys. Rev. B},
  volume = {110},
  issue = {4},
  pages = {045410},
  numpages = {8},
  year = {2024},
  month = {Jul},
  publisher = {American Physical Society},
  doi = {10.1103/PhysRevB.110.045410},
  url = {https://link.aps.org/doi/10.1103/PhysRevB.110.045410}
}

@article{Quasi1DCrSBr:2023,
author = {Klein, Julian and Pingault, Benjamin and Florian, Matthias and Hei{\ss}enb{\"u}ttel, Marie-Christin and Steinhoff, Alexander and Song, Zhigang and Torres, Kierstin and Dirnberger, Florian and Curtis, Jonathan B. and Weile, Mads and Penn, Aubrey and Deilmann, Thorsten and Dana, Rami and Bushati, Rezlind and Quan, Jiamin and Luxa, Jan and Sofer, Zden{\v{e}}k and Al{\`u}, Andrea and Menon, Vinod M. and Wurstbauer, Ursula and Rohlfing, Michael and Narang, Prineha and Lončar, Marko and Ross, Frances M.},
doi = {10.1021/acsnano.2c07316},
journal = {ACS Nano},
pages = {5316--5328},
title = {{The Bulk van der Waals Layered Magnet CrSBr is a Quasi-1D Material}},
volume = {17},
year = {2023}
}

@article{2022-Nature-Pengjie-1DTLL-in-2DMoire,
author = {Wang, Pengjie and Yu, Guo and Kwan, Yves H. and Jia, Yanyu and Lei, Shiming and Klemenz, Sebastian and Cevallos, F. Alexandre and Singha, Ratnadwip and Devakul, Trithep and Watanabe, Kenji and Taniguchi, Takashi and Sondhi, Shivaji L. and Cava, Robert J. and Schoop, Leslie M. and Parameswaran, Siddharth A. and Wu, Sanfeng},
doi = {10.1038/s41586-022-04514-6},
issn = {14764687},
journal = {Nature},
number = {7908},
pages = {57--62},
publisher = {Springer US},
title = {{One-dimensional Luttinger liquids in a two-dimensional moir{\'{e}} lattice}},
volume = {605},
year = {2022}
}

@article{Sanfeng2023:tWTe2,
author = {Yu, Guo and Wang, Pengjie and Uzan-Narovlansky, Ayelet J and Jia, Yanyu and Onyszczak, Michael and Singha, Ratnadwip and Gui, Xin and Song, Tiancheng and Tang, Yue and Watanabe, Kenji and Taniguchi, Takashi and Cava, Robert J and Schoop, Leslie M and Wu, Sanfeng},
doi = {10.1038/s41467-023-42821-2},
journal = {Nat. Commun.},
pages = {7025},
publisher = {Springer US},
title = {{Evidence for two dimensional anisotropic Luttinger liquids at millikelvin temperatures}},
volume = {14},
year = {2023}
}

@Article{AbbamonteEELS:2017SciPost,
	title={{Measurement of the dynamic charge response of materials using low-energy, momentum-resolved electron energy-loss spectroscopy (M-EELS)}},
	author={Sean Vig and Anshul Kogar and Matteo Mitrano and Ali A. Husain and Vivek Mishra and Melinda S. Rak and Luc Venema and Peter D. Johnson and Genda D. Gu and Eduardo Fradkin and Michael R. Norman and Peter Abbamonte},
	journal={SciPost Phys.},
	volume={3},
	pages={026},
	year={2017},
	publisher={SciPost},
	doi={10.21468/SciPostPhys.3.4.026},
	url={https://scipost.org/10.21468/SciPostPhys.3.4.026},
}

@article{AbbamonteEELS:2025AnnuRev,
   author = "Abbamonte, Peter and Fink, Jörg",
   title = {{Collective Charge Excitations Studied by Electron Energy-Loss Spectroscopy}}, 
   journal= {Annu. Rev. Condens. Matter Phys.},
   year = "2025",
   volume = "16",
   number = "Volume 16, 2025",
   pages = "465-480",
   doi = "https://doi.org/10.1146/annurev-conmatphys-032822-044125",
   url = "https://www.annualreviews.org/content/journals/10.1146/annurev-conmatphys-032822-044125"
  }

@article{KrambergerEELS_CNT:2008PRL,
  title = {Linear Plasmon Dispersion in Single-Wall Carbon Nanotubes and the Collective Excitation Spectrum of Graphene},
  author = {Kramberger, C. and Hambach, R. and Giorgetti, C. and R\"ummeli, M. H. and Knupfer, M. and Fink, J. and B\"uchner, B. and Reining, Lucia and Einarsson, E. and Maruyama, S. and Sottile, F. and Hannewald, K. and Olevano, V. and Marinopoulos, A. G. and Pichler, T.},
  journal = {Phys. Rev. Lett.},
  volume = {100},
  issue = {19},
  pages = {196803},
  numpages = {4},
  year = {2008},
  doi = {10.1103/PhysRevLett.100.196803},
  url = {https://link.aps.org/doi/10.1103/PhysRevLett.100.196803}
}

@article{Rondin_NV:2014ROPP,
doi = {10.1088/0034-4885/77/5/056503},
url = {https://doi.org/10.1088/0034-4885/77/5/056503},
year = {2014},
month = {may},
publisher = {IOP Publishing},
volume = {77},
number = {5},
pages = {056503},
author = {Rondin, L and Tetienne, J-P and Hingant, T and Roch, J-F and Maletinsky, P and Jacques, V},
title = {Magnetometry with nitrogen-vacancy defects in diamond},
journal = {Rep. Prog. Phys.}
}

@article{SchirhaglNV:2014AnnuRev,
   author = "Schirhagl, Romana and Chang, Kevin and Loretz, Michael and Degen, Christian L.",
   title = {{Nitrogen-Vacancy Centers in Diamond: Nanoscale Sensors for Physics and Biology}}, 
   journal= "Annu. Rev. Phys. Chem.",
   year = "2014",
   volume = "65",
   number = "Volume 65, 2014",
   pages = "83-105",
   doi = "https://doi.org/10.1146/annurev-physchem-040513-103659",
   url = "https://www.annualreviews.org/content/journals/10.1146/annurev-physchem-040513-103659"
  }

@article{DegenNV:2017RMP,
  title = {Quantum sensing},
  author = {Degen, C. L. and Reinhard, F. and Cappellaro, P.},
  journal = {Rev. Mod. Phys.},
  volume = {89},
  issue = {3},
  pages = {035002},
  numpages = {39},
  year = {2017},
  month = {Jul},
  publisher = {American Physical Society},
  doi = {10.1103/RevModPhys.89.035002},
  url = {https://link.aps.org/doi/10.1103/RevModPhys.89.035002}
}

@article{Eggert:1992PRB,
  title = {Magnetic impurities in half-integer-spin {Heisenberg} antiferromagnetic chains},
  author = {Eggert, Sebastian and Affleck, Ian},
  journal = {Phys. Rev. B},
  volume = {46},
  issue = {17},
  pages = {10866--10883},
  numpages = {0},
  year = {1992},
  month = {Nov},
  publisher = {American Physical Society},
  doi = {10.1103/PhysRevB.46.10866},
  url = {https://link.aps.org/doi/10.1103/PhysRevB.46.10866}
}

@article{Fabrizio:1995PRB,
  title = {Interacting one-dimensional electron gas with open boundaries},
  author = {Fabrizio, M. and Gogolin, Alexander O.},
  journal = {Phys. Rev. B},
  volume = {51},
  issue = {24},
  pages = {17827--17841},
  numpages = {0},
  year = {1995},
  month = {Jun},
  publisher = {American Physical Society},
  doi = {10.1103/PhysRevB.51.17827},
  url = {https://link.aps.org/doi/10.1103/PhysRevB.51.17827}
}

@article{HageEELS_CNT:2017PRL,
  title = {Momentum- and space-resolved high-resolution electron energy loss spectroscopy of individual single-wall carbon nanotubes},
  author = {Hage, F. S. and Hardcastle, T. P. and Scott, A. J. and Brydson, R. and Ramasse, Q. M.},
  journal = {Phys. Rev. B},
  volume = {95},
  issue = {19},
  pages = {195411},
  numpages = {11},
  year = {2017},
  month = {May},
  publisher = {American Physical Society},
  doi = {10.1103/PhysRevB.95.195411},
  url = {https://link.aps.org/doi/10.1103/PhysRevB.95.195411}
}

@article{Huynh:2DM2026,
doi = {10.1088/2053-1583/ae4ca3},
url = {https://doi.org/10.1088/2053-1583/ae4ca3},
year = {2026},
month = {mar},
publisher = {IOP Publishing},
volume = {13},
number = {2},
pages = {022002},
author = {Huynh, T and Lee, H Y and Kim, B S Y},
title = {{Emerging strategies for moiré systems: new knobs, dimensions, and materials}},
journal = { 2D Mater. },
}

@article{Efimkin-2018-TBG-network,
author = {Efimkin, Dmitry K. and MacDonald, Allan H.},
doi = {10.1103/PhysRevB.98.035404},
journal = {Phys. Rev. B},
pages = {035404},
publisher = {American Physical Society},
title = {{Helical network model for twisted bilayer graphene}},
volume = {98},
year = {2018}
}

@article{San-Jose-TBG-network,
author = {San-Jose, Pablo and Prada, Elsa},
doi = {10.1103/PhysRevB.88.121408},
journal = {Phys. Rev. B},
number = {12},
pages = {121408(R)},
title = {{Helical networks in twisted bilayer graphene under interlayer bias}},
volume = {88},
year = {2013}
}

@article{Cenke-PRB-moire-coupledWire,
author = {Wu, Xiao Chuan and Jian, Chao Ming and Xu, Cenke},
doi = {10.1103/PhysRevB.99.161405},
journal = {Phys. Rev. B},
number = {16},
pages = {161405(R)},
title = {{Coupled-wire description of the correlated physics in twisted bilayer graphene}},
volume = {99},
year = {2019}
}

@article{Resistive-NMR-Klitzing,
  title = {Electrical detection of nuclear magnetic resonance in {GaAs--Al$_x$Ga$_{1-x}$As} heterostructures},
  author = {Dobers, M. and Klitzing, K. v. and Schneider, J. and Weimann, G. and Ploog, K.},
  journal = {Phys. Rev. Lett.},
  volume = {61},
  issue = {14},
  pages = {1650--1653},
  numpages = {0},
  year = {1988},
  month = {Oct},
  publisher = {American Physical Society},
  doi = {10.1103/PhysRevLett.61.1650},
  url = {https://link.aps.org/doi/10.1103/PhysRevLett.61.1650}
}

@article{YCChou-PRB-moire-network,
author = {Chou, Yang Zhi and Lin, Yu Ping and {Das Sarma}, Sankar and Nandkishore, Rahul M},
doi = {10.1103/PhysRevB.100.115128},
journal = {Phys. Rev. B},
pages = {115128},
title = {{Superconductor versus insulator in twisted bilayer graphene}},
volume = {100},
year = {2019}
}

@article{Ren-PRB-Metallic-network,
  title = {Metallic network of topological domain walls},
  author = {Hou, Tao and Ren, Yafei and Quan, Yujie and Jung, Jeil and Ren, Wei and Qiao, Zhenhua},
  journal = {Phys. Rev. B},
  volume = {101},
  issue = {20},
  pages = {201403},
  numpages = {5},
  year = {2020},
  month = {May},
  publisher = {American Physical Society},
  doi = {10.1103/PhysRevB.101.201403},
  url = {https://link.aps.org/doi/10.1103/PhysRevB.101.201403}
}

@article{Fleischmann-NanoLett-2020,
author = {Fleischmann, Maximilian and Gupta, Reena and Wullschläger, Florian and Theil, Simon and Weckbecker, Dominik and Meded, Velimir and Sharma, Sangeeta and Meyer, Bernd and Shallcross, Samuel},
title = {Perfect and Controllable Nesting in Minimally Twisted Bilayer Graphene},
journal = {Nano Lett.},
volume = {20},
number = {2},
pages = {971-978},
year = {2020},
doi = {10.1021/acs.nanolett.9b04027},
URL = {https://doi.org/10.1021/acs.nanolett.9b04027}
}

@article{Xu-moireDW-GiantOscil,
author = {Xu, S. G. and Berdyugin, A. I. and Kumaravadivel, P. and Guinea, F. and {Krishna Kumar}, R. and Bandurin, D. A. and Morozov, S. V. and Kuang, W. and Tsim, B. and Liu, S. and Edgar, J. H. and Grigorieva, I. V. and Fal'ko, V. I. and Kim, M. and Geim, A. K.},
doi = {10.1038/s41467-019-11971-7},
journal = {Nat. Commun.},
number = {1},
pages = {4008},
title = {{Giant oscillations in a triangular network of one-dimensional states in marginally twisted graphene}},
volume = {10},
year = {2019}
}

@article{muraki-science-2006,
author = {Kumada, Norio and Muraki, Koji and Hirayama, Yoshiro},
doi = {10.1126/science.1127094},
journal = {Science},
number = {5785},
pages = {329--332},
title = {{Low-frequency spin dynamics in a canted antiferromagnet}},
volume = {313},
year = {2006}
}

@article{muraki:2012,
author = {L. Tiemann  and G. Gamez  and N. Kumada  and K. Muraki },
title = {{Unraveling the Spin Polarization of the $\nu = 5/2$ Fractional Quantum Hall State}},
journal = {Science},
volume = {335},
number = {6070},
pages = {828},
year = {2012},
doi = {10.1126/science.1216697}
}

@article{Peeters-PRMater,
  title = {{DC} conductivity of twisted bilayer graphene: Angle-dependent transport properties and effects of disorder},
  author = {An\dj{}elkovi\ifmmode \acute{c}\else \'{c}\fi{}, M. and Covaci, L. and Peeters, F. M.},
  journal = {Phys. Rev. Mater.},
  volume = {2},
  issue = {3},
  pages = {034004},
  numpages = {10},
  year = {2018},
  month = {Mar},
  publisher = {American Physical Society},
  doi = {10.1103/PhysRevMaterials.2.034004},
  url = {https://link.aps.org/doi/10.1103/PhysRevMaterials.2.034004}
}

@article{Carr-DW-hetero,
  title = {Relaxation and domain formation in incommensurate two-dimensional heterostructures},
  author = {Carr, Stephen and Massatt, Daniel and Torrisi, Steven B. and Cazeaux, Paul and Luskin, Mitchell and Kaxiras, Efthimios},
  journal = {Phys. Rev. B},
  volume = {98},
  issue = {22},
  pages = {224102},
  numpages = {7},
  year = {2018},
  month = {Dec},
  publisher = {American Physical Society},
  doi = {10.1103/PhysRevB.98.224102},
  url = {https://link.aps.org/doi/10.1103/PhysRevB.98.224102}
}

@article{Verbakel-PRB,
  title = {Valley-protected one-dimensional states in small-angle twisted bilayer graphene},
  author = {Verbakel, J. D. and Yao, Q. and Sotthewes, K. and Zandvliet, H. J. W.},
  journal = {Phys. Rev. B},
  volume = {103},
  issue = {16},
  pages = {165134},
  numpages = {5},
  year = {2021},
  month = {Apr},
  publisher = {American Physical Society},
  doi = {10.1103/PhysRevB.103.165134},
  url = {https://link.aps.org/doi/10.1103/PhysRevB.103.165134}
}

@article{Walet-2020,
doi = {10.1088/2053-1583/ab57f8},
url = {https://dx.doi.org/10.1088/2053-1583/ab57f8},
year = {2019},
month = {dec},
publisher = {IOP Publishing},
volume = {7},
number = {1},
pages = {015023},
author = {Niels R Walet and Francisco Guinea},
title = {The emergence of one-dimensional channels in marginal-angle twisted bilayer graphene},
journal = {2D Mater.},
}

@article{Koshino-moireDW-2020,
author = {Tsim, Bonnie and Nam, Nguyen N.T. and Koshino, Mikito},
doi = {10.1103/PhysRevB.101.125409},
issn = {24699969},
journal = {Phys. Rev. B},
number = {12},
pages = {125409},
title = {{Perfect one-dimensional chiral states in biased twisted bilayer graphene}},
volume = {101},
year = {2020}
}

@article{Lyon:2017PRL,
  title = {{Probing Electron Spin Resonance in Monolayer Graphene}},
  author = {Lyon, T. J. and Sichau, J. and Dorn, A. and Centeno, A. and Pesquera, A. and Zurutuza, A. and Blick, R. H.},
  journal = {Phys. Rev. Lett.},
  volume = {119},
  issue = {6},
  pages = {066802},
  numpages = {5},
  year = {2017},
  month = {Aug},
  publisher = {American Physical Society},
  doi = {10.1103/PhysRevLett.119.066802},
  url = {https://link.aps.org/doi/10.1103/PhysRevLett.119.066802}
}

@article{Klitzing:1991PRL,
  title = {{Magnetoquantum oscillations of the nuclear-spin-lattice relaxation near a two-dimensional electron gas}},
  author = {Berg, A. and Dobers, M. and Gerhardts, R. R. and Klitzing, K. v.},
  journal = {Phys. Rev. Lett.},
  volume = {64},
  issue = {21},
  pages = {2563--2566},
  numpages = {0},
  year = {1990},
  month = {May},
  publisher = {American Physical Society},
  doi = {10.1103/PhysRevLett.64.2563},
  url = {https://link.aps.org/doi/10.1103/PhysRevLett.64.2563}
}

@book{Book-Resistive-NMR,
author = {Fanciulli, Marco},
year = {2009},
month = {01},
pages = {},
title = {Electron Spin Resonance and Related Phenomena in Low-Dimensional Structures},
volume = {115},
isbn = {978-3-540-79364-9},
journal = {Electron Spin Resonance and Related Phenomena in Low-Dimensional Structures},
doi = {10.1007/978-3-540-79365-6},
publisher = {Springer}
}

@article{Single-CNT-NMR-PRL,
author = {Singer, P M and Wzietek, P and Alloul, H and Simon, F and Kuzmany, H},
doi = {10.1103/PhysRevLett.95.236403},
journal = {Phys. Rev. Lett.},
number = {23},
pages = {236403},
title = {{NMR evidence for gapped spin excitations in metallic carbon nanotubes}},
volume = {95},
year = {2005}
}

@article{Fischer2009,
author = {Fischer, Jan and Trauzettel, Bj{\"{o}}rn and Loss, Daniel},
doi = {10.1103/PhysRevB.80.155401},
journal = {Phys. Rev. B},
number = {15},
pages = {155401},
title = {{Hyperfine interaction and electron-spin decoherence in graphene and carbon nanotube quantum dots}},
volume = {80},
year = {2009}
}

@article{Alex:2022PRB,
  title = {{Anisotropic topological superconductivity in Josephson junctions}},
  author = { Pekerten, Bar{\i}{\c{s}} and Pakizer, Joseph D. and Hawn, Benjamin and Matos-Abiague, Alex},
  journal = {Phys. Rev. B},
  volume = {105},
  issue = {5},
  pages = {054504},
  numpages = {13},
  year = {2022},
  month = {Feb},
  publisher = {American Physical Society},
  doi = {10.1103/PhysRevB.105.054504},
  url = {https://link.aps.org/doi/10.1103/PhysRevB.105.054504}
}

@article{Wu:2015PRB,
  title = {{Charge and spin currents in ferromagnetic Josephson junctions}},
  author = {Halterman, Klaus and Valls, Oriol T. and Wu, Chien-Te},
  journal = {Phys. Rev. B},
  volume = {92},
  issue = {17},
  pages = {174516},
  numpages = {17},
  year = {2015},
  month = {Nov},
  publisher = {American Physical Society},
  doi = {10.1103/PhysRevB.92.174516},
  url = {https://link.aps.org/doi/10.1103/PhysRevB.92.174516}
}

@article{LeRoy-moire-DW-exp,
author = {Huang, Shengqiang and Kim, Kyounghwan and Efimkin, Dmitry K and Lovorn, Timothy and Taniguchi, Takashi and Watanabe, Kenji and MacDonald, Allan H and Tutuc, Emanuel and LeRoy, Brian J},
doi = {10.1103/PhysRevLett.121.037702},
journal = {Phys. Rev. Lett.},
number = {3},
pages = {037702},
title = {{Topologically Protected Helical States in Minimally Twisted Bilayer Graphene}},
volume = {121},
year = {2018}
}

@article{PhilipKim-moire-DW,
author = {Yoo, Hyobin and Engelke, Rebecca and Carr, Stephen and Fang, Shiang and Zhang, Kuan and Cazeaux, Paul and Sung, Suk Hyun and Hovden, Robert and Tsen, Adam W and Taniguchi, Takashi and Watanabe, Kenji and Yi, Gyu Chul and Kim, Miyoung and Luskin, Mitchell and Tadmor, Ellad B and Kaxiras, Efthimios and Kim, Philip},
doi = {10.1038/s41563-019-0346-z},
journal = {Nat. Mater.},
number = {5},
pages = {448--453},
publisher = {Springer US},
title = {{Atomic and electronic reconstruction at the van der Waals interface in twisted bilayer graphene}},
url = {http://dx.doi.org/10.1038/s41563-019-0346-z},
volume = {18},
year = {2019}
}

@article{YYC-2Dhelix-2025,
  title = {Quasi-two-dimensional spin helix and magnon-induced singularity in twisted bilayer graphene},
  author = {Chang, Yung-Yeh and Saito, Kazuma and Hsu, Chen-Hsuan},
  journal = {Phys. Rev. Res.},
  volume = {7},
  issue = {3},
  pages = {L032066},
  numpages = {9},
  year = {2025},
  month = {Sep},
  publisher = {American Physical Society},
  doi = {10.1103/h5sd-v51h},
  url = {https://link.aps.org/doi/10.1103/h5sd-v51h}
}

@article{YiMing2024:tWTe2,
  title = {Possible Sliding Regimes in Twisted Bilayer {${\mathrm{WTe}}_{2}$}},
  author = {Wu, Yi-Ming and Murthy, Chaitanya and Kivelson, Steven A.},
  journal = {Phys. Rev. Lett.},
  volume = {133},
  issue = {24},
  pages = {246501},
  numpages = {8},
  year = {2024},
  month = {Dec},
  publisher = {American Physical Society},
  doi = {10.1103/PhysRevLett.133.246501},
  url = {https://link.aps.org/doi/10.1103/PhysRevLett.133.246501}
}

@book{slichter2013principles,
  title={Principles of Magnetic Resonance},
  author={Slichter, C.P.},
  isbn={9783662094419},
  series={Springer Series in Solid-State Sciences},
  url={https://books.google.com.tw/books?id=jF3xCAAAQBAJ},
  year={2013},
  publisher={Springer Berlin Heidelberg}
}

@article{moire-DW-Ensslin,
author = {Rickhaus, Peter and Wallbank, John and Slizovskiy, Sergey and Pisoni, Riccardo and Overweg, Hiske and Lee, Yongjin and Eich, Marius and Liu, Ming Hao and Watanabe, Kenji and Taniguchi, Takashi and Ihn, Thomas and Ensslin, Klaus},
doi = {10.1021/acs.nanolett.8b02387},
journal = {Nano Lett.},
month = {nov},
number = {11},
pages = {6725--6730},
pmid = {30336041},
publisher = {American Chemical Society},
title = {{Transport Through a Network of Topological Channels in Twisted Bilayer Graphene}},
url = {https://pubs.acs.org/sharingguidelines},
volume = {18},
year = {2018}
}

@article{Churchill2009-naturephysics,
author = {Churchill, H. O. H. and Bestwick, A. J. and Harlow, J. W. and Kuemmeth, F. and Marcos, D. and Stwertka, C. H. and Watson, S. K. and Marcus, C. M.},
doi = {10.1038/nphys1247},
journal = {Nat. Phys.},
number = {5},
pages = {321--326},
publisher = {Nature Publishing Group},
title = {{Electron-nuclear interaction in $^{13}\mathrm{C}$ nanotube double quantum dots}},
volume = {5},
year = {2009}
}

@article{Pennington:1996RMP,
  title = {{Nuclear magnetic resonance of ${\mathrm{C}}_{60}$ and fulleride superconductors}},
  author = {Pennington, Charles H. and Stenger, Victor A.},
  journal = {Rev. Mod. Phys.},
  volume = {68},
  issue = {3},
  pages = {855--910},
  numpages = {0},
  year = {1996},
  month = {Jul},
  publisher = {American Physical Society},
  doi = {10.1103/RevModPhys.68.855},
  url = {https://link.aps.org/doi/10.1103/RevModPhys.68.855}
}

@article{Churchill:2009PRL,
  title = {{Relaxation and Dephasing in a Two-Electron $^{13}\mathbf{C}$ Nanotube Double Quantum Dot}},
  author = {Churchill, H. O. H. and Kuemmeth, F. and Harlow, J. W. and Bestwick, A. J. and Rashba, E. I. and Flensberg, K. and Stwertka, C. H. and Taychatanapat, T. and Watson, S. K. and Marcus, C. M.},
  journal = {Phys. Rev. Lett.},
  volume = {102},
  issue = {16},
  pages = {166802},
  numpages = {4},
  year = {2009},
  month = {Apr},
  publisher = {American Physical Society},
  doi = {10.1103/PhysRevLett.102.166802},
  url = {https://link.aps.org/doi/10.1103/PhysRevLett.102.166802}
}

@article{F-adatom-graphene,
author = {Hong, X and Zou, K and Wang, B and Cheng, S. H. and Zhu, J},
doi = {10.1103/PhysRevLett.108.226602},
journal = {Phys. Rev. Lett.},
number = {22},
pages = {226602},
title = {{Evidence for spin-flip scattering and local moments in dilute fluorinated graphene}},
volume = {108},
year = {2012}
}

@article{2018-CHH-nuclear-2DTI,
author = {Hsu, Chen-Hsuan and Stano, Peter and Klinovaja, Jelena and Loss, Daniel},
doi = {10.1103/PhysRevB.97.125432},
journal = {Phys. Rev. B},
number = {12},
pages = {125432},
publisher = {American Physical Society},
title = {{Effects of nuclear spins on the transport properties of the edge of two-dimensional topological insulators}},
url = {https://doi.org/10.1103/PhysRevB.97.125432},
volume = {97},
year = {2018}
}

@article{Emery-sLL-2000,
author = {Emery, V. J. and Fradkin, E. and Kivelson, S. A. and Lubensky, T. C.},
doi = {10.1103/PhysRevLett.85.2160},
journal = {Phys. Rev. Lett.},
number = {10},
pages = {2160},
title = {{Quantum theory of the smectic metal state in stripe phases}},
volume = {85},
year = {2000}
}

@article{Carpentier-sLL-2001,
author = {Vishwanath, Ashvin and Carpentier, David},
doi = {10.1103/PhysRevLett.86.676},
journal = {Phys. Rev. Lett.},
number = {4},
pages = {676--679},
primaryClass = {cond-mat},
title = {{Two-dimensional anisotropic non-Fermi-liquid phase of coupled Luttinger liquids}},
volume = {86},
year = {2001}
}

@article{Mukhopadhyay-sLL-PRB-2001,
author = {Mukhopadhyay, Ranjan and Kane, C. L. and Lubensky, T. C.},
doi = {10.1103/PhysRevB.64.045120},
journal = {Phys. Rev. B},
number = {4},
pages = {045120},
primaryClass = {cond-mat},
title = {{Sliding Luttinger liquid phases}},
volume = {64},
year = {2001}
}

@article{YPChiu:2026Nature,
author = {Yang, Zi-Liang and Huang, Bo-Chao and Lin, Yu-Kuan and Lin, Yan-Ruei and Hsu, Hung-Chang and Chen, Hao-Yu and Wan, Yi and Tseng, Kai-Wei and Yang, Shu-Ting and Lo, Han-Chieh and Huang, You-Jia and Zheng, Fangyuan and Yang, Ni and Meng, Wanqing and Min, Jiacheng and Huang, Po-Cheng and Lan, Yann-Wen and Li, Lain-Jong and Chiu, Ya-Ping},
doi = {10.1038/s41586-026-10707-0},
issn = {1476-4687},
journal = {Nature},
number = {8122},
pages = {350--356},
title = {{Directly probing the carrier transfer length in 2D-material transistors}},
url = {https://doi.org/10.1038/s41586-026-10707-0},
volume = {655},
year = {2026}
}

@article{Mukhopadhyay-csLL-PRB,
  title = {{Crossed sliding Luttinger liquid phase}},
  author = {Mukhopadhyay, Ranjan and Kane, C. L. and Lubensky, T. C.},
  journal = {Phys. Rev. B},
  volume = {63},
  issue = {8},
  pages = {081103},
  numpages = {4},
  year = {2001},
  month = {Feb},
  publisher = {American Physical Society},
  doi = {10.1103/PhysRevB.63.081103},
  url = {https://link.aps.org/doi/10.1103/PhysRevB.63.081103}
}

@article{Braunecker-2009,
  title = {{Nuclear magnetism and electron order in interacting one-dimensional conductors}},
  author = {Braunecker, Bernd and Simon, Pascal and Loss, Daniel},
  journal = {Phys. Rev. B},
  volume = {80},
  issue = {16},
  pages = {165119},
  numpages = {28},
  year = {2009},
  month = {Oct},
  publisher = {American Physical Society},
  doi = {10.1103/PhysRevB.80.165119},
  url = {https://link.aps.org/doi/10.1103/PhysRevB.80.165119}
}

@article{Bistritzer-PNAS-2011,
author = {Bistritzer, Rafi and MacDonald, Allan H.},
doi = {10.1073/pnas.1108174108},
journal = {Proc. Natl. Acad. Sci.},
number = {30},
pages = {12233--12237},
title = {{Moir{\'{e}} bands in twisted double-layer graphene}},
volume = {108},
year = {2011}
}

@article{Zhizhan:2021NatComm,
author = {Qiu, Zhizhan and Holwill, Matthew and Olsen, Thomas and Lyu, Pin and Li, Jing and Fang, Hanyan and Yang, Huimin and Kashchenko, Mikhail and Novoselov, Kostya S and Lu, Jiong},
doi = {10.1038/s41467-020-20376-w},
issn = {2041-1723},
journal = {Nat. Commun.},
number = {1},
pages = {70},
title = {{Visualizing atomic structure and magnetism of 2D magnetic insulators via tunneling through graphene}},
url = {https://doi.org/10.1038/s41467-020-20376-w},
volume = {12},
year = {2021}
}

@article{Hirayama:2009SST,

author = {Hirayama, Y and Yusa, G and Hashimoto, K and Kumada, N and Ota, T and Muraki, K},
doi = {10.1088/0268-1242/24/2/023001},
journal = {Semicond. Sci. Technol.},
month = {jan},
number = {2},
pages = {023001},
title = {{Electron-spin/nuclear-spin interactions and NMR in semiconductors}},
url = {https://doi.org/10.1088/0268-1242/24/2/023001},
volume = {24},
year = {2009}
}

@article{Yanyu:2022NatPhys,
author = {Jia, Yanyu and Wang, Pengjie and Chiu, Cheng-Li and Song, Zhida and Yu, Guo and J{\"{a}}ck, Berthold and Lei, Shiming and Klemenz, Sebastian and Cevallos, F Alexandre and Onyszczak, Michael and Fishchenko, Nadezhda and Liu, Xiaomeng and Farahi, Gelareh and Xie, Fang and Xu, Yuanfeng and Watanabe, Kenji and Taniguchi, Takashi and Bernevig, B Andrei and Cava, Robert J and Schoop, Leslie M and Yazdani, Ali and Wu, Sanfeng},
doi = {10.1038/s41567-021-01422-w},
issn = {1745-2481},
journal = {Nat. Phys.},
number = {1},
pages = {87--93},
title = {{Evidence for a monolayer excitonic insulator}},
url = {https://doi.org/10.1038/s41567-021-01422-w},
volume = {18},
year = {2022}
}

@article{Weber:2026NatureNews,
  author  = {Weber, Bent},
  title   = {{Ultrasmall electrodes can connect to 2D transistors}},
  journal = {Nature},
  year    = {2026},
  month   = jul,
  volume  = {655},
  pages   = {308--309},
  doi     = {10.1038/d41586-026-01807-y},
  url     = {https://doi.org/10.1038/d41586-026-01807-y},
  note    = {{News \& Views}}
}

@article{castro-TBG-network-PRB,
author = {Chen, Chuan and {Castro Neto}, A H and Pereira, Vitor M},
doi = {10.1103/PhysRevB.101.165431},
journal = {Phys. Rev. B},
number = {16},
pages = {165431},
title = {{Correlated states of a triangular net of coupled quantum wires: Implications for the phase diagram of marginally twisted bilayer graphene}},
volume = {101},
year = {2020}
}

@article{Braunecker-PRL-2009-C13,
  title = {Nuclear Magnetism and Electronic Order in {$^{13}\mathrm{C}$} Nanotubes},
  author = {Braunecker, Bernd and Simon, Pascal and Loss, Daniel},
  journal = {Phys. Rev. Lett.},
  volume = {102},
  issue = {11},
  pages = {116403},
  numpages = {4},
  year = {2009},
  doi = {10.1103/PhysRevLett.102.116403},
  url = {https://link.aps.org/doi/10.1103/PhysRevLett.102.116403}
}

@article{Hsu:2026,
author = {Hsu, Chen-Hsuan and Ohorodnyk, Anna},
title = {{Coupled-wire descriptions of unconventional quantum states in twisted nanostructures}},
journal = {Mod. Phys. Lett. B},
volume = {40},
number = {19},
pages = {2650143},
year = {2026},
doi = {10.1142/S0217984926501435},

URL = { 
    
        https://doi.org/10.1142/S0217984926501435
}
}

@article{Koshino:2026PRB_tWTe2,
  title = {{One-dimensional electronic states in a moir\'e superlattice of twisted bilayer ${\mathrm{WTe}}_{2}$}},
  author = {Kawakami, Takuto and Tateishi, Hayato and Yoshida, Daiki and Yang, Xiaohan and Nakatsuji, Naoto and Chen, Limi and Aso, Kohei and Yamada-Takamura, Yukiko and Oshima, Yoshifumi and Zhang, Yijin and Machida, Tomoki and Kato, Koichiro and Koshino, Mikito},
  journal = {Phys. Rev. B},
  volume = {114},
  issue = {10},
  pages = {105407},
  numpages = {17},
  year = {2026},
  month = {Aug},
  publisher = {American Physical Society},
  doi = {10.1103/bzwr-6m2d},
  url = {https://link.aps.org/doi/10.1103/bzwr-6m2d}
}

@article{Bockrath:1999Sci_nanotube,
author = {Bockrath, Marc and Cobden, David H and Lu, Jia and Rinzler, Andrew G and Smalley, Richard E and Balents, Leon and McEuen, Paul L},
doi = {10.1038/17569},
issn = {1476-4687},
journal = {Nature},
number = {6720},
pages = {598--601},
title = {{Luttinger-liquid behaviour in carbon nanotubes}},
url = {https://doi.org/10.1038/17569},
volume = {397},
year = {1999}
}

@article{Lian:2024PRB_tWTe2,
  title = {{Twisted coupled wire model for a moir\'e sliding Luttinger liquid}},
  author = {Hu, Yichen and Xu, Yuanfeng and Lian, Biao},
  journal = {Phys. Rev. B},
  volume = {110},
  issue = {20},
  pages = {L201106},
  numpages = {7},
  year = {2024},
  month = {Nov},
  publisher = {American Physical Society},
  doi = {10.1103/PhysRevB.110.L201106},
  url = {https://link.aps.org/doi/10.1103/PhysRevB.110.L201106}
}

@article{Kane:1997PRL_nanotube,
  title = {{Coulomb Interactions and Mesoscopic Effects in Carbon Nanotubes}},
  author = {Kane, Charles and Balents, Leon and Fisher, Matthew P. A.},
  journal = {Phys. Rev. Lett.},
  volume = {79},
  issue = {25},
  pages = {5086--5089},
  numpages = {0},
  year = {1997},
  month = {Dec},
  publisher = {American Physical Society},
  doi = {10.1103/PhysRevLett.79.5086},
  url = {https://link.aps.org/doi/10.1103/PhysRevLett.79.5086}
}

@article{Egger:1998EPJ_nanotube,
author = {Egger, R and Gogolin, A O},
doi = {10.1007/s100510050315},
issn = {1434-6036},
journal = {Eur. Phys. J. B - Condens. Matter Complex Syst.},
number = {3},
pages = {281--300},
title = {{Correlated transport and non-Fermi-liquid behavior in single-wall carbon nanotubes}},
url = {https://doi.org/10.1007/s100510050315},
volume = {3},
year = {1998}
}

@article{Egger:1997PRL_nanotube,
  title = {{Effective Low-Energy Theory for Correlated Carbon Nanotubes}},
  author = {Egger, Reinhold and Gogolin, Alexander O.},
  journal = {Phys. Rev. Lett.},
  volume = {79},
  issue = {25},
  pages = {5082--5085},
  numpages = {0},
  year = {1997},
  month = {Dec},
  publisher = {American Physical Society},
  doi = {10.1103/PhysRevLett.79.5082},
  url = {https://link.aps.org/doi/10.1103/PhysRevLett.79.5082}
}

@article{
Auslaender:2005Science,
author = {O. M. Auslaender  and H. Steinberg  and A. Yacoby  and Y. Tserkovnyak  and B. I. Halperin  and K. W. Baldwin  and L. N. Pfeiffer  and K. W. West },
title = {{Spin-Charge Separation and Localization in One Dimension}},
journal = {Science},
volume = {308},
number = {5718},
pages = {88-92},
year = {2005},
doi = {10.1126/science.1107821},
URL = {https://www.science.org/doi/abs/10.1126/science.1107821}
}

@article{Tserkovnyak:2003PRB,
  title = {{Interference and zero-bias anomaly in tunneling between Luttinger-liquid wires}},
  author = {Tserkovnyak, Yaroslav and Halperin, Bertrand I. and Auslaender, Ophir M. and Yacoby, Amir},
  journal = {Phys. Rev. B},
  volume = {68},
  issue = {12},
  pages = {125312},
  numpages = {17},
  year = {2003},
  month = {Sep},
  publisher = {American Physical Society},
  doi = {10.1103/PhysRevB.68.125312},
  url = {https://link.aps.org/doi/10.1103/PhysRevB.68.125312}
}

@article{Tserkovnyak:2003PRL,
  title = {{Finite-Size Effects in Tunneling between Parallel Quantum Wires}},
  author = {Tserkovnyak, Yaroslav and Halperin, Bertrand I. and Auslaender, Ophir M. and Yacoby, Amir},
  journal = {Phys. Rev. Lett.},
  volume = {89},
  issue = {13},
  pages = {136805},
  numpages = {4},
  year = {2002},
  month = {Sep},
  publisher = {American Physical Society},
  doi = {10.1103/PhysRevLett.89.136805},
  url = {https://link.aps.org/doi/10.1103/PhysRevLett.89.136805}
}

@article{Nagao:2006PRL,
  title = {One-Dimensional Plasmon in an Atomic-Scale Metal Wire},
  author = {Nagao, Tadaaki and Yaginuma, Shin and Inaoka, Takeshi and Sakurai, Toshio},
  journal = {Phys. Rev. Lett.},
  volume = {97},
  issue = {11},
  pages = {116802},
  numpages = {4},
  year = {2006},
  month = {Sep},
  publisher = {American Physical Society},
  doi = {10.1103/PhysRevLett.97.116802},
  url = {https://link.aps.org/doi/10.1103/PhysRevLett.97.116802}
}

@incollection{Schulz:2000TLL,
  author    = {Schulz, Heinz J. and Cuniberti, Gianaurelio and Pieri, Pierbiagio},
  title     = {{Fermi Liquids and Luttinger Liquids}},
  booktitle = {Field Theories for Low-Dimensional Condensed Matter Systems:
               Spin Systems and Strongly Correlated Electrons},
  editor    = {Morandi, Giuseppe and Sodano, Pasquale and
               Tagliacozzo, Arturo and Tognetti, Valerio},
  series    = {Springer Series in Solid-State Sciences},
  volume    = {131},
  pages     = {9--81},
  year      = {2000},
  publisher = {Springer},
  address   = {Berlin, Heidelberg},
  doi       = {10.1007/978-3-662-04273-1_2}
}

@article{Yao:1999Nature,
author = {Yao, Zhen and Postma, Henk W Ch. and Balents, Leon and Dekker, Cees},
doi = {10.1038/46241},
issn = {1476-4687},
journal = {Nature},
number = {6759},
pages = {273--276},
title = {{Carbon nanotube intramolecular junctions}},
url = {https://doi.org/10.1038/46241},
volume = {402},
year = {1999}
}

@article{Ishii:2003Science,
author = {Ishii, Hiroyoshi and Kataura, Hiromichi and Shiozawa, Hidetsugu and Yoshioka, Hideo and Otsubo, Hideo and Takayawa, Yasuhiro and Miyahara, Tsuneaki and Suzuki, Shinzo and Achiba, Yohji and Nakatake, Masahi and Narimura, Takamasa and Higashiguchi, Mitsuharu and Shimada, Kenya and Namatame, Hirofumi and Taniguchi, Maasaki},
doi = {10.1038/nature02074},
issn = {00280836},
journal = {Nature},
number = {6966},
pages = {540--544},
pmid = {14654836},
title = {{Direct observation of Tomonaga-Luttinger-liquid state in carbon nanotubes at low temperatures}},
volume = {426},
year = {2003}
}

@article{Jompol:2009Science,
author = {Y. Jompol  and C. J. B. Ford  and J. P. Griffiths  and I. Farrer  and G. A. C. Jones  and D. Anderson  and D. A. Ritchie  and T. W. Silk  and A. J. Schofield },
title = {Probing Spin-Charge Separation in a {Tomonaga}-{Luttinger} Liquid},
journal = {Science},
volume = {325},
number = {5940},
pages = {597-601},
year = {2009},
doi = {10.1126/science.1171769},
URL = {https://www.science.org/doi/abs/10.1126/science.1171769}
}

@incollection{Eggert2007Bosonization,
  author       = {Eggert, Sebastian},
  title        = {One-Dimensional Quantum Wires: A Pedestrian Approach to Bosonization},
  booktitle    = {Theoretical Survey of One Dimensional Wire Systems},
  editor       = {Kuk, Y. and Hasegawa, S. and Xue, Q.-K.},
  chapter      = {2},
  publisher    = {Sowha Publishing},
  address      = {Seoul},
  year         = {2007},
  url          = {https://arxiv.org/abs/0708.0003}
}

@article{Flensberg:2016PRB,
  title   = {Effects of spin-orbit coupling and spatial symmetries on the {Josephson} current in {SNS} junctions},
  author  = {Rasmussen, Asbj{\o}rn and Danon, Jeroen and Suominen, Henri and Nichele, Fabrizio and Kjaergaard, Morten and Flensberg, Karsten},
  journal = {Phys. Rev. B},
  volume  = {93},
  pages   = {155406},
  year    = {2016},
  doi     = {10.1103/PhysRevB.93.155406}
}

@article{FlensbergPRB:2017_AnomalousFraunhoferPattern,
  title   = {Anomalous {Fraunhofer} interference in epitaxial superconductor-semiconductor {Josephson} junctions},
  author  = {Suominen, H. J. and Danon, J. and Kjaergaard, M. and Flensberg, K. and Shabani, J. and Palmstr{\o}m, C. J. and Nichele, F. and Marcus, C. M.},
  journal = {Phys. Rev. B},
  volume  = {95},
  pages   = {035307},
  year    = {2017},
  doi     = {10.1103/PhysRevB.95.035307}
}

@article{KTLawPRB:2018AJE,
  title   = {Asymmetric {Josephson} effect in inversion symmetry breaking topological materials},
  author  = {Chen, Chui-Zhen and He, James Jun and Ali, Mazhar N. and Lee, Gil-Ho and Fong, Kin Chung and Law, K. T.},
  journal = {Phys. Rev. B},
  volume  = {98},
  pages   = {075430},
  year    = {2018},
  doi     = {10.1103/PhysRevB.98.075430}
}

@article{FurusakiPhysicaB:1994DirtyJJ,
  title   = {{DC} {Josephson} effect in dirty {SNS} junctions: Numerical study},
  author  = {Furusaki, Akira},
  journal = {Physica B: Condensed Matter},
  volume  = {203},
  number  = {3--4},
  pages   = {214--218},
  year    = {1994},
  doi     = {10.1016/0921-4526(94)90061-2}
}

@article{AsanoPRB:2002,
  title   = {Numerical method for dc {Josephson} current between {$d$}-wave superconductors},
  author  = {Asano, Yasuhiro},
  journal = {Phys. Rev. B},
  volume  = {63},
  pages   = {052512},
  year    = {2001},
  doi     = {10.1103/PhysRevB.63.052512}
}

@article{diez2023symmetry,
  title   = {Symmetry-broken {Josephson} junctions and superconducting diodes in magic-angle twisted bilayer graphene},
  author  = {D{\'i}ez-M{\'e}rida, J. and D{\'i}ez-Carl{\'o}n, A. and Yang, S. Y. and Xie, Y.-M. and Gao, X.-J. and Senior, J. and Watanabe, K. and Taniguchi, T. and Lu, X. and Higginbotham, A. P. and Law, K. T. and Efetov, Dmitri K.},
  journal = {Nat. Commun.},
  volume  = {14},
  pages   = {2396},
  year    = {2023},
  doi     = {10.1038/s41467-023-38005-7}
}

@article{Klinovaja:2013,
  author  = {Klinovaja, Jelena and Stano, Peter and Yazdani, Ali and Loss, Daniel},
  title   = {Topological Superconductivity and {Majorana} Fermions in {RKKY} Systems},
  journal = {Phys. Rev. Lett.},
  volume  = {111},
  pages   = {186805},
  year    = {2013},
  doi     = {10.1103/PhysRevLett.111.186805}
}

@article{Braunecker:2013,
  author  = {Braunecker, Bernd and Simon, Pascal},
  title   = {Interplay between Classical Magnetic Moments and Superconductivity in Quantum One-Dimensional Conductors: Toward a Self-Sustained Topological {Majorana} Phase},
  journal = {Phys. Rev. Lett.},
  volume  = {111},
  pages   = {147202},
  year    = {2013},
  doi     = {10.1103/PhysRevLett.111.147202}
}

@misc{YCTsai:2026arxiv,
      title={{Field-controlled breaking and restoration of parity-time symmetry in Josephson interference}}, 
      author={Yi-Chen Tsai and Yung-Yeh Chang and Tao-Yi Hsu and Thomas Kuo and Chia-Nung Kuo and Chin-Shan Lue and Kuei-Lin Chiu and Chen-Hsuan Hsu and Chung-Ting Ke},
      year={2026},
      eprint={2608.16414},
      archivePrefix={arXiv},
      url={https://arxiv.org/abs/2608.16414}, 
}

@article{Sato:2019PRB,
  title = {{Strong electron-electron interactions of a Tomonaga-Luttinger liquid observed in InAs quantum wires}},
  author = {Sato, Yosuke and Matsuo, Sadashige and Hsu, Chen-Hsuan and Stano, Peter and Ueda, Kento and Takeshige, Yuusuke and Kamata, Hiroshi and Lee, Joon Sue and Shojaei, Borzoyeh and Wickramasinghe, Kaushini and Shabani, Javad and Palmstr\o{}m, Chris and Tokura, Yasuhiro and Loss, Daniel and Tarucha, Seigo},
  journal = {Phys. Rev. B},
  volume = {99},
  issue = {15},
  pages = {155304},
  numpages = {14},
  year = {2019},
  month = {Apr},
  publisher = {American Physical Society},
  doi = {10.1103/PhysRevB.99.155304},
  url = {https://link.aps.org/doi/10.1103/PhysRevB.99.155304}
}

@book{BaronePaterno:1982,
  author    = {Barone, Antonio and Patern{\`o}, Gianfranco},
  title     = {Physics and Applications of the {Josephson} Effect},
  publisher = {John Wiley \& Sons},
  address   = {New York},
  year      = {1982},
  doi       = {10.1002/352760278X}
}

@article{DynesFulton:1971,
  author  = {Dynes, R. C. and Fulton, T. A.},
  title   = {Supercurrent Density Distribution in {Josephson} Junctions},
  journal = {Phys. Rev. B},
  volume  = {3},
  pages   = {3015--3023},
  year    = {1971},
  doi     = {10.1103/PhysRevB.3.3015}
}

@article{Buzdin:2005RMP,
  author  = {Buzdin, A. I.},
  title   = {Proximity effects in superconductor-ferromagnet heterostructures},
  journal = {Rev. Mod. Phys.},
  volume  = {77},
  pages   = {935--976},
  year    = {2005},
  doi     = {10.1103/RevModPhys.77.935}
}

@article{Bergeret:2005RMP,
  author  = {Bergeret, F. S. and Volkov, A. F. and Efetov, K. B.},
  title   = {Odd triplet superconductivity and related phenomena in superconductor-ferromagnet structures},
  journal = {Rev. Mod. Phys.},
  volume  = {77},
  pages   = {1321--1373},
  year    = {2005},
  doi     = {10.1103/RevModPhys.77.1321}
}

@article{Nakatsuji:2023PRX_TTG,
  title = {Multiscale Lattice Relaxation in General Twisted Trilayer Graphenes},
  author = {Nakatsuji, Naoto and Kawakami, Takuto and Koshino, Mikito},
  journal = {Phys. Rev. X},
  volume = {13},
  issue = {4},
  pages = {041007},
  numpages = {20},
  year = {2023},
  month = {Oct},
  publisher = {American Physical Society},
  doi = {10.1103/PhysRevX.13.041007},
  url = {https://link.aps.org/doi/10.1103/PhysRevX.13.041007}
}

@article{Gimarchi:2007ChargeCorrelation,
  title = {{Fourier transform of the $2{k}_{F}$ Luttinger liquid density correlation function with different spin and charge velocities}},
  author = {Iucci, An\'{\i}bal and Fiete, Gregory A. and Giamarchi, Thierry},
  journal = {Phys. Rev. B},
  volume = {75},
  issue = {20},
  pages = {205116},
  numpages = {7},
  year = {2007},
  month = {May},
  publisher = {American Physical Society},
  doi = {10.1103/PhysRevB.75.205116},
  url = {https://link.aps.org/doi/10.1103/PhysRevB.75.205116}
}

@article{Manna:2026WTe2TK,
title = {{Disorder-driven Weyl-Kondo semimetal phase in WTe$_2$}},
journal = {Mater. Today Quantum},
volume = {10},
pages = {100070},
year = {2026},
issn = {2950-2578},
doi = {https://doi.org/10.1016/j.mtquan.2026.100070},
url = {https://www.sciencedirect.com/science/article/pii/S2950257826000120},
author = {Arpan Manna and Sunit Das and Amit Agarwal and Soumik Mukhopadhyay}
}

@article{Nagao:2008Nanotechnology,
doi = {10.1088/0957-4484/19/35/355204},
url = {https://doi.org/10.1088/0957-4484/19/35/355204},
year = {2008},
month = {jul},
publisher = {},
volume = {19},
number = {35},
pages = {355204},
author = {Liu, Canhua and Inaoka, Takeshi and Yaginuma, Shin and Nakayama, Tomonobu and Aono, Masakazu and Nagao, Tadaaki},
title = {{The excitation of one-dimensional plasmons in Si and Au–Si complex atom wires}},
journal = {Nanotechnology}
}

@inproceedings{Voit:2000AIPConf,
  author    = {Johannes Voit},
  title     = {A Brief Introduction to {Luttinger} Liquids},
  booktitle = {AIP Conf. Proc.},
  volume    = {544},
  pages     = {309--318},
  year      = {2000},
  publisher = {American Institute of Physics},
  doi       = {10.1063/1.1342524}
}

@article{Fink:1998PRL_scSeparation,
  title = {{Manifestation of Spin-Charge Separation in the Dynamic Dielectric Response of One-Dimensional ${\mathrm{Sr}}_{2}\mathrm{Cu}{O}_{3}$}},
  author = {Neudert, R. and Knupfer, M. and Golden, M. S. and Fink, J. and Stephan, W. and Penc, K. and Motoyama, N. and Eisaki, H. and Uchida, S.},
  journal = {Phys. Rev. Lett.},
  volume = {81},
  issue = {3},
  pages = {657--660},
  numpages = {0},
  year = {1998},
  month = {Jul},
  publisher = {American Physical Society},
  doi = {10.1103/PhysRevLett.81.657},
  url = {https://link.aps.org/doi/10.1103/PhysRevLett.81.657}
}

@article{Schulz:2020PRB,
  title = {{Decoherence of charge density waves in beam splitters for interacting quantum wires}},
  author = {Schulz, Andreas and Schneider, Imke and Anglin, James},
  journal = {Phys. Rev. B},
  volume = {101},
  issue = {23},
  pages = {235136},
  numpages = {22},
  year = {2020},
  month = {Jun},
  publisher = {American Physical Society},
  doi = {10.1103/PhysRevB.101.235136},
  url = {https://link.aps.org/doi/10.1103/PhysRevB.101.235136}
}

@article{KunYang:2017PRB,
  title = {{Coulomb interaction driven instabilities of sliding Luttinger liquids}},
  author = {Sur, Shouvik and Yang, Kun},
  journal = {Phys. Rev. B},
  volume = {96},
  issue = {7},
  pages = {075131},
  numpages = {9},
  year = {2017},
  month = {Aug},
  publisher = {American Physical Society},
  doi = {10.1103/PhysRevB.96.075131},
  url = {https://link.aps.org/doi/10.1103/PhysRevB.96.075131}
}

@article{Elio:2020PRBNV,
  title = {{Spin magnetometry as a probe of stripe superconductivity in twisted bilayer graphene}},
  author = {K\"onig, E. J. and Coleman, Piers and Tsvelik, A. M.},
  journal = {Phys. Rev. B},
  volume = {102},
  issue = {10},
  pages = {104514},
  numpages = {12},
  year = {2020},
  month = {Sep},
  publisher = {American Physical Society},
  doi = {10.1103/PhysRevB.102.104514},
  url = {https://link.aps.org/doi/10.1103/PhysRevB.102.104514}
}

@article{Fominykh:2026FedopedWTe2,
    author = {Fominykh, B. M. and Perevalova, A. N. and Irkhin, V. Yu. and Naumov, S. V. and Stepanov, A. and Marchenkova, E. B. and Marchenkov, V. V.},
    title = {{Signatures of the Kondo effect in Fe-diluted tungsten ditelluride single crystal}},
    journal = {J. Appl. Phys.},
    volume = {139},
    number = {21},
    pages = {215104},
    year = {2026},
    month = {06},
    issn = {0021-8979},
    doi = {10.1063/5.0326413},
    url = {https://doi.org/10.1063/5.0326413}
}

@article{Jia:2022HLL,
  author  = {Jia, Junxiang and Marcellina, Elizabeth and Das, Anirban and Lodge, Michael S. and Wang, BaoKai and Ho, Duc-Quan and Biswas, Riddhi and Pham, Tuan Anh and Tao, Wei and Huang, Cheng-Yi and Lin, Hsin and Bansil, Arun and Mukherjee, Shantanu and Weber, Bent},
  title   = {{Tuning the many-body interactions in a helical Luttinger liquid}},
  journal = {Nat. Commun.},
  volume  = {13},
  pages   = {6046},
  year    = {2022},
  doi     = {10.1038/s41467-022-33676-0}
}

@article{Bena:2012PRB,
  title = {{Spectral properties of Luttinger liquids: A comparative analysis of regular, helical, and spiral Luttinger liquids}},
  author = {Braunecker, Bernd and Bena, Cristina and Simon, Pascal},
  journal = {Phys. Rev. B},
  volume = {85},
  issue = {3},
  pages = {035136},
  numpages = {16},
  year = {2012},
  month = {Jan},
  publisher = {American Physical Society},
  doi = {10.1103/PhysRevB.85.035136},
  url = {https://link.aps.org/doi/10.1103/PhysRevB.85.035136}
}

@article{Jongjun:2020PRL,
  title = {{Stable Flatbands, Topology, and Superconductivity of Magic Honeycomb Networks}},
  author = {Lee, Jongjun M. and Geng, Chenhua and Park, Jae Whan and Oshikawa, Masaki and Lee, Sung-Sik and Yeom, Han Woong and Cho, Gil Young},
  journal = {Phys. Rev. Lett.},
  volume = {124},
  issue = {13},
  pages = {137002},
  numpages = {7},
  year = {2020},
  month = {Apr},
  publisher = {American Physical Society},
  doi = {10.1103/PhysRevLett.124.137002},
  url = {https://link.aps.org/doi/10.1103/PhysRevLett.124.137002}
}

@article{Jongjun:2021PRL,
  title = {{Non-Fermi Liquids in Conducting Two-Dimensional Networks}},
  author = {Lee, Jongjun M. and Oshikawa, Masaki and Cho, Gil Young},
  journal = {Phys. Rev. Lett.},
  volume = {126},
  issue = {18},
  pages = {186601},
  numpages = {7},
  year = {2021},
  month = {May},
  publisher = {American Physical Society},
  doi = {10.1103/PhysRevLett.126.186601},
  url = {https://link.aps.org/doi/10.1103/PhysRevLett.126.186601}
}

@misc{zenodo_data,
    note = {Y.-Y. Chang and C.-H. Hsu (2026). Data for research on ``Carrier-mediated spin helices in low-dimensional systems: Josephson-interference features and correlation renormalization'',  Zenodo. \url{https://doi.org/10.5281/zenodo.22872659}.}  
}

\end{document}